\documentclass[a4paper]{article}

\usepackage{amsfonts,amsmath}
\usepackage{latexsym,epic,graphicx}
\usepackage{amssymb}
\usepackage{enumerate}

\makeatletter
\renewcommand{\theequation}{\thesection.\arabic{equation}}
\@addtoreset{equation}{section}
\makeatother

\newcommand{\be}{\begin{eqnarray}}
\newcommand{\ee}{\end{eqnarray}}
\newcommand{\beg}{\begin{eqnarray*}}
\newcommand{\eeg}{\end{eqnarray*}}

\newcommand{\aWeylN}{\mathcal{D}^{\rho}}
\newcommand{\Weyl}{\mathcal{W}}
\newcommand{\aWeyl}{\mathcal{W}^{\rho}}
\newcommand{\DWeylN}{\mathcal{D}^{\textit{s}}}
\newcommand{\DWeyl}{\mathcal{W}^{\textit{s}}}
\newcommand{\SWeylN}{\mathcal{D}^{\textit{vs}}}
\newcommand{\SWeyl}{\mathcal{W}^{\textit{vs}}}
\newcommand{\cN}{\mathcal{N}}
\newcommand{\cL}{\mathcal{L}}
\newcommand{\cV}{\mathcal{V}}
\newcommand{\cP}{\mathcal{P}}
\newcommand{\cQ}{\mathcal{Q}}
\newcommand{\cR}{\mathcal{R}}

\newcommand{\qed}{\hfill $\square$}

\newcommand{\nn}{\nonumber}

\renewcommand{\phi}{\varphi}
\newcommand{\e}{\varepsilon}
\newcommand{\ep}{\epsilon}
\newcommand{\p}{\partial}
\newcommand{\G}{\Gamma}
\newcommand{\g}{\gamma}

\newcommand{\Z}{\mathbb{Z}}
\newcommand{\R}{\mathbb{R}} 
\newcommand{\N}{\mathbb{N}} 
 
\newcommand{\Oc}{\mathbb{O}}

\newcommand{\id}{\mathbf{1\hspace{-2.9pt}l}}
\begin{document}
{\flushright IHES/P/09/31\\[9mm]}

\renewcommand{\thefootnote}{\fnsymbol{footnote}}

\begin{center}
{\LARGE \bf Fermionic Kac-Moody Billiards and Supergravity}\\[1cm]
Thibault Damour\footnote[1]{E-mail: \tt damour@ihes.fr} \\
Christian Hillmann\footnote[3]{E-mail: \tt hillmann@ihes.fr} 
\\[5mm]
{\sl  Institut des Hautes \'Etudes Scientifiques\\
           35, Route de Chartres\\
      91440 Bures-sur-Yvette, France} \\[15mm]

\begin{tabular}{p{12cm}}
\hline\\\hspace{5mm}{\bf Abstract:}
We study the ``fermionic billiards'', i.e. the chaotic dynamics of the gravitino, that arise in the near-spacelike-singularity limit of eleven-dimensional supergravity and of its dimensional truncations (notably four-dimensional simple supergravity). By exploiting the gravity-coset correspondence, we show that the billiard dynamics of the gravitino is described by a `spin extension' of the Weyl group of the hyperbolic Kac--Moody algebra $E_{10}$. This spin extension is a discrete subgroup of (a spin covering of) the maximal compact subgroup $K(E_{10})$ of $E_{10}$ that is generated by ten (simple-root-related) idempotent elements of order 8. The `super-billiard' that combines the bosonic and fermionic billiards is found to have a remarkably simple structure, which exhibits a striking analogy with a polarized photon propagating in the ten-dimensional Lorentzian Weyl chamber of $E_{10}$.
\\\\\hline
\end{tabular}\\[9mm]
\end{center}

\renewcommand{\thefootnote}{\arabic{footnote}}
\setcounter{footnote}{0}

\begin{section}{Introduction}
Some time ago, the study \`a la Belinski, Khalatnikov, Lifshitz (BKL) \cite{BKL} of the chaotic behaviour \cite{DH00b} of the general solution of the \textit{bosonic} sector of $11$-dimensional supergravity (SUGRA$_{11}$) near a spacelike singularity revealed a hidden connection with the hyperbolic Kac--Moody algebra $E_{10}$ \cite{DH00}. More precisely, it was found that, in the near-singularity limit, most bosonic degrees of freedom ``freeze'' (i.e. have some finite limit), except for the $10$ `diagonal' components of the spatial metric  (parametrized by their logarithms; $\beta^a\sim -\ln g_{aa}$) which undergo, at each spatial point, a chaotic \textit{billiard motion} within a conical polyhedron in an auxiliary $10$-dimensional Lorentzian space, which can be identified with the Weyl chamber of $E_{10}$. Further work has shown that this $E_{10}$-related billiard motion was the lowest approximation (`height 1') of a hidden correspondence, which was checked up to height $29$, between the SUGRA$_{11}$ dynamics and the dynamics of a massless particle on the infinite dimensional coset space $E_{10}/K(E_{10})$ \cite{DHN02b}. [Here, $K(E_{10})$ denotes the maximal compact subgroup of $E_{10}$, see below.] More evidence for the presence of a hidden $E_{10}$ symmetry in SUGRA$_{11}$ came from the consideration of the fermionic sector of supergravity which could also be put into correspondence (up to some approximation roughly equivalent to the height $29$ of the bosonic sector) with a fermionic extension of the $E_{10}/K(E_{10})$ coset model \cite{DKN05,BHP05b,DKN06}. For more work on the correspondences between gravity theories and Lorentzian Kac--Moody algebras see \cite{DN04,DN05,DKN07,dBD07,DHJN01}. For a recent review, see \cite{HPS07}. For other works suggesting a hidden role of Lorentzian Kac--Moody algebras in supergravity or string theory, see \cite{DHN02,BGGH04,W01,CV08}.\\

The aim of the present paper is to study the \textit{fermionic} side of the billiard dynamics taking place in the near-singularity limit. Instead of testing the conjectural gravity-coset correspondence to higher levels of approximation, we shall here go back to the lowest level of approximation (`height 1') and consider the Kac--Moody-related mathematical structure behind the billiard dynamics of the SUGRA$_{11}$-gravitino. We shall find that this structure exhibits a remarkably simple interplay between $E_{10}$ (hyperbolic) and $K(E_{10})$ (elliptic) aspects.
\\

In this work, we shall study the fermionic side of cosmological billiards within the standard algebraic framework where the `classical' supergravity dynamics makes mathematical sense. Namely, we assume that the fermionic variables $(\psi)$ take values in the odd-part of a Grassmann algebra, while the bosonic variables $(g,A)$ take values in the even part of the same algebra. Then, with respect to the grading defined by a basis in the underlying Grassmann algebra, one has $\psi=\psi_1+\psi_3+\dots$, while $g=g_0+g_2+\dots$, $A=A_0+A_2+\dots$. The `body' $(g_0,A_0)$ then satisfies the Einstein-3-form equations without any `back-reaction' of $\psi$ (which only enters the `soul' parts $g_2,A_2,\dots$), while the lowest component $\psi_1$ of $\psi$ follows a dynamics driven by $g_0$ and $A_0$. We are aware that \cite{BK76,BHP05} have shown that, in the case of an Einstein--Dirac system, the replacement of the $\mathcal{O}(\psi^2)$ source terms for $g_2$ by classical VEV's (of grading zero instead of two) strongly modifies the bosonic dynamics (then taking entirely place at grading zero), and, in particular, destroys the chaotic nature of the near-singularity limit. However, we do not think that such a replacement is mathemetically or physically justified. Pending a truly quantum analysis (in Hilbert space) of the back reaction of fermions, we think our analysis below is mathematically sound.\\

We shall also consider the truncation of this fermionic billiard to $4$-dimensional $\cN=1$ supergravity (SUGRA$_4$). In this case, the relevant (hidden) hyperbolic Kac--Moody symmetry is $AE_3$ (or $A_1^{++}$, the hyperbolic extension of $A_1\simeq sl_2$), and the relevant coset model involves $AE_3/K(AE_3)$ \cite{DHN02}. In both cases ($E_{10}/K(E_{10})$ and $AE_3/K(AE_3)$), we shall find that the `fermionic billiards' exhibit a remarkable factorized structure involving both Lorentzian ($SO(9,1)$ or $SO(2,1)$) and Euclidean ($SO(10)$ or $SO(3)$) structures. These results may suggest new ways of thinking about the fermionic extension of the $E_{10}/K(E_{10})$ coset model. Some of our results may also be of interest for mathematics as they involve what we shall call ``spin extensions of the Weyl groups'' of hyperbolic Kac--Moody algebras.
\end{section}

\begin{section}{Cosmological billiards and the conjectured gravity/coset correspondence}
In this section, we shall present a brief survey of the so-called ``cosmological billiards'' \cite{DHN02} and describe some elements of the bosonic side of the gravity-coset correspondence. This will allow us to fix the conventions and notations used in this article.

\begin{subsection}{Maximal supergravity in the BKL-limit}\label{maxBKL}
We start with a short review of the billiard limit of the bosonic sector of $D=11$ supergravity.\footnote{See Appendix \ref{appendix0} for our conventions in the Cremmer-Julia-Scherk action \cite{CJS78}.} In other words, we consider the behaviour of a general solution close to a spacelike singularity in the leading BKL-like `gradient approximation', where one keeps only time derivatives and the dominant terms involving spatial gradients. We will adopt a space-time slicing such that the singularity ``occurs'' on the coordinate time slice $t=+\infty$. This slicing is built by using pseudo-Gaussian coordinates defined by a vanishing shift leading to a metric of the form
	\be\label{metric11}
	ds^2 &=& -N^2(t,x) dt^2 + \sum\limits_{m,n=1}^{10} g_{mn}(t,x)\xi^m(x) \xi^n(x)
	.
	\ee
	Here, $\xi^m(x)=\sum_{i=1}^{10}\xi^m_i(x)dx^i$ could be an arbitrary spatial frame.	For simplicity, we will use a coordinate frame $\xi^m(x)=dx^m$. Note that, in order to avoid ambiguities in later formul\ae{} that involve repeated indices but no sum, we shall often dispense with the Einstein summation convention and explicitly indicate the needed sums. Furthermore, we choose the time coordinate\footnote{The proper time $T$ is related to $t$ by $dT=N(t,x)dt$ at each spatial point $x$ \cite{DHN02}.} $t$ so that the lapse function is linked to the spatial volume density via
	\be\label{lapse}
	N(t,x)&=&\sqrt{\det{g_{mn}}(t,x)}
	\ee
	and we adopt a generalized temporal gauge for the three-form potential, i.e. $A_{0mn}=0$. Without loss of generality, we describe the independent degrees of freedom of the spatial metric $g_{mn}$ by an Iwasawa decomposition of the associated vielbein $e^a_m$ ($g_{mn}=\sum_{a=1}^{10}e^a_m e^a_n$), 
	\be\label{vielbein}
	e_m^a&=:&e^{-\beta^a}\cN^a{}_m\\
	\label{Iwasawa}
	\text{implying}\quad g_{mn} &=& \sum\limits_{a=1}^{10}e^{-2\beta^a}\cN^a{}_m\cN^a{}_n.
	\ee
Here the $\cN^a{}_m$'s are upper triangular matrices which have $1$ as diagonal entries. The logarithmic scale factors $\beta^a$ describing the `diagonal' components of the metric (see Eq. (\ref{Iwasawa})) play a crucial role in the `near-spacelike-singularity limit' (which corresponds to $T\rightarrow 0$, $t\rightarrow +\infty$ and $\sum_{a=1}^{10}\beta^a\rightarrow +\infty$, so that the local volume $\sqrt{\det{g_{mn}}}=e^{-\sum_{a=1}^{10}\beta^a}$ collapses at each spatial point). This limit shall also be referred to as the `BKL-limit' in the following. Let us first recall that the part of the supergravity Lagrangian that only depends on the time derivatives $\dot{\beta}:=\frac{\p \beta}{ \p t}$ of the scale factors $\beta$ reads
\be\label{supermetric}
	\cL_{\dot{\beta}} &=& \sum\limits_{a=1}^{10} \Big(\dot{\beta}^a\Big)^2 -\Big(\sum\limits_{a=1}^{10} \dot{\beta}^a\Big)^2\,=:\, \sum\limits_{a,b=1}^{10} G_{ab}\dot{\beta}^a \dot{\beta}^b.
	\ee
This defines a flat Lorentzian metric $G$ in the $d=D-1=10$ dimensional space of scale factors $\beta^a$ with signature $(-+\cdots +)$. This Lorentzian metric $G$ plays a crucial role in the gravity-Kac--Moody-coset correspondence.\footnote{$G$ is part of the deWitt supermetric.} It will also be of prime importance for the discussion of the gravitino in section \ref{Ferm}. If the dynamics were completely described by the Lagrangian $\cL_{\dot{\beta}}$ (\ref{supermetric}), the scale factors $\beta$ would follow a geodesic in the flat Lorentzian space  $(\R^{10},G)$ (or $\beta$-space):
\be\label{Kasner}
\beta^a&=& v^a t +\beta^a_0.
\ee
In addition, the leading $\dot{\beta}$-terms in the Hamiltonian constraint impose the condition
\be\label{null}
 0&=& \sum_{a,b=1}^{10} G_{ab}v^av^b
\ee
which means that the geodesic (\ref{Kasner}) is a null geodesic.\\

The null geodesic (\ref{Kasner},\,\ref{null}) is the $\beta$-space description of a Kasner solution. However, such a Kasner-like solution is profoundly altered when one takes into account the additional contributions to the Lagrangian (beyond (\ref{supermetric})) coming from the off-diagonal components $\cN^a{}_m$ (\ref{Iwasawa}) of the metric, the electric and magnetic energy of the three-form $A_{mnp}(t,x)$, and from the spatial gradients of the metric. The detailed analysis in \cite{DHN02} shows, after having changed to the Hamiltonian formalism, that all the other degrees of freedom of supergravity result in adding to $\cL_{\dot{\beta}}$ (\ref{supermetric}) several potential densities $\cV_A$ (labelled by an index $A$) which have an exponential dependence upon the $\beta$'s:
\be\label{pot}
\cV_A&=&c_A e^{-2\alpha_A(\beta)}.
\ee
Here, the coefficients $c_A$ depend on other degrees of freedom ($\cN_m^a$, $A_{mnp}$,$\dots$ and their conjugate momenta) and the $\alpha_A(\beta)$ are linear functionals $\alpha_A(\beta)=\sum_{a=1}^{10}\alpha_{A,\,a}\beta^a$. The sum of (\ref{supermetric}) and (\ref{pot}) leads to a kind of Toda model for the dynamics of the $\beta$'s. In the near-singularity limit $\sum_{a=1}^{10}\beta^a\rightarrow +\infty$, one can order the various potentials in terms of their importance for altering the monotonic zeroth-order geodesic Kasner motion (\ref{Kasner}) into a chaotic dynamics. There are $10$ \textit{dominant} Toda terms with exponents $\alpha_i(\beta)$ with $i=0,\dots,9$ (see Eq. (\ref{dominant}) below), which all have positive coefficients $c_i$. The ten terms $\propto e^{-2\alpha_i(\beta)}$ conventionally define the `height 1' in an approximation scheme where the higher order terms are made of various products of the ten dominant terms (e.g. $e^{-2\alpha_1}e^{-2\alpha_2}=e^{-2(\alpha_1+\alpha_2)}$ is of `height 2', $e^{-2(\alpha_1+\alpha_2+\alpha_3)}$ of `height 3', etc.).\\

Nine of the dominant Toda terms come from terms (\ref{pot}) in the Hamiltonian where the coefficients are quadratic in the conjugate momenta of $\cN_m^a$ (\ref{Iwasawa}). They are called `symmetry walls'. The choice of parametrizing the spacetime metric $g$ by upper-triangular matrices $\cN$ (\ref{Iwasawa}) implies that the functionals $\alpha_A$ corresponding to symmetry walls have the form \cite{DHN02}
\be\label{symmW}
\alpha^{\text{\textit{s}}}_{(ab)}(\beta) &=& \beta^b-\beta^a\quad \text{with }a<b=1,\dots,10.
\ee
Among the $(10\times 9)/2=45$ symmetry walls, the dominant ones are
\begin{subequations}\label{dominant}
\be\label{dominant1}
\alpha_{i}\,:=\,\alpha^{\text{\textit{s}}}_{(ii+1)}&=& \beta^{i+1}-\beta^{i}\quad\text{for } i\,=\,1,\dots,9.
\ee
The tenth dominant Toda term comes from the potential (\ref{pot}) in the Hamiltonian which arises from the `electric' energy density of the $3$-form $A_{mnp}(t,x)$. The antisymmetry of $A_{mnp}$ implies that the corresponding `electric walls' have the form
\beg
\alpha^{\text{\textit{el}}}_{abc}(\beta)=\beta^a+\beta^b+\beta^c
\eeg
with $a,b,c\in \{1,\dots,10\}$ being all different. Among these, the dominant `electric wall' is
\be\label{electric}
\alpha_{0}(\beta)\,:=\,\alpha^{\text{\textit{el}}}_{123}(\beta)&=&\beta^1+\beta^2+\beta^3.
\ee
\end{subequations}
All the other potential terms, which include the subdominant symmetry and electric walls, the `magnetic walls' $\alpha^{\text{\textit{mag}}}_{a_1\dots a_6}(\beta)=\beta^{a_1}+\dots+\beta^{a_6}$, and the `gravitational walls' $\alpha^{(\text{\textit{g}})}$ (see e.g. (\ref{gravwalls}) below) can be written as linear combinations with positive integer coefficients of the $10$ dominant walls $\alpha_i$ (\ref{dominant}). It has been shown in \cite{DHN02b,DN04} that, up to height $29$ in an expansion in dominant walls, the SUGRA$_{11}$-dynamics (for all the bosonic degrees of freedom) following from the sum of (\ref{supermetric}) and all the other terms (\ref{pot}) can be identified with the geodesic dynamics of a particle on the infinite-dimensional coset space $E_{10}/K(E_{10})$ (which can be written as the sum of (\ref{supermetric}) and of an infinite number of potential terms (\ref{pot}) where the linear forms $\alpha_A(\beta)$ now label the positive roots of $E_{10}$).\\

In the following, we shall focus on the `height 1' approximation (which is common to SUGRA$_{11}$ and to the $E_{10}/K(E_{10})$ coset) that consists of keeping only the $10$ dominant Toda walls, i.e. of adding to (\ref{supermetric}) a potential term of the form (with $c_i>0$)
\be\label{height1}
\cV_1&=&\sum_{i=0}^9 c_i e^{-2\alpha_i(\beta)}.
\ee
One can even further approximate the $10$ Toda walls (\ref{height1}) in the BKL-limit $\sum_{a=1}^{10}\beta^a\rightarrow +\infty$ by a sum of `sharp walls'
\be\label{sharppot}
\cV_{\text{sharp}}&=&\sum_{i=0}^9 \Theta[-2\alpha_i(\beta)],
\ee
where $\Theta[x]$ denotes an `infinite step function' $\Theta[x]=0$ for $x<0$ and $\Theta[x]=+\infty$ for $x>0$. The `sharp potential' (\ref{sharppot}) has the effect of modifying the monotonic geodesic-Kasner solution  (\ref{Kasner}) of the free action (\ref{supermetric}) into a ``zigzag'' of straight line segments, of the form (\ref{Kasner}), interrupted by `collisions' with the $10$ sharp dominant walls 
\be\label{hyper}
\alpha_i(\beta)&=&0\quad \text{with }i\,=\,0,\dots,9.
\ee
Thus, the motion of the scale factors is restricted to the polywedge $\alpha_i(\beta)\geq 0$ for $i=0,\dots,9$. In addition, the instantaneous `velocity' $v^a=\dot{\beta}^a$ (\ref{Kasner}) of the $\beta$-motion is still constrained by the BKL-limit of the Hamiltonian constraint to satisfy Eq. (\ref{null}), i.e. to be lightlike. The effect of each collision on a particular wall, say $\alpha_i$ (\ref{hyper})\footnote{All the dominant walls are found to be timelike, i.e. to have spacelike gradients \cite{DHN02}.} is to transform the incoming Kasner velocity $v$ (\ref{Kasner}) into the outgoing one $v'{}=r_{i}(v)$ by the usual formula for a geometric reflection $r_{i}$ in the hyperplane (\ref{hyper}) in $\beta$-space $(\R^{10},G)$ \cite{DKN06}:
\be\label{Reflection}
r_{i}(v) &=& v -\frac{2\alpha_i(v)}{(\alpha_i|\alpha_i)}\alpha_i^{\#}.
\ee
Here $\alpha_i^{\#}$ denotes the contravariant version of the covariant vector (or linear form) $\alpha_i(\beta)=\sum_{[a=1}^{10}\alpha_{i,\,a}\beta^a$, i.e. 
\be\label{coroot}
\alpha_i^{\#\, a}&:=&\sum_{b=1}^{10}G^{ab}\alpha_{i,\,b}
\ee
where $G^{ab}$ denotes the inverse of the basic $\beta$-space metric (\ref{supermetric}). The round brackets $(\cdot|\cdot)$ appearing in the denominator in Eq. (\ref{Reflection}) denote the scalar product defined by the metric $G_{ab}$, or its inverse (depending on whether one considers covariant or contravariant vectors). For instance, $(\alpha_i|\alpha_j)=\sum_{a,b=1}^{10}G^{ab}\alpha_{i\,,a} \alpha_{j\,,b}$, which is equal to $(\alpha_i^{\#}|\alpha_j^{\#})=\sum_{a=1}^{10}G_{ab}\alpha_i^{\#\, a}\alpha_{j}^{\#\,b}$ as well as to $\alpha_{i}(\alpha_j^{\#})=\sum_{a=1}^{10}\alpha_{i\,,a}\alpha_j^{\#\, a}$.\footnote{Note that the reflection (\ref{Reflection}) preserves the null constraint (\ref{Kasner}) on the velocities $v^a$.} \\

From the scalar products between the dominant walls $\alpha_{i}$ (\ref{dominant}), one defines the following matrix:
\be\label{cartan}
A_{ij}&:=&
\frac{2(\alpha_{i}|\alpha_{j})}{(\alpha_{i}|\alpha_{i})}.
\ee
It is found that this matrix is integer valued: with $A_{ii}=2$ and $-A_{ij}\in \N$ when $i\neq j$. In addition, for SUGRA$_{11}$ (i.e. $E_{10}$) as well as for SUGRA$_4$ (i.e. $AE_3$), all the dominant walls have the same `squared length' $(\alpha_{i}|\alpha_{i})=2$, so that $A_{ij}$ is \textit{symmetric}. One can associate to $A$ a Dynkin diagram by drawing a node for the $10$ dominant walls $\alpha_{i}$ (\ref{dominant}) and link them by $-A_{ij}$ lines, which results in:
\begin{center}
\begin{tabular}{cl}
$E_{10}$
&
\scalebox{.7}{
\begin{picture}(280,50)(0,0)

\put(245,5){ \circle*{10}}

\put(244,-8){$\alpha_{9}$}

\put(215,5){ \circle*{10}}
\put(245,5){\line(-1,0){25}}
\put(214,-8){$\alpha_{8}$}

\put(185,5){ \circle*{10}}
\put(215,5){\line(-1,0){25}}
\put(184,-8){$\alpha_{7}$}

\put(155,5){ \circle*{10}}
\put(185,5){\line(-1,0){25}}
\put(154,-8){$\alpha_{6}$}

\put(125,5){ \circle*{10}}
\put(155,5){\line(-1,0){25}}
\put(124,-8){$\alpha_{5}$}

\put(95,5){ \circle*{10}}
\put(125,5){\line(-1,0){25}}
\put(94,-8){$\alpha_{4}$}

\put(65,5){ \circle*{10}}
\put(95,5){\line(-1,0){25}}
\put(64,-8){$\alpha_{3}$}

\put(35,5){ \circle*{10}}
\put(65,5){\line(-1,0){25}}
\put(34,-8){$\alpha_{2}$}

\put(5,5){ \circle*{10}}
\put(35,5){\line(-1,0){25}}
\put(4,-8){$\alpha_{1}$}

\put(65,35){ \circle*{10}}
\put(68,35){\line(0,-1){25}}
\put(79,35){$\alpha_{0}$}

\end{picture}
}
\end{tabular}\\
\vspace{6.0pt}
\small{figure 1}
\end{center}
This is the Dynkin diagram of the hyperbolic Kac--Moody algebra $E_{10}$ \cite{DH00}. The matrix $A$ (\ref{cartan}) is the corresponding Cartan matrix \cite{K95} of $E_{10}$ for the present case of $D=11$ supergravity. In addition, the restriction of the $E_{10}$ invariant metric to the Cartan subalgebra coincides (when using suitable coordinates, also denoted $\beta^a$ in the Cartan subalgebra) with the basic $\beta$-space metric $G_{ab}$ introduced in Eq. (\ref{supermetric}).\\

Summarizing so far: the SUGRA$_{11}$ billiard picture in $\beta$-space, arising as leading approximation in the sharp wall BKL-limit, can be identified with a `Kac--Moody billiard' via a dictionary that maps: 
\begin{enumerate}
	\item the ten gravity scale factors to coordinates $\beta^a$ in the Cartan subalgebra of $E_{10}$
	\item the ten SUGRA$_{11}$ dominant walls $\alpha_i$ to the ten simple roots of $E_{10}$.
\end{enumerate}
In other words, the billiard table, defined by the ten inequalities $\alpha_i(\beta)\geq 0$ is identified with the Weyl chamber of $E_{10}$, and the bosonic billiard dynamics, comprising successive reflections (\ref{Reflection}) in the dominant walls, is mapped into a product of Weyl reflections $r_{i_k}$, i.e. an element $w$ of the Weyl group $\Weyl_{E_{10}}$ of $E_{10}$, say
\be\label{wb}
w&=&r_{i_1}r_{i_2}\cdots r_{i_n}\cdots
\ee
where the length of the Weyl `word' $w\in \mathcal{W}_{E_{10}}$ (describing the billiard dynamics towards the singularity) grows indefinitely as $t\rightarrow+\infty$. It is this aspect of the bosonic gravity-coset correspondence that we will generalize to the fermionic sector in section \ref{Ferm}.
\end{subsection}

\begin{subsection}{Truncating SUGRA$_{11}$ to SUGRA$_4$, and correlatively reducing $E_{10}$ to $AE_3$}\label{subsystem}
Before doing so, let us explain a result that we shall need below: how the inclusion of $\cN=1$ SUGRA$_{4}$ within SUGRA$_{11}$ corresponds to an embedding of $AE_3$ within $E_{10}$. Here, we shall indicate how a sub-billiard of the (full) SUGRA$_{11}$ billiard gives rise to the Cartan matrix of $AE_3$ in the BKL-limit.\\

To obtain $\cN=1$ supergravity in $D=4$ from SUGRA$_{11}$, we have to perform two steps. First, we reduce the theory down to $4$ dimensions by compactifying on a flat seven torus $T^7$. Second, we truncate the degrees of freedom of SUGRA$_{11}$ down to those of $\cN=1$ SUGRA$_{4}$. In other words, we discard the three-form and in our pseudo-Gaussian gauge (\ref{metric11}), we can restrict the indices $m,n$ to the values $1,2,3$ only. The $\beta$-space now becomes $3$-dimensional, and endowed with a `reduced' Lorentzian metric $G_r$ (of signature ($-++$)) defined by restricting the sums in Eq. (\ref{supermetric}) to $a,b=1,2,3$. Among the $9$ symmetry walls (\ref{dominant1}), only two survive: $\alpha_{1}:=\alpha_{(12)}^{\text{\textit{s}}}$ and $\alpha_{2}:=\alpha_{(23)}^{\text{\textit{s}}}$. The previously tenth dominant wall (\ref{electric}) disappears due to the truncation of the $3$-form. However, we must now take into account the gravitational walls, which we could neglect above, because they were hidden `behind' other walls (i.e. subdominant), but which now will affect the dynamics of the $\beta$-motion near a singularity.\\

In $D=11$ supergravity, the gravitational walls are given by \cite{DHN02}
\be\label{gravwalls}
\alpha_{cde}^{\text{\textit{g}}}(\beta)&=& \beta^c- \beta^d-\beta^e+\sum\limits_{a=1}^{10}\beta^a
\ee 
with $c,d,e\in\{1,\dots,10\}$ being all different. They are linked to the structure functions $C^c{}_{de}$ ($d\theta_{\text{Iwa}}^c=\frac12 C^c{}_{de}\theta_{\text{Iwa}}^d\wedge\theta_{\text{Iwa}}^e$) of the Iwasawa frame $\theta_{\text{Iwa}}^c:=\cN^c{}_m dx^m$. The dimensional reduction on $T^7$ to $D=3+1$ dimensions with the gauge fixing of the metric as in (\ref{metric11}) implies that only the three spatial one forms $\theta_{\text{Iwa}}^1,\theta_{\text{Iwa}}^2,\theta_{\text{Iwa}}^3$ have a non-trivial coordinate dependence. Therefore, all structure functions $C^c{}_{de}$ apart from the ones with $c,d,e\in\{1,2,3\}$ are zero. This entails that only the gravitational walls (\ref{gravwalls}) with these indices are present, because the corresponding potential (\ref{pot}) vanishes for zero structure functions $C^c{}_{de}$ \cite{DHN02}. Among the surviving gravitational walls, one easily finds (making use of the dominant symmetry walls $\alpha_{(12)}^{\text{\textit{s}}}(\beta)=\beta^2-\beta^1>0$ and $\alpha_{(23)}^{\text{\textit{s}}}(\beta)=\beta^3-\beta^2>0$) that the dominant surviving gravitational wall is:\footnote{Note that this dominant surviving gravitational wall in $D=4$ differs from the gravitational wall that would dominate the other gravitational walls in a direct analysis of SUGRA$_{11}$, namely $\alpha_{1\,9\,10}^{(g)}$.}
\be
\alpha_{*}:=\alpha_{123}^{\text{\textit{g}}}=2\beta^1.
\ee
Then following the prescription presented in the previous section, one computes the scalar products between the three dominant walls $\alpha_{(12)}^{\text{\textit{s}}},\,\alpha_{(23)}^{\text{\textit{s}}},\,\alpha_{123}^{\text{\textit{g}}}$, with respect to the inverse of the reduced metric $G_r$, thus obtaining a $3\times 3$ matrix $A_{ij}$. One finds that this matrix is the Cartan matrix of $AE_3$, whose Dynkin diagram is:
\begin{center}
\begin{tabular}{cl}
&
\scalebox{.5}{
\begin{picture}(90,40)(0,0)

\put(94,5){ \circle*{15}}

\put(52,5){ \circle*{15}}
\put(60,5){\line(1,0){35}}

\put(50,5){\line(-1,1){10}}
\put(50,5){\line(-1,-1){10}}
\put(18,5){\line(1,1){10}}
\put(18,5){\line(1,-1){10}}

\put(20,3){\line(1,0){35}}
\put(20,7){\line(1,0){35}}

\put(10,5){ \circle*{15}}

\end{picture}
}
\\
&\scalebox{.7}{
\begin{picture}(90,10)(0,0)

\put(64,8){$\alpha_{2}$}
\put(34,8){$\alpha_{1}$}
\put(4,8){$\alpha_{*}$}
\end{picture}
}
\end{tabular}\\

\qquad\small{figure 2}
\end{center}

\end{subsection}

\begin{subsection}{Kac--Moody coset model and the role of the connection $\cQ$}\label{coset1}
After having briefly reviewed how the BKL-limit of the bosonic sector of supergravity gives rise to a billiard motion taking place in a Lorentzian polywedge, which can be identified with the Weyl chamber of a Kac--Moody algebra, let us review some of the elements of the definition of the $E_{10}/K(E_{10})$ (respectively $AE_3/K(AE_3)$) coset model which has been conjectured to capture the hidden $E_{10}$ (resp. $AE_3$) symmetry of SUGRA$_{11}$ (resp. SUGRA$_4$). First, we recall that the relevant Kac--Moody algebra is inductively constructed by its Chevalley--Serre presentation from the set of generators $(e_i,f_i,h_i)$ each corresponding to a simple root or to a node in the Dynkin diagrams of $E_{10}$ or $AE_{3}$. The elements $h_i$ span the Cartan subalgebra $\mathfrak{h}$ ($[h_i,h_j]=0$). The basic Lie brackets among $(e_i,f_i,h_i)$ are $[h_i,e_j]=A_{ij}e_j$, $[h_i,f_j]=-A_{ij}f_j$ and $[e_i,f_j]=\delta_{ij}h_i$ together with Jacobi identities and Serre relations \cite{K95}. The maximal compact subalgebra $K(\mathfrak{g})$ of an algebra $\mathfrak{g}$ is defined as the fixed point set of the Chevalley involution defined by $\omega(e_i)=-f_i$, $\omega(f_i)=-e_i$ and $\omega(h_i)=-h_i$.\footnote{Note that for $\mathfrak{g}=gl_d$, the automorphism $-\omega$ corresponds to the transposition in the standard matrix representation of $\textit{GL}(d)$ and hence the corresponding subgroup $K(G)=SO(d)$ indeed is the maximally compact subgroup of the group $G=\textit{GL}(d)$ in the topological sense.} Hence, $K(\mathfrak{g})$ is spanned by the elements 
\be\label{JDefi}
J_{\alpha,s}&:=&E_{\alpha,s} -E_{-\alpha,s},
\ee
where $E_{\alpha,s}$ is a general `raising operator', i.e. a multiple commutator of the simple ``raising operators'' $e_i$, and $E_{-\alpha,s}:=-\omega(E_{\alpha,s})$. Each general raising operator is labelled by: (1) a positive root\footnote{We recall that a root $\alpha$ is a linear form on Cartan space $\mathfrak{h}$ and that a positive root $\alpha$ is a sum $\sum_{i}n_i\alpha_i$ where $n_i\in \N$ and $\alpha_i$ is a simple root.} $\alpha\in \Delta^+$ (i.e. $[h,E_{\alpha,s}]=\alpha(h)E_{\alpha,s}$ for $h\in \mathfrak{h}$) and (2) a degeneracy index $s=1,\dots,\text{mult}(\alpha)$ labelling the different elements of $\mathfrak{g}$ having the same root $\alpha$. Due to the similarity of the coset descriptions for hyperbolic Kac--Moody algebras, in particular for $E_{10}$ and $AE_3$, we will for clarity focus on the former. It is straightforward to generalize the construction to other algebras, which will also be addressed in Appendix \ref{appendix1}.
\\

Assuming the existence of an Iwasawa decomposition, we parametrize the coset $\cV\in E_{10}/K(E_{10})$ for the Lie algebra $\mathfrak{g}=E_{10}$ by\footnote{We use the same symbol for groups and algebras. It should be clear from the context what is meant.}
\be\label{coset}
\cV&=& \exp\big({\sum\limits_{a=1}^{10}\beta^a(t)H_a}\big)
\exp\big({\sum\limits_{\alpha\in \Delta^+}\sum\limits_{s=1}^{\text{mult}(\alpha)}\nu_{\alpha,s}(t)E_{\alpha,s}}\big).
\ee
Here $H_a$ ($a=1,\dots,10$) is a general basis in the Cartan subalgebra $\mathfrak{h}$ that was identified with $\beta$-space and is hence endowed with the metric $G_{ab}$ (\ref{supermetric}) via $(H_a|H_b)=G_{ab}$. The link between the general basis $H_a$ of $\mathfrak{h}$ and the specific ten Cartan elements $h_i$ entering the Serre--Chevalley presentation involves the simple roots $\alpha_i$ (identified with the ten simple walls (\ref{symmW},\,\ref{electric}) via $h_i=\sum_{a=1}^{10}\check{\alpha}_{i}^{a} H_a$ where
\be\label{hNorm}
\check{\alpha}_i^a &:=& \frac{2}{(\alpha_i|\alpha_i)}\alpha_i^{\#\,a}\,\equiv\,\frac{2}{(\alpha_i|\alpha_i)}\sum\limits_{b=1}^{10}G^{ab}\alpha_{i\,b}.
\ee
Note that the `co-root' $\check{\alpha}_i\equiv \sum_{a=1}^{10}\check{\alpha}_{i}^{a} H_a\equiv h_i$ (which belongs to the Cartan subalgebra, i.e. the dual of the root space to which $\alpha_i$ belongs) enters both in the (Weyl) reflections (\ref{Reflection}) in the hyperplanes defined by the roots, which can be written as
\beg
r_i(v)&=&v-\alpha_i(v)\check{\alpha}_i,
\eeg
and in the definition of the Cartan matrix (\ref{cartan}) tantamount to
\beg
A_{ij}&=&\alpha_j(h_i)\,=\, \alpha_j(\check{\alpha}_i).
\eeg
Let us also recall that a Kac--Moody algebra is called `simply laced' if the Cartan matrix $A_{ij}$ is symmetric. This implies that all the simple roots (and more generally all the real roots) have the same length, which is conventionally normalized to $(\alpha_i|\alpha_i)=2$. In that case, which applies to the two specific Kac--Moody algebras $E_{10}$ and $AE_3$ that we shall consider here, the co-root $ h_i=\check{\alpha}_i$ becomes equal to the contravariant version of the corresponding root $\alpha_i^{\#}$ (\ref{coroot}) and the metric in Cartan space simply yields $(\alpha_i|\alpha_j)=(h_i|h_j)=(\check{\alpha}_i|\check{\alpha}_j)=\alpha_j(\check{\alpha}_i)=A_{ij}$. For simplicity, we shall assume in the main text that we are in this simply-laced case. The general case of a non-symmetric Cartan matrix will be discussed in Appendix \ref{appendix1}.\\

The decomposition of the Lie algebra valued `Maurer--Cartan velocity'
\be\label{maurer}
\dot{\cV}\cV^{-1} &=:& \cP+\cQ\in E_{10}
\ee
into an ``antisymmetric'' part $\cQ\in K(E_{10})$ and a ``symmetric'' one $\cP\in E_{10}\ominus K(E_{10})$ allows us to define the coset Lagrangian
\be\label{Lagra}
\mathcal{L}&=& \frac{1}{2n(t)}\left( \cP|\cP\right).
\ee
The bilinear form $(\cdot|\cdot)$ entering the coset Lagrangian (\ref{Lagra}) is the (unique) $E_{10}$-invariant bilinear form \cite{K95}. In the simply-laced case, it is such that $(H_a|H_b)=G_{ab}$, $( h_i|h_j)=A_{ij}$ and $(E_{\alpha,s}|E_{\alpha',s'})=\delta_{\alpha+\alpha'}\delta_{ss'}$. See Appendix \ref{appendix1} for the modifications that are necessary for the non-simply laced case.\\

The \textit{bosonic} coset equations of motion derived from the Lagrangian (\ref{Lagra}) read (in the gauge $n(t)=1$ that we shall henceforth use for simplicity)
\be\label{eomCoset2}
\p_t\cP(t) &=&\big[\cQ(t),\cP(t)\big].
\ee
In addition, the variation of the coset `lapse' $n(t)$ in (\ref{Lagra}) yields the constraint
\be\label{eomCoset1}
0&\approx& \left( \cP|\cP\right).
\ee
In these equations, $\cP\in E_{10}\ominus K(E_{10})$ represents the `velocity' of the coset `particle' moving on the (infinite dimensional) coset space $E_{10}/K(E_{10})$. See \cite{DN04} for the explicit proof that the equations of motion (\ref{eomCoset2}) can be identified (up to height $29$) with the equations of motion of SUGRA$_{11}$ via a `dictionary' which relates the (first few rungs of the infinite ladder of) coset variables $(\beta^a,\nu_\alpha)$ to the bosonic gravity variables $(\beta^a,\cN^a{}_m,A_{mnp}, \p_{[m}A_{npq]},C^a{}_{bc})$.\\

The fermionic extension of the coset model \cite{DKN05,BHP05b,DKN06} has similarly shown that the bosonic gravity-coset `dictionary' could be extended (up to an approximation that corresponds to the bosonic one) to a map relating the gravitino field $\psi_M^{(11)}(t,x)$ of SUGRA$_{11}$ to a coset fermionic variable $\psi(t)$ which belongs to the so-called `vector-spinor' \textit{vs} representation of $K(E_{10})$. Then, the \textit{fermionic} coset equation of motion simply reads
\be\label{eomCoset3}
\p_t\psi(t)&=&\cQ^{\textit{vs}}(t)\psi(t)
\ee
where $\cQ^{\textit{vs}}$ denotes the `vector-spinor' representation of the object $\cQ\in K(E_{10})$ that entered the basic decomposition (\ref{maurer}).\\

When comparing Eq. (\ref{eomCoset2}) to (\ref{eomCoset3}), we see that the `antisymmetric' part $\cQ\in K(E_{10})$ of the Lie algebra valued velocity $\dot{\cV}\cV^{-1}\in E_{10}$ (\ref{maurer}) plays the role of a $K(E_{10})$-connection, realizing the `parallel transport' of both the bosonic (coset-valued) `velocity' $\cP\in E_{10}\ominus K(E_{10})$ and the fermionic coset variable $\psi\in S^{\textit{vs}}_{K(E_{10})}$ (where $S^{\textit{vs}}_{K(E_{10})}$ denotes the $K(E_{10})$ `vector-spinor' representation space). In more intuitive terms, we can think of $\cQ(t)\in K(E_{10})$ as being a continuous \textit{rotational angular velocity} under which both $\cP(t)$ and $\psi(t)$ `turn' within their respective linear representation spaces of $K(E_{10})$. The common occurrence of the same $\cQ(t)$ in the bosonic and fermionic dynamics will play a central role in the present work.
\end{subsection}

\begin{subsection}{`Spiky' structure of the $K(E_{10})$ `angular velocity' $\cQ(t)$ in the near-singularity limit}\label{one}
Having reviewed the crucial role of the $K(E_{10})$ `angular velocity' $\cQ(t)$ in both the bosonic and the fermionic equations of motion (\ref{eomCoset2},\,\ref{eomCoset3}), let us study the structure of $\cQ(t)\in K(E_{10})$ in the BKL-type, near-singularity limit. In that limit, previous work on cosmological billiards has shown that, as a function of the coset time $t$ (which tends to $+\infty$ rougly proportionally to minus the logarithm of the proper time) $\cQ(t)$ was exponentially small most of the time, except around the time of a `collision' with a dominant wall, when $\cQ(t)$ rose up to a finite value \cite{DHN02,dBD07}. To study the precise structure of these `spikes' in $\cQ(t)$, it is enough to consider the `collision' with a \textit{single} wall. [Indeed, as $t\rightarrow +\infty$, the coset time intervalls between wall collisions grow linearly in $t$. This `one-wall collision' process is anyway interesting in its own right (``S-brane solution'', see below).]\\

For simplicity, as this will be the case for the simply-laced algebras $E_{10}$ and $AE_3$ that we focus on here, we consider a wall form $\alpha(\beta)$ where $\alpha$ is a real (positive) root $\alpha\in \Delta^+$ with norm $( \alpha|\alpha)=2$.\footnote{Note that the final result of this section is not altered for the more general case that occurs e.g. for Kac--Moody algebras with non-symmetric Cartan matrices, which is treated in Appendix \ref{appendix1} in detail.} In this one-wall approximation, the Iwasawa form of the coset element $\cV$ (\ref{coset}) only contains one term, $\nu_{\alpha}(t)E_{\alpha}$, in the second exponent of (\ref{coset}).\footnote{Let us recall in passing that real roots are non-degenerate \cite{K95}. Hence, we can drop the degeneracy index $s$ in (\ref{coset}) without loss of generality.} Then, one can easily evaluate the `Maurer--Cartan velocity' (\ref{maurer}) by using the basic commutation relation $[h,E_{\alpha}]=\alpha(h)E_\alpha$ involving any Cartan subalgebra element $h\in \mathfrak{h}$:
\be\label{MC2}
\left.\dot{\cV}\cV^{-1}\right|_{\alpha}
&=&
\sum\limits_{a=1}^{10}\dot{\beta}^a(t)H_a + \dot{\nu}_{\alpha}e^{\sum\limits_{a=1}^{10}\beta^a(t)H_a} E_{\alpha} e^{-\sum\limits_{a=1}^{10}\beta^a(t)H_a}
\nn\\
&=&
\sum\limits_{a=1}^{10}\dot{\beta}^a(t)H_a + e^{\alpha(\beta)}\dot{\nu}_{\alpha}E_{\alpha}.
\ee
Its splitting into a `symmetric' part $\cP\in E_{10}\ominus K(E_{10})$ and an `antisymmetric' one $\cQ\in K(E_{10})$ as in (\ref{maurer})
\begin{subequations}\label{PQtrunc0}
\be\label{Ptrunc0}
\left.\cP\right|_{\alpha}&=&\sum\limits_{a=1}^{10}\dot{\beta}^a(t)H_a +\frac12 e^{\alpha(\beta)}\dot{\nu}_{\alpha}\big(E_{\alpha}+E_{-\alpha}\big)\\
\left.\cQ\right|_{\alpha}&=&\frac12 e^{\alpha(\beta)}\dot{\nu}_{\alpha}J_{\alpha},
\label{Qtrunc0}
\ee
\end{subequations}
where $J_\alpha=E_\alpha-E_{-\alpha}\equiv E_\alpha +\omega(E_\alpha)$ (\ref{JDefi}) allows one to evaluate the Lagrangian (\ref{Lagra}), using the gauge $n(t)=1$ for simplicity again:
\be\label{Lagra2}
\left.\mathcal{L}\right|_{\alpha}&=& 
\frac12 \sum\limits_{a,b=1}^{10}G_{ab}\dot{\beta}^a\dot{\beta}^b + \frac14 e^{2\alpha(\beta)}\dot{\nu}_{\alpha}^2.
\ee
Note that the normalization of the kinetic term of $\nu_\alpha$ has come from the normalization $(E_\alpha|E_{-\alpha})=1$ of the generators entering the $K(E_{10})$ generator $J_\alpha$. The `cyclicity' of the variable $\nu_{\alpha}$ implies that its conjugate momentum 
\be\label{constmom}
\Pi_{\alpha}\,\,:=\,\,\frac{\p \mathcal{L}|_{\alpha}}{\p \dot{\nu}_{\alpha}}&=&\frac{1}{2}e^{2\alpha(\beta)}\dot{\nu}_\alpha
\ee
is constant. This conservation law will prove important for analyzing the time evolution of the Cartan variables $\beta^a$ and of the $K(E_{10})$-connection $\cQ$ (\ref{Qtrunc0}). The analysis of the $\beta$-dynamics is greatly simplified by adapting the notation to the geometry. This is achieved by decomposing $\beta$ into a component parallel $(\beta_{||}^a)$ and a component orthogonal $(\beta_\bot^a)$ to the considered wall $\alpha$.
\be\label{decomp}
\beta^a
&=&\beta_\bot^a +\beta_{||}^a\\
\text{with}\quad \beta_\bot^a&:=&\frac{\alpha(\beta)}{(\alpha|\alpha)}\alpha^{\# a}
\nn\\
\text{and}\quad 0&=&\alpha(\beta_{||})\,\equiv\, (\alpha^{\#}|\beta_{||})
\nn
\ee
with $\alpha(\beta)=\sum_{a=1}^{10}\alpha_a\beta^a$ and the metric $(\cdot|\cdot)$ as used in (\ref{Reflection}). Note that this decomposition uses the Lorentzian metric $G$ (\ref{coroot}). As the wall potential in (\ref{Lagra2}) only depends on the one-dimensional orthogonal component $\beta_\bot\propto\alpha(\beta)$, it is easily seen that the motion parallel to the wall proceeds with a constant velocity $\dot{\beta}^a_{||}$ as in (\ref{Kasner}). By contrast, the motion orthogonal to the wall undergoes a non-trivial `Toda collision'. The corresponding dynamics can be exactly solved by using the conservation of energy, which reads
\be\label{energy2}
E_{||}&=&\frac14\alpha(\dot{\beta})^2 + e^{-2\alpha(\beta)}\Pi_\alpha^2,
\ee
where $E_{||}=-\frac12\sum_{a,b=1}^{10}G_{ab}\dot{\beta}_{||}^a\dot{\beta}_{||}^b$ denotes \textit{minus} the conserved parallel kinetic energy (which must be positive in view of the null constraint (\ref{null}) valid far from the wall). Note that the normalization factor $\frac14$ in the kinetic term of $\alpha(\beta)$ (\ref{energy2}) has made use of $(\alpha|\alpha)=2$. The conservation law (\ref{energy2}) allows one to explicitly find the time evolution of the `orthogonal' distance $\alpha(\beta)$. Besides the constants $\Pi_\alpha$ and $E_{||}$, it contains a time shift $t_c\in \R$ as a constant of integration:
\be\label{solution}
e^{\alpha(\beta)}&=&\frac{|\Pi_{\alpha}|}{\sqrt{E_{||}}}\cosh\left(2\sqrt{E_{||}}(t-t_c)\right).
\ee
Note that, when inserting the definition of $\Pi_\alpha$ (\ref{constmom}) and the result (\ref{solution}) into the expression of $\cQ$ (\ref{Qtrunc0}), we obtain
\be\label{Qt}
\left.\cQ(t)\right|_{\alpha}\,\,=\,\, e^{-\alpha(\beta)}\Pi_{\alpha}J_{\alpha}
&=&
\e_{\alpha}\frac{\sqrt{E_{||}}}{\cosh(2\sqrt{E_{||}}(t-t_c))}J_{\alpha}
\ee
where $\e_\alpha=\frac{\Pi_{\alpha}}{|\Pi_{\alpha}|}=\pm 1$ is the sign of $\Pi_\alpha$. Equation (\ref{Qt}) exhibits the precise structure of the `spike' of the `angular velocity' $\cQ\in K(E_{10})$, which occurs at each collision with a Toda wall: it is centered at $t=t_c$ and given by an inverse hyperbolic cosine multiplying the particular $K(E_{10})$ generator $J_\alpha=E_{\alpha}-E_{-\alpha}$ (\ref{JDefi}) associated to the considered root.\\

It will be useful in the following to also consider the integrated effect of $\cQ(t)$ through the collision. For this, we define a finite `angle of rotation' $\theta(t)$ about the fixed $K(E_{10})$ `rotation axis' $J_\alpha$ such that the instantaneous value of the `angular velocity' $\cQ(t)$ reads
\be\label{PQtrunc}
\left.\cQ(t)\right|_{\alpha}&=&\dot{\theta}(t)J_{\alpha}.
\ee
In other words, comparing with (\ref{Qt}) we have
\be
\label{theta}
\dot{\theta}\,\,:=\,\,e^{-\alpha(\beta)}\Pi_{\alpha}
&=&
\e_{\alpha}\frac{\sqrt{E_{||}}}{\cosh(2\sqrt{E_{||}}(t-t_c))}.
\ee
Integrating this differential equation yields the following expression for the time evolution of the `angle of rotation':
\be\label{solution2}
\theta(t) &=& \e_{\alpha}\arctan\left(e^{2\sqrt{E_{||}}(t-t_c)}\right)+\theta_{-\infty}
\ee
where $\theta_{-\infty}=\theta(-\infty)\in \R$ is another constant of integration.\\

We clearly see on (\ref{solution2}) the nature of the $K(E_{10})$-`rotation' that takes place at each wall: it consists of an $\arctan$ `kink' (or `antikink'), localized around the central time $t_c$ of the collision, and taking place on a finite time scale $\Delta t\approx 1/\sqrt{E_{||}}$, during which most of the rotation occurs. It is remarkable that the `total $K(E_{10})$-rotation' linking the initial state at $t=-\infty$ to the final one\footnote{The limits $t\rightarrow \pm \infty$ are to be physically interpreted as referring to the `billiard ball' $\beta$ being far away from the considered wall $\alpha$ (incoming or outgoing), but also far away from the other dominant walls that it encountered before, or will encounter later. In other words, the BKL-limit allows us to use a `dilute kink' approximation.
} at $t=+\infty$ is universally given by
\be\label{pi2}
\theta^{\textit{tot}}_{\e_\alpha}\,\,:=\,\,\theta(\infty)-\theta(-\infty) &=& \e_\alpha\frac{\pi}{2},
\ee
and is essentially independent of the initial conditions and of the characteristics of the considered wall. [We show in Appendix \ref{appendix1} that, if one normalizes the generators $E_\alpha$ in a standard Kac--Moody way \cite{K95}, the result (\ref{pi2}) is also independent of the norm of the root $\alpha$.] The only trace of the initial data is the sign $\e_\alpha$ of the momentum $\Pi_\alpha$ conjugate to $\nu_\alpha$. In other words, the sign of $\Pi_\alpha$ determines whether $\theta(t)$ undergoes a kink or an antikink.\\

We have checked that the universal result (\ref{solution2}), that we have derived here from the coset model, can also be obtained by using gravity variables directly. For instance, in the case where the `collision' occurs on an electric wall, we can derive (\ref{solution2}) from the electric-wall bounce solution written in \cite{DHHS85} (using a ``Freund--Rubin ansatz''), while in the case of a symmetry wall, Eq. (\ref{solution2}) follows from the formul\ae{} given in \cite{DHN02}. We have here another example where the coset model provides a common basis for expressing, in a uniform way, a priori unconnected special solutions of maximal supergravity that are sometimes referred to as ``S-brane solutions'' \cite{KN05}.
\end{subsection}

\begin{subsection}{Rederivation of the bosonic wall Weyl reflection from the connection `kink'}\label{kink}
Summarizing so far: We have shown that the $K(E_{10})$-valued `angular velocity' $\cQ(t)$ which entered (in different representations) both in the bosonic coset equation of motion (\ref{eomCoset2}) and in the fermionic one (\ref{eomCoset3}) was made of a sequence of well separated inverse-$\cosh$ `spikes', proportional to the $K(E_{10})$ generators $J_\alpha=E_\alpha-E_{-\alpha}$ (\ref{JDefi}) associated to the wall $\alpha$ on which the considered collision takes place. Before discussing, in the next section, the effect of these spikes on the fermionic variables, let us see how their effect on the bosonic variables reproduces the well-known BKL-type result (\ref{Reflection}) saying that the $\beta$-space `velocity' of the incoming billiard ball is Weyl-reflected in each wall.\\

To do this, we have to notice that, in view of Eq. (\ref{eomCoset2}), the action of the $K(E_{10})$-connection $\cQ$ on the coset velocity $\cP$ (\ref{maurer}) is given by a commutator, i.e. by a (formal) adjoint action.\footnote{Mathematically speaking, one should reserve the name `adjoint action' to the adjoint action of $K(E_{10})$ on itself. Here, $\cP$ belongs to a different representation space: the coset one $E_{10}\ominus K(E_{10})$. The action of $\cQ$ on $\cP$ should therefore be called the `coset action'. Note, however, that this coset action is canonically deduced from the adjoint $E_{10}$-action, given the decomposition (\ref{maurer}).} As the `rotation axis' $J_\alpha$ is fixed, one can formally integrate the equation of motion (\ref{eomCoset2}) in the one-wall approximation to obtain
\be\label{RDefi0}
\cP(t)&=&e^{\theta(t) J_{\alpha}}\cdot \cP(t_{0})\cdot e^{-\theta(t) J_{\alpha}}.
\ee
We see here how, at any intermediate time, the effect of a collision with one wall $\alpha$ on $\cP$ can be understood as a continuous `rotation' given by the adjoint action (within $E_{10}$) of the associated $K(E_{10})$-group element $\cR_{\alpha}(t)$:
\be\label{RDefi}
\cP(t)
&=&\cR_{\alpha}^{\textit{Ad}}(t)\big(\cP(t_{0})\big)
\\
\text{with}\quad
\label{Rcont}
\cR_{\alpha}(t)&:=& e^{\theta(t) J_{\alpha}}
\ee
where the superscript $\textit{Ad}$ indicates the adjoint group action of $\cR_{\alpha}(t)\in K(E_{10})$ on the coset element $\cP$ and where the ``rotation'' $\cR_{\alpha}(t)$ about the axis $J_{\alpha}$ is obtained from a formal exponentiation of the algebra element $\theta(t) J_{\alpha}$. One might worry whether the formal exponentiation (\ref{RDefi}) is a well-defined procedure. It is in the present case, as we are going to see. The basic reason is that, for a given real root $\alpha$, the entire computation takes place in an $\textit{SL}(2,\R)$ subgroup.\footnote{Note that the adjoint representation is integrable and that hence any $sl_2$-subalgebra generated by the Chevalley triple $(e_i,f_i,h_i)$ for a simple root $\alpha=\alpha_i$ can be integrated to the corresponding group \cite{K95}. As far as the action on the Cartan subalgebra $\mathfrak{h}$ is concerned, this statement extends to any real root of $E_{10}$, which can also be considered as a simple root of a subalgebra.}\\

Let us evaluate the action of the total $K(E_{10})$-rotation linking the initial configuration $\cP(-\infty)$ to the final one $\cP(+\infty)$:
\be\label{RDefi1}
\cP(\infty)&=&\cR_{\alpha,\e_\alpha}^{\textit{Ad}}\big(\cP(-\infty)\big)
\\
\text{with}\quad\cR_{\alpha,\e_\alpha}&:=&e^{\e_\alpha\frac{\pi}{2} J_{\alpha}}.
\label{RDefi2}
\ee
In order to evaluate this `total rotation' $\cR_{\alpha,\e_\alpha}$ on $\cP(-\infty)$, one starts by noticing that, far away from the wall ($t\rightarrow\pm \infty$), the $\dot{\nu}$-terms in (\ref{Ptrunc0}) become negligible due to $e^{-\alpha(\beta)}\rightarrow 0$ (\ref{constmom},\,\ref{solution}), which leads to a simple expression for the initial datum
\be\label{Pexpl}
\cP|_{\alpha}(-\infty)&=& \sum\limits_{a=1}^{10}v^a H_a,
\ee
where $v^a=\dot{\beta}^a$ (\ref{Kasner}) is the `incoming' $\beta$-velocity. Hence, the initial coset `velocity' (\ref{Pexpl}) belongs to the Cartan subalgebra. This drastically simplifies the calculation. The easiest way to evaluate the action (\ref{RDefi1}) is to work within the $\textit{SL}(2,\R)$ subgroup generated by the algebra elements $E_\alpha$, $E_{-\alpha}$ and $[E_\alpha,E_{-\alpha}]= h_\alpha\equiv\check{\alpha}\in \mathfrak{h}$. As in equation (\ref{decomp}), we decompose the Cartan-subalgebra valued initial datum $\cP(-\infty)$ (\ref{Pexpl}) into two parts $\cP_\bot$ and $\cP_{||}$ where $\alpha(\cP_{||})=0$. The component $\cP_{||}$ is found to commute with $E_{\alpha}$ and $E_{-\alpha}$, while the action of the $SO(2)$ rotation $\cR_{\alpha,\e_\alpha}$ (\ref{RDefi2}) on $\cP_\bot \propto h_\alpha$ is obtained by exponentiating $J_\alpha=E_\alpha-E_{-\alpha}$ using $[h_\alpha,E_\alpha]=2E_\alpha$. This is easily done by noting that the commutation relations of $h_\alpha$, $E_\alpha$ and $E_{-\alpha}$ are those of an $sl_2$-subalgebra. In the canonical basis of this subalgebra, we can represent $E_\alpha\approx \binom{0\,\,\,\,1}{0\,\,\,\,0}$, $E_{-\alpha}\approx \binom{0\,\,\,\,0}{1\,\,\,\,0}$, $h_\alpha=[E_\alpha,E_{-\alpha}]\approx \binom{1\,\,\,\,\,0}{0\,\,-\!1}$ so that $J_\alpha=E_{\alpha}-E_{-\alpha}\approx\binom{\,0\,\,\,\,1}{-\!1\,\,0}$. Within this 3-dimensional representation of $sl_2$, we can exponentiate $J_\alpha\approx\binom{\,0\,\,\,\,1}{-\!1\,\,0}$ into $e^{\theta J_\alpha}=\id_2\cos\theta+J_\alpha\sin\theta$, so that the adjoint action (\ref{RDefi1}) acting on the perpendicular component of $\cP$, $\cP_\bot\propto h_\alpha \approx \binom{1\,\,\,\,\,0}{0\,\,-\!1}$ evaluates to
\be\label{RTH}
&&\left(\id_2\cos\theta+J_\alpha\sin\theta\right)h_\alpha \left(\id_2\cos\theta-J_\alpha\sin\theta\right)
\nn\\
&=&
\left(\begin{tabular}{cc} $\cos^2\theta-\sin^2\theta$&$-2\sin\theta\cos\theta$\\
$-2\sin\theta\cos\theta$&$-\cos^2\theta+\sin^2\theta$\end{tabular}\right)
\nn\\
&=&
\cos(2\theta)h_\alpha-\sin(2\theta)(E_\alpha+E_{-\alpha})
.
\ee
We see that, while for intermediate angles of rotation, $\cP_\bot$ is mapped out of the Cartan subalgebra $\mathfrak{h}$ and acquires a component $\propto \sin(2\theta)(E_\alpha+E_{-\alpha})$, the full rotation by $\theta=\pm \frac{\pi}{2}$ associated to a wall $\alpha$ transforms $h_\alpha$ into $-h_\alpha$ (i.e. back into $\mathfrak{h}$).\\

In conclusion, this `$K(E_{10})$ $\frac{\pi}{2}$-rotation'-calculation says that $\cP_{||}(+\infty)=+\cP_{||}(-\infty)$ whereas $\cP_\bot(\infty)=-\cP_{\bot}(-\infty)$, which is precisely a reflection in the $\alpha(\beta)=0$ hyperplane, as generally expressed by the reflection formula (\ref{Reflection}). Thus, the equation (\ref{RDefi1}) links the `incoming' velocity $v^a$ (\ref{Pexpl}) for $t\rightarrow -\infty$ to the `outgoing' one $v'{}^a$ for $t\rightarrow \infty$:
\be\label{Reflection2}
v'\,\,=\,\,r_{\alpha}(v) &=& v -\alpha(v)\check{\alpha}.
\ee
Note also that the sign $\e_\alpha=\pm1$ of the $\frac{\pi}{2}$-rotation has dropped from this bosonic calculation, essentially because the relevant variable $\cP$ involve the doubled angle $2\theta$ as we see in Eq. (\ref{RTH}) (i.e. they live in the three-dimensional `spin $1$' representation of $sl_2$ rather than in the fundamental `spin $\frac12$' one).\\
 
In order to keep the notation simple, we have confined the analysis to the simply-laced case from the start by assuming $(\alpha|\alpha)=2$. It is explicitly checked in Appendix \ref{appendix1} that, actually, the norm of the (real) root $\alpha$ drops out of the final result and always yields a geometrical reflection (\ref{Reflection2}) for the $\beta$-velocity. The fact that a general Weyl reflection of a finite-dimensional group $G$ can be expressed by a $\frac{\pi}{2}$-adjoint rotation about the `axis' $J_\alpha$ in the compact subgroup $K\subset G$ was recently noticed in \cite{FS07}, too. We also note that a different exponential representation of a Weyl reflection (in the general Kac--Moody case) is used in \cite{K95} (p. 36). Its relation to (\ref{RDefi2}) will be discussed below.\\

In this section, we have shown explicitly that the Kac--Moody billiard associated to the BKL-limit of $D=11$ supergravity can be understood as a sequence of $90^\circ$-$K(E_{10})$-rotations about the different axes $J_{\alpha_i}$ that are defined by the dominant walls $\alpha_i$ (\ref{dominant},\,\ref{JDefi}). Note that the ten `axes' $J_\alpha$ are generators in $K(E_{10})$ which are orthogonal to each other with respect to the $E_{10}$-metric $(\cdot|\cdot)$. We will see in the next section that the same picture holds for the fermions. Namely, the same formal rotation $\cR_{\alpha,\e_\alpha}$ (\ref{RDefi2}), now evaluated in the spinor representation of $K(E_{10})$ \cite{DKN05,BHP05b,DKN06}, acts on the fermions of supergravity, which can be considered as the induced effect of the ``collision of the scale factors $\beta^a$ with a wall $\alpha$''. This will lead us to introducing some `spin extension' of the Weyl group.
\end{subsection} 
 \end{section}

\begin{section}{Fermionic billiard of $D=11$ supergravity and its $K(E_{10})$ structure}\label{Ferm}
\begin{subsection}{Gravitino dynamics: gravity versus coset description}\label{Ferm2}
As a prelude to our study of fermionic billiards, let us recall the link between the standard supergravity description of fermions and the coset one, which reveals its hidden $K(E_{10})$-structure.\\

The gravity form of the fermionic equations of motion (neglecting non-linear fermionic terms) derived from the Cremmer--Julia--Scherk action \cite{CJS78} reads (in the conventions recalled in Appendix \ref{appendix0}, using Einstein's summation convention and flat indices $A,\ldots=0,\dots,10$, e.g. $\p_A=E_A^M\frac{\p}{\p x^M}$.)
\be
\G^B\nabla_{[A}\psi^{(11)}_{B]}&=&-\frac{1}{144}\G^B\left(\G_{[A}{}^{CDEF}-8\delta_{[A}^C\G^{DEF}\right)\psi^{(11)}_{B]}F_{CDEF}.
	\label{fermieom}
	\ee
	In this form, the coupling to the $3$-form $A$ (or rather $F=dA$) appears explicitly on the RHS, while the coupling to gravity is contained in the spin connection $\omega_{A[BC]}$ appearing on the LHS of (\ref{fermieom}):
	\beg
	\nabla_A\psi^{(11)}_B&=&\p_A\psi^{(11)}_B+\frac14\omega_{ACD}\G^{CD}\psi^{(11)}_B +\omega_{AB}{}^C\psi^{(11)}_C.
	\eeg
	In the coset formulation, all the $F$-dependent terms on the RHS can be merged with the spin connection to form a $K(E_{10})$-connection $\cQ=\cQ^{(0)} +\cQ^{(1)} +\cQ^{(2)} +\cQ^{(3)}+\dots$ where the zero level term $\cQ^{(0)}$ involves the part of the spin connection which contains time derivatives of the metric $g_{mn}(t,x)$ (\ref{metric11}), while the level $1$ term $\cQ^{(1)}$ involves time derivatives of the three-form $A_{mnp}$, the level $2$ term $\cQ^{(2)}$ space derivatives of $A_{mnp}$ and the level $3$ term $\cQ^{(3)}$ space derivatives of the metric (in the form of the structure constants $C^a{}_{bc}$). To reveal this structure, one needs to gauge-fix both the metric and the gravitino, and to work with some redefined fermionic variables: namely the following `coset' combinations of the original $D=11$ gravitino field $\psi_{(11)}$ and of the supersymmetry transformation parameter $\ep_{(11)}$
\begin{subequations}\label{fermTrafo}
\be
\psi^0&:=&\det(g_{mn})^{\frac14}\Big(\psi_{(11)}^0-\G^0\sum\limits_{a=1}^{10}\G_a\psi^a_{(11)}\Big)\\
\psi^a&:=&\det(g_{mn})^{\frac14}\psi^a_{(11)}\\
\ep&:=&\det(g_{mn})^{-\frac14}\ep_{(11)}.
\ee
\end{subequations}
One fixes the supersymmetry gauge freedom by setting
\be\label{susyfix}
\psi^0&=&0.
\ee
Note that we are not exhibiting the $\mathbf{32}$-valued spinor index of neither the supersymmetry parameter $\ep$, nor the vector-spinor gravitino field $\psi^a$ at this stage. As already mentioned above, the gravitino equations of motion in the gauge (\ref{susyfix}) and up to `level 3' in the sense indicated above, were found to be expressible as a \textit{parallel transport} of an element of a `vector-spinor' representation of $K(E_{10})$ by the \textit{same} connection $\cQ(t)\in K(E_{10})$ that entered the bosonic equation of motion (\ref{eomCoset3}):
\be\label{ke10ferm}
 \p_t\psi(t) &=&\cQ^{\textit{vs}}(t)\psi(t).
\ee
Let us again emphasize the analogies between fermions and bosons. The time evolution of the gravitino (\ref{ke10ferm}) is completely analogous to the bosonic one (\ref{eomCoset2}): at each moment, $\psi$ transforms by an infinitesimal $K(E_{10})$-`rotation'  with the same `angular velocity' $\cQ(t)\in K(E_{[10})$ as for the bosonic case, except that now, $\cQ$ is acting in a different representation. The reader should keep in mind that the `fermionic' representation of $\cQ(t)$ will have, modulo the fact that it lives in a different space, the same time-structure that we studied in section \ref{one}, namely a sequence of well-separated inverse-$\cosh$ `spikes'.\footnote{Note that the time-evolution of the $K(E_{10})$-algebra element $\cQ(t)$ is unambiguously fixed by the bosonic dynamics (\ref{eomCoset2}). Due to the different grading in the classical Grassmann algebra, any fermionic ``backreaction'' of the fermions only affects higher orders in the Grassmann algebra.}\\

In the near-singularity limit, we need only to consider the `spikes' associated to collisions on the ten dominant walls $\alpha_{i}$ (\ref{dominant}) for $i=0,\dots,9$. We need now to recall the definition of the vector-spinor representation $\textit{vs}$ of $K(E_{10})$. It lives in a $320$ dimensional vector space and was defined in \cite{DKN05,BHP05b,DKN06}. Here, it will be enough for our purpose to know the action of the ten generators $J_{\alpha_i}$ (\ref{JDefi}) corresponding to the simple roots of $E_{10}$. They act on $\psi$ in the vector-spinor representation $\textit{vs}$ as $320\times 320$-matrices. The explicit form of these matrices follows from the results of \cite{DKN06} (see. Eq. (2.26) there) or \cite{BHP05b} and read
\begin{subequations}\label{Jexpl}
\be
J_{\alpha_{0}}^{\textit{vs}}(\psi)^a&=& \frac12 \G^{123}\psi^a +4\delta^{a[1}\G^2\psi^{3]} -\G{}^{a[12}\psi^{3]}\\
J_{\alpha_{i}}^{\textit{vs}}(\psi)^a&=& \frac12 \G^{i}\G^{i+1}\psi^a +2\delta^{a[i}\psi^{i+1]} \quad\text{for }i\,=\,1,\dots,9.
\ee
\end{subequations}
Here, the $\G^a$'s, $a=1,\dots,10$, are $SO(10)$ real gamma matrices ($\{\G^a,\G^b\}=\delta^{ab}$). As a next step, recall from section \ref{one} that the effects of the walls on the dynamics are obtained by considering, separately and successively, the effect of each `spike' in the connection $\cQ$. We then start by integrating Eq. (\ref{ke10ferm}) over one spike of the form
\beg
\cQ^{\textit{vs}}(t)\,\,=\,\, \dot{\theta}J_{\alpha}^{\textit{vs}}
&=&
\e_{\alpha}\frac{\sqrt{E_{||}}}{\cosh(2\sqrt{E_{||}}(t-t_c))}J_{\alpha}^{\textit{vs}}.
\eeg
The only difference with the previous case (\ref{Qt}) is that $J_{\alpha}^{\textit{vs}}$ now acts in a $320$ dimensional space, according to (\ref{Jexpl}). As in the bosonic case (\ref{RDefi}), one can integrate the sequence of infinitesimal $K(E_{10})$-rotations (\ref{ke10ferm}) with angular velocity $\dot{\theta}J_{\alpha}^{\textit{vs}}$ into
\be\label{ke10ferm2}
\psi(t) &=& \cR_\alpha^{\textit{vs}}(t)\psi(t_0)\\
\text{where}\quad \cR_\alpha^{\textit{vs}}(t)&=& e^{\theta(t)J_{\alpha}^{\textit{vs}}}.
\nn
\ee
Finally, the total effect of the collision on a wall $\alpha$ will be an integrated $K(E_{10})$ rotation between the incoming gravitino $\psi$ and the outgoing one $\psi'$, given by
\begin{subequations}\label{ke10ferm3}
\be
\psi'{} &=&\cR_{\alpha,\e_\alpha}^{\textit{vs}}\psi\\
\text{with}\quad \cR_{\alpha,\e_\alpha}^{\textit{vs}}&=&e^{\e_\alpha\frac{\pi}{2}J_{\alpha}^{\textit{vs}}}.
\ee
\end{subequations}
Therefore, if we consider, as in Eq. (\ref{wb}) above, the billiard dynamics towards a singularity, its effect on the gravitino variable $\psi$ will be represented by a growing `word' made of the product of the induced Weyl vector-spinor rotation (\ref{ke10ferm3}) corresponding to the Weyl reflections in the bosonic Weyl word $w$ (\ref{wb}), i.e.
\be\label{wf}
w^{\textit{vs}}&=&\cR_{\alpha_{i_1},\e_{\alpha_{i_1}}}^{\textit{vs}}
\cR_{\alpha_{i_2},\e_{\alpha_{i_2}}}^{\textit{vs}}
\cdots \,\,\cR_{\alpha_{i_n},\e_{\alpha_{i_n}}}^{\textit{vs}}\cdots
\ee
At this stage, our task to describe the motion of the gravitino is reduced to exponentiating the operators (\ref{Jexpl}) within the $320$ dimensional representation space.\\

For a further investigation of this action, we now introduce the following (second) redefinition of fermionic variables
	\be\label{phiDefi}
	\phi^a&:=&\G_*\G^a\psi^a\quad \text{(no sum)}\\
	\text{with}\quad \G_*&:=& \G^1\cdots \G^{10}.
	\nn
	\ee
	The definition (\ref{phiDefi}) would seem to drastically violate the $SO(10)$ symmetry (the unbroken subgroup of the space-time Lorentz symmetry $SO(10,1)$) which was present in the general action of the vector-spinor generators $J_{\alpha}^{\textit{vs}}$ (see their general definition in \cite{DKN06}). However, we shall see that this loss of explicit symmetry is compensated by the appearance of simple structures linked to the $SO(9,1)$ Lorentzian metric $G_{ab}$ (the restriction of the invariant form $(\cdot|\cdot)$ to the Cartan subalgebra of $E_{10}$) rather than to the $SO(10)$ Euclidean metric $\delta_{ab}$ built in the zeroth level of $K(E_{10})$.
	\end{subsection}

\begin{subsection}{Induced Weyl group action on the redefined gravitino}\label{induced}
The usefulness of the redefinition (\ref{phiDefi}) will appear in streamlining the structure of the wall-induced `rotation' $\cR_{\alpha,\e_\alpha}^{\textit{vs}}$ (\ref{ke10ferm3}) acting on the gravitino. Adapting the notation to the geometry as for the $\dot{\beta}^a$ reflection (\ref{decomp}), we split the `vector' index\footnote{In doing so, the hidden spinor index of $\phi^a$ is left untouched.} of the gravitino $\phi^a$ (\ref{phiDefi}) into components parallel and orthogonal to the wall $\alpha$:
\be\label{decomp2}
\phi^a&=:&\phi^a_\bot +\phi^a_{||}\\
\text{with}\quad \phi^a_\bot&=&\frac{\alpha(\phi)}{(\alpha|\alpha)}\alpha^{\# a}
\nn\\
\text{and}\quad 0&=&\alpha(\phi_{||})
\nn
\ee
where $\alpha(\phi)$ is a short-hand notation for $\sum_{a=1}^{10}\alpha_a\phi^a$, modelled on the canonical pairing $\alpha(\beta)$ that appeared in the bosonic case (\ref{decomp}). A quick calculation then reveals that the action of the generators $J_{\alpha_{i}}$ (\ref{Jexpl}) drastically simplifies to
\begin{subequations}\label{Jgut}
\be
J^{\textit{vs}}_{\alpha_{i}}\phi_{||}^a &=& J^{\textit{s}}_{\alpha_{i}}\phi_{||}^a\\
J^{\textit{vs}}_{\alpha_{i}}\phi_\bot^a &=& -3J^{\textit{s}}_{\alpha_{i}}\phi_\bot^a
\ee
\end{subequations}
where $J^{\textit{s}}_{\alpha_{i}}$ denote the Dirac spinor actions \cite{BHP05} (see also Eq. (2.23) in \cite{DKN06}) defined by
\begin{subequations}\label{SDefi}
\be
J^{\textit{s}}_{\alpha_{0}}&=&\frac{1}{2}\G^{123}\\
J^{\textit{s}}_{\alpha_{i}}&=&\frac{1}{2}\G^{i}\G^{i+1} \quad\text{for }i=1,\dots,9.
\ee
\end{subequations}
It is convenient to summarize the structure of the Clifford algebra valued generators (\ref{SDefi}) by writing them simply as
\be\label{JGa}
J^{\textit{s}}_{\alpha_{i}}&=&\frac{1}{2}\G^{\alpha_i} \quad\text{for }i=0,\dots,9.
\ee
where one introduces the short-hand notation
\begin{subequations}\label{GaW}
\be
\text{electric wall:}\quad \alpha_0=\beta^1+\beta^2+\beta^3&\rightarrow&\G^{\alpha_0}\equiv\G^{1}\G^{2}\G^{3}\\
\text{symmetry walls:}\quad \alpha_i=\beta^{i+1}-\beta^i&\rightarrow&\G^{\alpha_i}\equiv\G^{i}\G^{i+1}
\ee
\end{subequations}
Note that these ten $\G^{\alpha_i}$'s all satisfy the relation
\be\label{Ga2}
\left(\G^{\alpha_i}\right)^2&=&-\id_{32}
\ee
and therefore can be considered as imaginary units $\mathbf{i}$ (with $\mathbf{i}^2=-1$). A remarkable feature of the result (\ref{Jgut},\,\ref{SDefi},\,\ref{Ga2}) besides their simplicity (compared to the original form (\ref{Jexpl})), is the symmetry they reveal between a symmetry wall $\alpha_i$, $i=1,\dots,9$, and an electric one, $\alpha_0$.\\

The simplicity of the structure (\ref{Jgut},\,\ref{SDefi}) now allows us to compute the needed exponentiated action of $\e_\alpha\frac{\pi}{2}J_\alpha^{\textit{vs}}$, namely
\be\label{RDefi4VS}
\cR_{\alpha,\e_\alpha}^{\textit{vs}}&=& e^{\e_\alpha\frac{\pi}{2}J^{\textit{vs}}_{\alpha}}.
\ee
Note that equations (\ref{Jgut}) mean that the parallel and the orthogonal parts $\phi_{||}^a$ and $\phi_\bot^a$ of the gravitino $\phi^a$ are `eigendirections' of the vector part of the vector-spinor generator $J^{\textit{vs}}_{\alpha_{i}}$. More precisely, it is interesting to remark that, with the notation (\ref{JGa}), the action of the vector-spinor generators reads
\beg
J^{\textit{vs}}_{\alpha_{i}}\phi_{||}&=& \frac12\G^{\alpha_i} \phi_{||},
\\
J^{\textit{vs}}_{\alpha_{i}}\phi_{\bot}&=& -\frac32\G^{\alpha_i} \phi_{\bot},
\eeg
thereby exhibiting the fact that the gravitino is a vector-spinor (spin~$1$~$\otimes$~spin~$\frac12$) that algebraically contains both a spin $\frac12$ and a spin $\frac32$ part. This eigencharacter, together with the structure of the Dirac spinor action (\ref{JGa}) implies that $\phi_{||}$ will rotate, under (\ref{RDefi4VS}), by $\exp(\e_\alpha\frac{\pi}{4}\G^{\alpha})$ (where $\alpha$ is any of the ten simple roots $\alpha_i$ (\ref{GaW})) while $\phi_\bot$ will rotate by $\exp(-\e_\alpha\frac{3\pi}{4}\G^{\alpha})$. Using the fact that each $\G^\alpha$ (or $\e_\alpha\G^\alpha$) can be thought of as being an imaginary unit $\mathbf{i}$ (\ref{Ga2}), so that we can write $\exp(-\frac{3\pi}{4}\mathbf{i})=-\exp(\frac{\pi}{4}\mathbf{i})$, we find that the exponentiated action (\ref{RDefi4VS}) boils down to
\beg
\cR_{\alpha,\e_\alpha}^{\textit{vs}}\phi_{||}&=& \cR_{\alpha,\e_\alpha}^{\textit{s}}\phi_{||},
\\
\cR_{\alpha,\e_\alpha}^{\textit{vs}}\phi_{\bot}&=& -\cR_{\alpha,\e_\alpha}^{\textit{s}}\phi_{\bot}.
\eeg
Now, we recognize in these formul\ae{} the general definition of a geometric reflection in the $10$-dimensional $\beta^a$-vector space (which was identified with the Cartan subalgebra $\mathfrak{h}$): the orthogonal component changes sign, whereas the parallel one remains unaltered. In other words, if we recombine the two `eigenspaces', we find that the exponentiated vector-spinor rotation associated to any dominant wall $\alpha$ is simply given by
\be\label{ke10ferm4}
\phi'{}^{a\,A} &=&\sum\limits_{b=1}^{10}\sum\limits_{B=1}^{32}\left[r_{\alpha}\right]^a{}_b\left[\cR_{\alpha,\e_\alpha}^{\textit{s}}\right]^A{}_B \phi^{b\,B}\\
\text{where}\quad \left[r_{\alpha}(\phi)\right]^a&=&\phi^a -\frac{2\alpha(\phi)}{(\alpha|\alpha)}\alpha^{\# a}
\nn
\ee
is a vector reflection (in $\beta$-space, i.e. $\mathfrak{h}$) in the hyperplane $\alpha(\beta)=0$, and where
\be\label{RDefi4}
\cR_{\alpha,\e_\alpha}^{\textit{s}}&=& e^{\e_\alpha\frac{\pi}{2}J^{\textit{s}}_{\alpha}}\,\,=\,\,e^{\e_\alpha\frac{\pi}{4}\G^{\alpha}}
\ee
is a $32\times 32$-matrix representing the $K(E_{10})$ action on a Dirac spinor. To be clear, we have made explicit all the indices on the redefined gravitino $\phi$ in Eq. (\ref{ke10ferm4}): the vector indices $a,b=1,\dots,10$ (`contravariant' in $\beta^a$ space, i.e. living in $\mathfrak{h}$), and the (heretofore suppressed) spinor indices $A,B=1,\dots,32$ (living in a real, Majorana representation of the Clifford algebra associated to $SO(10)$).\\

The crucial point to note in Eq. (\ref{ke10ferm4}) is that the induced action on the fermionic variable $\phi$ (\ref{phiDefi}) of what was a (Weyl) vector reflection on the bosonic velocity $v\in \mathfrak{h}$ is the \textit{tensor product} of a Dirac spinor action $\cR_{\alpha,\e_\alpha}^{\textit{s}}$ (\ref{RDefi4}) and a vector action $r_{\alpha}$ (\ref{Reflection2}). In other words, the redefinition of the gravitino (\ref{phiDefi}) revealed that the vector-spinor action $\cR_{\alpha,\e_\alpha}^{\textit{vs}}$ (\ref{Jexpl},\,\ref{ke10ferm4}) in fact \textit{factorizes} into a Dirac spinor and a vector action. As is clear from the explicit expression (\ref{ke10ferm4}), this factorization implies that, when considering no longer one isolated collision, but a succession of several of them, we can write, say for a consecutive collision with two walls $\beta,\alpha$:
\be\label{product}
\cR_{\alpha,\e_\alpha}^{\textit{vs}}\cR_{\beta,\e_\beta}^{\textit{vs}}(\phi)
&=&
\big(r_{\alpha}\circ r_{\beta}\big)\otimes 
\big(\cR_{\alpha,\e_\alpha}^{\textit{s}}\cR_{\beta,\e_\beta}^{\textit{s}}\big)
(\phi).
\ee
Applying this factorization to the word representing the fermionic dynamics (\ref{wf}), we see that the fermionic word $w^{\textit{vs}}$ (\ref{wf}) similarly factorizes in the tensor product of the bosonic Weyl word (\ref{wb}) with a corresponding word $w^{\textit{s}}$ in the Dirac-spinor representation:
\be\label{ws}
w^{\textit{s}}&=& \cR_{\alpha_{i_1},\e_{\alpha_{i_1}}}^{\textit{s}}
\cR_{\alpha_{i_2},\e_{\alpha_{i_2}}}^{\textit{s}}
\cdots \,\,\cR_{\alpha_{i_n},\e_{\alpha_{i_n}}}^{\textit{s}}\cdots.
\ee
In the next section, we will show that the redefinition of the gravitino (\ref{phiDefi}) also simplifies the analysis of its degrees of freedom, before investigating the Dirac spinor action $\cR_{\alpha,\e_\alpha}^{\textit{s}}$ in more detail in section \ref{Weyl}.
\end{subsection}

\begin{subsection}{Supersymmetry constraint as a tranversality constraint on a `polarized' billiard particle}\label{redefi}
A further instance of the simplifications brought by the redefinition of the gravitino (\ref{phiDefi}) occurs when considering the ``supersymmetry constraint'' $\mathcal{S}\approx 0$ \cite{DKN06}. This is a constraint on the initial data for the gravitino, which originates from the equation of motion of the gravitino component $\psi^0$. Its explicit expression has been written down in equation (5.14) in \cite{DKN06}. The first observation is that it considerably simplifies in the BKL-limit when considered away from any wall. This is due to the fact that the higher level contributions $P^{(1)}_{a_1a_2a_3},\, P^{(2)}_{a_1\dots a_6},\,\ldots$ as well as the off-diagonal components of $P^{(0)}_{ab}$ become exponentially negligible far away from the wall (i.e. they are proportional to positive powers of $e^{-\alpha(\beta)}$ that vanishes in the BKL limit $\sum_{a=1}^{10}\beta^a\rightarrow + \infty$ (\ref{solution})). Thus, after a rescaling, we are left with a simplified supersymmetry constraint $\mathcal{S}_{\text{BKL}}\approx 0$, with
\be\label{susy1}
\mathcal{S}_{\text{BKL}} &=& \sum\limits_{a=1}^{10}\big(\dot{\beta}^a-\sum\limits_{c=1}^{10}\dot{\beta}^c\big)\G_*\G^a\psi^a.
\ee
If we now replace the original (coset) fermionic variables $\psi^a$ in terms of the redefined gravitino $\phi^a$ (\ref{phiDefi}), we find that $\mathcal{S}_{\text{BKL}}$ acquires a remarkably simple form:
\be\label{susyC}
\mathcal{S}_{\text{BKL}}&=& \sum\limits_{a,b=1}^{10} G_{ab}\dot{\beta}^a\phi^b.
\ee
Note that, contrary to the original form (\ref{susy1}), this form is now \textit{factorized} in the sense that the hidden spinor index on $\phi^b$ is left untouched. The constraint $\mathcal{S}_{\text{BKL}}$ now affects only the vector index of $\phi^a$. More precisely, we see that the constraint $\mathcal{S}_{\text{BKL}}\approx 0$ says that the contravariant billiard velocity $v^a=\dot{\beta}^a$ (\ref{Kasner}) is orthogonal (with respect to the $\beta$-space metric $G$ (\ref{supermetric})) to the `contravariant vector' $\phi^a$ (\ref{phiDefi}). The latter could be thought of as a `polarization direction' orthogonal to the lightlike velocity $v^a$ (\ref{null}) of the `billiard particle' (as if we were talking about a Maxwell-like vector of helicity $\propto \phi^a$).\\

Note the remarkable fact that the ($K(E_{10})$-related) Euclidean $SO(10)$ structure of the original, exact supersymmetry constraint (exhibited in Eq. (5.14) in \cite{DKN06}) has given rise to a constraint which formally looks $SO(9,1)$ invariant, in the sense that it now involves the $(E_{10}$ related) Lorentzian structure $G_{ab}$ (\ref{supermetric}). [One should however keep in mind that the hidden, factorized spinor index on $\phi$ maintains its $SO(10)$-related nature.]\\

Furthermore, notice that the constraint $\mathcal{S}$ transforms as a Dirac spinor under a collision with a wall $\alpha$:
\be\label{susyConTrafo}
\mathcal{S}_{\text{BKL}}'&=& \cR_{\alpha,\e_\alpha}^{\textit{s}}\mathcal{S}_{\text{BKL}}.
\ee 
This follows from combining the transformation properties of $v^a$ (\ref{Reflection2}) and of $\phi^a$ (\ref{ke10ferm4}) and it proves in particular that the constraint $\mathcal{S}_{\text{BKL}}\approx 0$ is preserved under the evolution in the BKL-limit. \\

Note finally how the billiard limit has analagously simplified the Hamiltonian constraint (\ref{null}) and the supersymmetry constraint (\ref{susyC}). They now seem to unite in expressing the `gauge invariance' (in $\beta$-space) of an effective Maxwell-like particle: the billiard limit of the Hamiltonian constraint expresses the massless nature of this $\beta$-space `photon', while the billiard limit of the supersymmetry constraint expresses the transversality constraint $\eta^{\mu\nu}p_\mu A_\nu=0$ linked to the Maxwell gauge invariance. As both constraints originally followed from other gauge invariances (diffeomorphisms and local supersymmetry), this intriguing transformation of gauge constraints in the billiard limit may be the tip of the iceberg of an improved understanding of the \textit{symmetries} of supergravity, and of their possible $E_{10}/K(E_{10})$ coset reformulation.
\end{subsection}

\begin{subsection}{Conserved bilinear form and `gauge transformations'}\label{BilinH}
It was found in \cite{DKN05,BHP05b,DKN06} that there existed a symmetric\footnote{The form $(\cdot|\cdot)$ is symmetric when evaluated on commuting variables, but antisymmetric for Grassmann variables.}, bilinear form $(\cdot|\cdot)_{\textit{vs}}$ on the $\mathbf{320}$-dimensional vector-spinor representation space which was $K(E_{10})$-invariant, and therefore invariant under the exact, coset fermionic equations of motion (\ref{ke10ferm}). This bilinear form was, in particular, invariant under the $SO(10)$ subgroup of $K(E_{10})$ corresponding to $SO(10)$ rotations of the vector index $a$ of $\psi^a$. Indeed, its original expression in terms of two gravitini $\psi_1,\psi_2$ was \cite{DKN06}:
\be\label{form}
(\psi_1|\psi_2)_{\textit{vs}} &:=&-\sum\limits_{a,b=1}^{10}\psi_1^a{}^T\G_{ab}\psi_2^b.
\ee
Let us see how this form is transformed when substituting $\psi^a$ by its expression in terms of the redefined gravitino $\phi^a:=\G_*\G^a\psi^a$ (\ref{phiDefi}). An easy calculation (using $\G_{ab}=\G_a\G_b-\delta_{ab}$) yields
\be\label{bilin}
(\phi_1|\phi_2)_{\textit{vs}} &=&\sum\limits_{a,b=1}^{10}G_{ab}\phi^a_1{}^T\phi_2^b\,\,=\,\,\sum\limits_{a,b=1}^{10}\sum\limits_{A=1}^{32}G_{ab}\phi^{a\, A}_1\phi_2^{b\,A}.
\ee
Note again how the use of the $\phi^a$ variables has revealed a hidden `Lorentzian' structure in the form $(\cdot|\cdot)_{\textit{vs}}$. More precisely, the $K(E_{10})$-covariant bilinear form (\ref{form}) of two gravitini $\phi_1^a$ and $\phi_2^b$ amounts to contracting their $\mathbf{32}$-valued Dirac-spinor indices in the standard Euclidean way, whereas their vector indices are contracted with the Lorentzian metric $G_{ab}$ that was defined on the space of scale factors $\beta$ (\ref{supermetric}). Let us also consider the corresponding norm (when evaluated on real, commuting variables)
\be\label{norm}
 Q^{\textit{vs}}(\phi)&:=&(\phi|\phi)_{\textit{vs}}\,\,=\,\,\sum\limits_{a,b=1}^{10}G_{ab}\phi^a{}^T\phi^b.
\ee
On the full $\phi$-space, this norm has a hyper-Lorentzian signature with $9\times 32$ pluses and $32$ minuses. However, we note that, when the supersymmetry constraint $\mathcal{S}_{\text{BKL}} =0$ is satisfied, the norm $Q^{\textit{vs}}(\phi)$ becomes positive semi-definite. Indeed the supersymmetry constraint $\mathcal{S}$ is saying that (when forgetting about the spinor index) the vector $\phi^a$ is $G_{ab}$-orthogonal to the billiard velocity $\dot{\beta}^b$ (\ref{susyC}). As, by the Hamiltonian constraint (\ref{null}), $\dot{\beta}^a$ is a $G_{ab}$-lightlike vector, $\phi^a$ must then be either $G_{ab}$-spacelike or parallel to $\dot{\beta}^a$, so that $ Q^{\textit{vs}}(\phi)\geq 0$ (when evaluated on real, commuting variables).\\

In addition to this positivity property, the supersymmetry constraint $\mathcal{S}$ (\ref{susyC}) also guarantees that the norm $Q^{\textit{vs}}$ (\ref{norm}) is not affected by the following Maxwell-like (longitudinal) `gauge transformation' of $\phi^a$ along the lightlike direction $\dot{\beta}^a$ (\ref{null}) with any Dirac spinor $\xi\in \mathbf{32}$:
\be\label{gauge}
 Q^{\textit{vs}}(\phi')&=&Q^{\textit{vs}}(\phi)\\
 \text{for}\quad\phi'{}^a&=& \phi^a + \dot{\beta}^a\xi.
\nn
\ee
As another example of the intriguing appearance of a Maxwell-type structure in the fermionic billiard, it is interesting to observe that supersymmetry is a particular case of this gauge transformation (\ref{gauge}). To show this, we insert the redefinition (\ref{phiDefi}) into the supersymmetry variation $\delta\phi^a$ (Eq. (3.9) in \cite{DKN06}). Then in the BKL-limit, due to simplifications similar to those that occurred in the evaluation of the supersymmetry constraint $\mathcal{S}$ (\ref{susy1}), the only surviving term on the RHS of Eq. (3.9) in \cite{DKN06} comes from the term $\propto \frac12N\Omega_{0(ab)}\G^b\G^0\e$ which yields (when keeping the only billiard-surviving contribution to $\Omega_{0(ab)}$, which comes from the time derivative of the diagonal part of the metric)
\be\label{susy2}
\delta_\ep\phi^a&=&\frac12\dot{\beta}^a\ep.
\ee
Note that since $\dot{\beta}^a$ is a lightlike direction (\ref{null}), a supersymmetry variation leaves invariant the orthogonality imposed by the supersymmetry constraint $\mathcal{S}$ (\ref{susyC}). However, in contradistinction to the arbitrary Dirac spinor $\xi$ that we could use in our definition (\ref{gauge}) of a general `coset Maxwell-like gauge transformation', the supersymmetry transformation parameter $\ep$ (\ref{susy2}) is not arbitrary. Since local supersymmetry has been used to gauge-fix $\psi^0$ to vanish (\ref{susyfix}), only residual ``quasi-rigid'' transformations are admissible. In other words, $\ep$ is subject to the constraint $\delta_\ep\psi^0=0$, which yields (see Eq. (5.2) of \cite{DKN06})
\be\label{epDyn}
\p_t\ep &=& \cQ^{\textit{s}}(t)\ep
\ee
where $\cQ$ (\ref{maurer}) is again the same connection that entered the evolution equation for both the bosons (\ref{eomCoset2}) and the gravitino (\ref{ke10ferm}), except that now it acts in the Dirac spinor representation $\textit{s}$ of $K(E_{10})$ (\ref{SDefi}). The previous reasoning (based on the spiky nature of $\cQ(t)$) then shows that the dynamics of $\ep$ (\ref{epDyn}) in the BKL-limit also follows a chaotic billiard motion. In this `Dirac billiard', any collision of the scale factors $\beta^a$ with a wall $\alpha$ induces a $K(E_{10})$-rotation $\cR_{\alpha,\e_\alpha}^{\textit{s}}$ (\ref{RDefi4}) in the Dirac spinor representation $\textit{s}$ (i.e. a $32\times 32$-matrix) in complete analogy to the transformation of the supersymmetry constraint $\mathcal{S}$ (\ref{susyConTrafo}):
\be\label{RDefi3}
 \ep'&=&\cR_{\alpha,\e_\alpha}^{\textit{s}}\ep.
\ee
Note how this result, together with the $r_\alpha$ transformation law of $\dot{\beta}^a$ upon collision on a wall (\ref{Reflection}), implies that the supersymmetry variation term $\delta_\ep\phi^a$ (\ref{susy2}) varies, under a collision, in exactly the same factorized way as the full $\phi^a$ (\ref{ke10ferm4}). This is another check of the consistency of our findings. Due to the importance of the rotations $\cR_{\alpha,\e_\alpha}^{\textit{s}}$ for the dynamics of all fermions, we will continue their analysis in the following section. To conclude, let us remark that Eq. (\ref{susy2}) shows that the effective Maxwell-like gauge transformations of the `polarization vector' added to the $\beta$-particle by the fermionic degrees of freedom comes from supersymmetry transformations. This shows again how the billiard limit exhibits some intriguing metamorphoses of gauge symmetries, as well as of group symmetries ($SO(10)\leftrightarrow SO(9,1)$).
\end{subsection}

\begin{subsection}{Spin extension $\mathcal{W}^{\textit{spin}}$ of the Weyl group $\mathcal{W}$ of $E_{10}$}\label{Weyl}
Summarizing the review of section \ref{maxBKL}, the bosonic dynamics reduce in the BKL-limit to the evolution of the scale factors $\beta^a$ (\ref{Iwasawa}) that follow a billiard motion: Hitting the cushion of the billiard table corresponds to a collision with a dominant wall $\alpha_i$ (\ref{dominant}) which interrupts the intermediate Kasner behaviour (\ref{Kasner}) and whose effect consists of a Lorentzian reflection of the Kasner velocity $v^a=\dot{\beta}^a$ (\ref{Reflection}). The link to the Kac--Moody picture is established by identifying the dominant walls $\alpha_i$ (\ref{dominant}) with the simple roots of the Kac--Moody algebra $E_{10}$ and the space of scale factors $\beta^a$ with its Cartan subalgebra, spanned by the co-roots $\check{\alpha}_i=h_i$ (\ref{hNorm}). Any Lorentzian reflection due to a collision with a dominant wall $\alpha_i$ then corresponds to a fundamental Weyl reflection $r_{\alpha_i}\in \mathcal{W}$ \cite{K95}. Hence, any sequence of collisions during the chaotic evolution of the bosonic billiard corresponds to a \textit{word} $w$ in the Weyl group $\mathcal{W}$ of $E_{10}$ in Eq. (\ref{wb}), and the full evolution towards a spacelike singularity corresponds to a semi-infinite Weyl word: $\cdots w_n\cdots w_2w_1$.\\
 
We have shown in section \ref{Ferm2} that the gravitino can be understood as a (spinor valued) `polarization vector' $\phi^a$ associated to the scale factors $\beta^a$. In particular, any collision induces a discrete $K(E_{10})$-rotation $\cR_{\alpha,\e_\alpha}^{\textit{vs}}$ (\ref{ke10ferm3}) in the (unfaithful) vector-spinor representation. When acting on this `polarization vector' $\phi^a$, the discrete $K(E_{10})$-rotation $\cR_{\alpha,\e_\alpha}^{\textit{vs}}$ factorizes as the tensor product of an ordinary Weyl reflection $r_\alpha$ (acting on the vector index of the suitably redefined gravitino variable $\phi^a$ (\ref{phiDefi})), and of a Dirac spinor rotation $\cR_{\alpha,\e_\alpha}^{\textit{s}}$ (\ref{ke10ferm4},\,\ref{RDefi4}) (acting on the hidden spinor index of $\phi^a$).\\

These results suggest that to describe the general behaviour of fermions in the billiard limit, one should consider a certain abstract `spin' extension of the Weyl group of $E_{10}$ (and more generally of any Kac--Moody algebra), say $\mathcal{W}^{\textit{spin}}$. This group can be abstractly `defined' as the `discrete' subgroup of (a suitable covering of) the `group' $K(E_{10})$, which is (multiplicatively) generated by the formal exponentials
\be\label{Rabstract}
\cR_{\alpha_i}^{\pm}&:=&e^{\pm\frac{\pi}{2}J_{\alpha_i}},
\ee
where $i$ labels the simple roots ($i=1,\dots,10$ in the $E_{10}$-case) and, where $J_\alpha\equiv E_\alpha+\omega(E_\alpha)=E_{\alpha}-E_{-\alpha}$ (\ref{JDefi}) is an element of the $K(E_{10})$ Lie algebra. We have put some quotation marks in this definition, because the precise mathematical setting within which one could define, in full generality, the `group' $K(E_{10})$, its needed `spin' covering (analogous to the $\textit{Spin}(n)$ covering of $\textit{SO}(n)$), and the formal exponentials (\ref{Rabstract}), is unclear to us.\footnote{This covering can probably be viewed as the amalgamation of $\textit{Spin}(2)$ groups associated to each $SO(2)$ group generated by $J_{\alpha_i}$ for any simple root $\alpha_i$. This interpretation will be supported by the observation in Eq. (\ref{GenCoxA}) below that a rotation by $2\pi$, i.e. $\cR_{\alpha_i}^{4}=e^{2\pi J_{\alpha_i}}$ (\ref{Rabstract}), indeed corresponds to a change of sign acting in a spinorial representation of $K(E_{10})$.} On the other hand, our study above has given us two concrete realizations of the abstract generators (\ref{Rabstract}), namely $\cR_{\alpha,\e_\alpha}^{\textit{vs}}$ (\ref{RDefi4VS}) and $\cR_{\alpha,\e_\alpha}^{\textit{s}}$ (\ref{RDefi4}). We shall then focus in the following on the two corresponding matrix groups, say $\SWeyl$ and $\DWeyl$. We define $\SWeyl$ as the subgroup of $\textit{GL}(320)$ multiplicatively generated by (for $i=1,\dots,10$)
\be\label{Wvs}
\cR_{\alpha_i}^{\textit{vs}} &:=& e^{\frac{\pi}{2}J_{\alpha_i}^{\textit{vs}}}
\ee
and $\DWeyl$ as the subgroup of $SO(32)$\footnote{We recall that the $J_{\alpha_i}^{\textit{vs}}$ preserve the Euclidean norm $Q^{\textit{s}}(\ep)=\ep^T\ep$ \cite{DKN06}.} multiplicatively generated by the matrix exponentials of $J_{\alpha_i}^{\textit{s}}$ (\ref{SDefi})
\be\label{Ws}
\cR_{\alpha_i}^{\textit{s}} &:=& e^{\frac{\pi}{2}J_{\alpha_i}^{\textit{s}}}.
\ee
Here, we have dispensed with the signs $\e_\alpha=\pm$ that entered our study above for two reasons:
\begin{enumerate}
	\item We shall see that the matrix generators $\cR_{\alpha_i}^{\textit{vs}}$ and $\cR_{\alpha_i}^{\textit{s}}$ are idempotent (so that a suitable power of, say, $\cR_{\alpha_i}^{\textit{vs}}$ equals $(\cR_{\alpha_i}^{\textit{vs}})^{-1}=e^{-\frac{\pi}{2}J_{\alpha_i}^{\textit{vs}}}$).
	\item The sign of $E_{\alpha_i}$ and therefore of $J_{\alpha_i}=E_{\alpha_i}-E_{-\alpha_i}$ (\ref{JDefi}), is conventional and can therefore be changed to absorb any sequence of $\e_{\alpha_i}$, without loss of generality.\footnote{Note in this respect that $\e_{\alpha_i}$ is the sign of a conserved quantity in the BKL-limit. Hence, the sign $\e_{\alpha_i}$ for any dominant wall $\alpha_i$ cannot flip during the asymptotic Billiard dynamics.}
\end{enumerate}
With this notation, we have proven some mathematical results on the concrete, matrix groups $\SWeyl$ and $\DWeyl$ that we can summarize as follows, before indicating the main elements entering the proofs (the mathematical details of which are given in Appendix \ref{appendix2}).

\begin{subsubsection}*{Proposition 1}
\textit{
\begin{itemize}
	\item The vector-spinor realization $\SWeyl$ of $\mathcal{W}^{\textit{spin}}$, i.e. the group multiplicatively generated by the ten $320\times 320$ matrices $\cR_{\alpha_i}^{\textit{vs}}$ (\ref{Wvs}), is infinite.
	\item The generators $\cR_{\alpha_i}^{\textit{vs}}$ of $\SWeyl$ satisfy the following generalized Coxeter relations:
	\begin{enumerate} [ (a)]
	\item For all nodes $\alpha_i$ in the Dynkin diagram of $E_{10}$ (figure 1), we find
	\begin{subequations}\label{GenCox}
		\be\label{GenCoxA}
\big(\cR_{\alpha_i}^{\textit{vs}}\big)^4&=&-\id.
\ee
\item For adjacent nodes $\alpha_i,\alpha_j$, the corresponding generators fulfill
\be
\big(\cR_{\alpha_i}^{\textit{vs}}\cR_{\alpha_j}^{\textit{vs}}\big)^3&=&-\id.
\ee
\item For non-adjacent nodes $\alpha_i,\alpha_j$, the corresponding generators commute.
\end{subequations}
\end{enumerate}
	\item The squares of the matrices $(\cR_{\alpha_i}^{\textit{vs}})^2$ generate a normal subgroup $\SWeylN$ of $\SWeyl$, which is \textbf{non-abelian} and whose cardinality is $2048$.
	\item The Weyl group $\mathcal{W}$ of $E_{10}$ is \textbf{isomorphic} to the quotient group $\SWeyl/\SWeylN$:
	\be\label{homo2}
	\mathcal{W}&\simeq&\SWeyl/\SWeylN.
	\ee
	\end{itemize}
}
\end{subsubsection}
The corresponding statement concerning the Dirac-spinor realization of $\mathcal{W}^{\textit{spin}}$ looks similar, but differs in a very important aspect: $\DWeyl$ is a finite group, while $\SWeyl$ was infinite (like the Weyl group of $E_{10}$):
\begin{subsubsection}*{Proposition 2}
\textit{
\begin{itemize}
	\item The spinor realization $\DWeyl$ of $\mathcal{W}^{\textit{spin}}$ (i.e. the group multiplicatively generated by the ten $32\times 32$ matrices $\cR_{\alpha_i}^{\textit{s}}$ (\ref{SDefi},\,\ref{Ws}) with $i=0,\dots,9$) is of finite cardinality.
	\item The generators $\cR_{\alpha_i}^{\textit{s}}$ of $\DWeyl$ fulfill the same generalized Coxeter relations (\ref{GenCox}) as those of $\SWeyl$.	
	\item The squares of the matrices $(\cR_{\alpha_i}^{\textit{vs}})^2$ generate a normal subgroup $\DWeylN$ of $\DWeyl$, which is isomorphic to $\SWeylN$ defined for the vector-spinor representation.
	\item There exists a homomorphism $\Phi$ mapping the Weyl group $\mathcal{W}$ of $E_{10}$ to the quotient group $\DWeyl/\DWeylN$:
	\be\label{homo4}
	\Phi:\mathcal{W}&\rightarrow&\DWeyl/\DWeylN.
	\ee
	The non-trivial kernel of $\Phi$ forms an (infinite cardinality) \textbf{normal subgroup} of the Weyl group $\mathcal{W}$ of $E_{10}$.
	\end{itemize}
}
\end{subsubsection}
It is interesting to contrast these results on what we called representations of the `spin' extension $\mathcal{W}^{\textit{spin}}$ of the Weyl group, to the results of Kac \cite{K95} (p.36) on some representations of the usual Weyl group $\mathcal{W}$ of a Kac--Moody algebra, namely:
\begin{subsubsection}*{Proposition 3 [Kac]}
\textit{
Let $\rho$ be an integrable\footnote{A representation $\rho$ of a Kac--Moody algebra on a vector space is called integrable, if the raising and lowering operators $e_i$ and $f_i$, defined in section \ref{coset1}, are represented as locally nilpotent endomorphisms \cite{K95}.} representation of $E_{10}$ on a vector space $V$. To any Chevalley triple $(e_i,f_i,h_i)$ with $i=0,\dots,9$, linked to the ten nodes in the Dynkin diagram of $E_{10}$ (figure 1), associate an element in the space of endomorphisms $\textit{End}(V)$ by
\be\label{rKac}
r_{\alpha_i}^{\rho}&:=&\exp(f_i^\rho)\exp(-e_i^\rho)\exp(f_i^\rho).
\ee
These ten elements $r_{\alpha_i}^{\rho}$ generate a group $\aWeyl\subset \textit{End}(V)$. Furthermore, their squares form an \textbf{abelian} normal subgroup $\aWeylN$ of $\aWeyl$
\be\label{HDefi}
 \aWeylN&:=&\langle\langle (r_{\alpha_i}^{\rho})^2|i=0,\dots,9\rangle\rangle
\ee
such that the quotient group is \textbf{isomorphic} to the Weyl group $\mathcal{W}$ of $E_{10}$
\be\label{Kac}
\mathcal{W}&\simeq&\aWeyl/\aWeylN
\ee
}
\end{subsubsection}
Referring to Appendix \ref{appendix2} for details of the proofs of proposition 1 and 2, let us comment on some of their salient aspects:
\begin{itemize}
	\item The fact that for $\SWeyl$ and $\DWeyl$, we get the specific \textit{generalized} Coxeter relations of propositions 1 and 2 (i.e. $(\cR_\alpha^{\textit{vs}})^4=-1$ and therefore $(\cR_\alpha^{\textit{vs}})^8=1$ instead of $r_\alpha^2=1$ for the usual Weyl group) comes from the spinorial nature of the representations $\textit{vs}$ and \textit{s}. Technically, this is due to the factor $\frac12$ in $J_\alpha^{\textit{s}}=\frac12\G^\alpha$ which leads to factors $e^{\pm \frac{i\pi}{4}}$ in $\cR_\alpha^{\textit{vs}}$ and $\cR_\alpha^{\textit{s}}$. By contrast, the action of $\cR_\alpha=e^{\frac{\pi}{2}J_\alpha}$ on a Cartan valued $\cP$ in the coset representation led (as in the setting of Kac's proposition 3) to phase factors $e^{\pm i\pi}$ in $\cR_\alpha^{\textit{coset}}$ (\ref{pi2},\,\ref{RTH}). More generally, note that an eigenvalue $\pm is$ of $J_\alpha$ (that we shall refer to as being `spin $s$') entails a phase factor $e^{\pm\frac{i\pi}{2}s}$. In this language, the phase factors $e^{\pm i\pi}$ in $\cR_\alpha^{\textit{coset}}$ can be thought of as coming from the fact that $\cP_{ab}^{(0)}$ was a symmetric $SO(10)$-tensor, hence `spin $2$' versus `spin $\frac12$' of the basic Dirac-spinor representation $J_\alpha^{\textit{s}}=\frac12\G^\alpha$ with $(\G^\alpha)^2=-\id$. [Note in this respect that the type of generalized Coxeter relations, and associated extensions of the Weyl group  that have been studied in \cite{K95,KP85} would, in the present language, be connected to \textit{integer} `spin $s$'. The integral nature of $s$ guarantees for these cases that the Weyl group is extended in such a way that the generators $e^{\frac{\pi}{2}J_{\alpha_i}}$ ($i=0,\dots,9$) are represented by fourth roots of unity.]
	\item The finiteness of $\DWeyl$ was far from being a priori evident. This finite cardinality does not follow from the unfaithfulness of the spinor representation because the vector spinor representation \textit{vs} is also unfaithful (it leads to a finite-dimensional representation space), but $\SWeyl$ is infinite. This finite cardinality does not either follow alone from the fact that $\DWeyl$ preserves a positive definite norm, namely $Q^{\textit{s}}(\ep)=\ep^T\ep$, because the action of the infinite Weyl group on the compact unit spere $\ep^T\ep=1$ in $\R^{32}$ could have led to a `chaotic' non-periodic action with many accumulation points.
	\item The technical difficulty in proving the finiteness of $\DWeyl$ came from the fact that (using $(\G^\alpha)^2=-1$, like $i$)
	\be\label{Ra2}
	\cR_\alpha^{\textit{s}}\,\,=\,\,e^{\frac{\pi}{2}J_\alpha^{\textit{s}}}&=&e^{\frac{\pi}{4}\G^\alpha}\,\,=\,\,\frac{1+\G^\alpha}{\sqrt{2}}
	\ee
	contains $\sqrt{2}$ in the \textit{denominator}. Therefore, a product of $n$ exponentials or, as we shall say, a word of length $n$, $\cR_{\alpha_1}^{\textit{s}}\cR_{\alpha_2}^{\textit{s}}\cdots \cR_{\alpha_n}^{\textit{s}}$ a priori contains a factor $(1/\sqrt{2})^n$ which could decrease without bound as $n\rightarrow \infty$ and generate an infinite number of group elements in $\DWeyl$. Appendix \ref{appendix2} explains in detail how we surmounted this difficulty. Let us here only say that we used a four-pronged approach:
\begin{enumerate}[(1)]
	\item We crucially use the fact that we are working within an \textit{algebra}, which follows from the representation in Eq. (\ref{Ra2}). The anticommutation properties of Clifford matrices $\G^a$ then imply some definite (anti-) commutation relations among the $\cR_\alpha^{\textit{s}}$ which involve coefficients of the form $a+b\sqrt{2}$ with $a,b\in \Z$. At this first step, the use of the algebraic relations among elementary products $\cR_{\alpha_1}^{\textit{s}}\cR_{\alpha_2}^{\textit{s}}\cdots$ has allowed us to push the dangerous $\sqrt{2}$'s up to the \textit{numerator}.
	\item Then using repeatedly, as in the proof of Wick's theorem, the (anti-) commutation relations among the $\cR_\alpha^{\textit{s}}$, we can reorder any element of $\mathcal{W}^{\textit{s}}$ and express it as a \textit{finite} linear combination of \textit{ordered} $\cR_\alpha^{\textit{s}}$-products with $\Z[\sqrt{2}]$-valued coeffcients. [Here we use the fact that the Galois extension $\Z[\sqrt{2}]=\{a+b\sqrt{2};a,b\in\Z\}$ is a \textit{ring}, i.e. that it is stable under addition and multiplication.] This leads to the following result: any word $w$ in $\DWeyl$ uniquely corresponds to an element in a $\Z[\sqrt{2}]$-lattice spanned by the \textit{finite basis of ordered products} $\cR_{\alpha_1}^{\textit{s}}\cR_{\alpha_2}^{\textit{s}},\dots$.
		\item At this stage, we can substitute the expression (\ref{Ra2}) of $\cR_{\alpha_i}^{\textit{s}}$ in terms of $\G$-matrices in the general expression of any word $w$ in $\DWeyl$. This change of basis introduces $\sqrt{2}$-denominators, \textit{but only up to a finite power} (which turns out to be $(1/\sqrt{2})^{10}$), because the basis is finite.
	\item Expressing the `orthogonality' $(w^T w=\id_{32})$ of a general word $w\in \DWeyl\subset SO(32)$ then leads to many $\Z[\sqrt{2}]$-valued diophantine equations for the coefficients $d_A\in \Z[\sqrt{2}]$ of the expansion of $w$ (with $A$ labelling the finitely many basis elements). One of these $\Z[\sqrt{2}]$-diophantine equations reads
	\be\label{sphere}
	32&=&\sum\limits_{A}d_A^2.
	\ee	
	Although the lattice of generalized integers $a+b\sqrt{2}$ is \textit{dense} on the real line, there is a \textit{finite number} of $\Z[\sqrt{2}]$-valued points on the `sphere' (\ref{sphere}) (as can be easily seen by decomposing Eq. (\ref{sphere}) in its two components along the basis $\{1,\sqrt{2}\}$). 
\end{enumerate}
	\end{itemize}
	It might be interesting to add that our proof of proposition 2 of the generalized Coxeter properties and of the finiteness of the spinor representation $\DWeyl$ of the abstract $\mathcal{W}^{\textit{spin}}$ on the one hand can be straightforwardly generalized to other Kac--Moody algebras, and, on the other hand, can be reformulated within a more abstract setting which does not make use of the gamma-matrix realization (\ref{JGa},\,\ref{GaW}) of the `compact' generators $J_{\alpha_i}^{\textit{s}}$. Indeed, it seems that the main structure behind the proof is the following abstract, Kac--Moody related `graded Clifford algebra'. First, independently of any explicit representation, we can always \textit{define} some abstract generators $\G^{\alpha_i}$ (associated to each node of a Dynkin diagram) as $\G^{\alpha_i}:=2J^{\textit{s}}_{\alpha_i}$, i.e. such that Eq. (\ref{JGa}) holds by definition. An abstract `Dirac-spinor' realization can then be defined by requiring that: 
	\begin{enumerate}[(1)]
	\item all the `spin eigenvalues' of the basic (`Hermitian') rotation generators $iJ^{\textit{s}}_{\alpha_i}$ are equal to $j^{\textit{s}}_{\alpha_i}=\pm\frac12$. This implies that $(\G^{\alpha_i})^2=-\id$;
	\item the $\G^{\alpha_i}$'s corresponding to different nodes of the Dynkin diagram satisfy \textit{either anticommutation or commutation} relations $[\G^{\alpha_i},\G^{\alpha_i}]_{\pm}=-\delta_{ij}$. E.g. for the $E_{10}$ case, we have \textit{anticommutation} for adjoint nodes (when $A_{ij}=-1$), and \textit{commutation} for disconnected nodes $(A_{ij}=0)$.
\end{enumerate}
The same relation between this `graded' Clifford structure\footnote{It seems that this structure, i.e. the properties (1),(2) above, follow from the requirement that the eigenvalues of a general $K(E_{10})$-\textit{normalized} ($\sum_i n_{\alpha_i}^2=1$ in the simply-laced case) linear combination $\sum_i n_{\alpha_i}J^{\alpha_i}$ has eigenvalues $\pm \frac{i}{2}$.} (anticommutation versus commutation) and the Dynkin diagram ($A_{ij}=-1$ or $A_{ij}=0$) holds (in the generic case $D>4$) for the $AE_{D-1}$ Kac--Moody algebra associated to pure gravity in space-time dimension $D$ \cite{DHJN01}. [The case $D=4$, which will be treated below, is non-generic in that the special `gravity-wall' generator $\G^{\alpha_*}$ commutes with the symmetry-wall generators.] Our proof applies in such a general setting (provided the strictly ordered products of $\G^{\alpha_i}$'s define a linearly independent basis), even if the representation space were infinite (though necessarily made of finite, reducible components).\\

	Concerning the links between the content of the propositions 1 and 2, and Kac's proposition 3, we think that the definition of $r_{\alpha_i}^{\rho}$ (\ref{rKac}), though it looks very different from that of $\cR_{\alpha_i}^{-}=e^{-\frac{\pi}{2}J_{\alpha_i}}$ (\ref{Rabstract}), will be formally equivalent to it in any $E_{10}$-representation $\rho$ in which the exponential $e^{-\frac{\pi}{2}J_{\alpha_i}^\rho}$ is well defined.\footnote{Any $E_{10}$ representation $\rho$, defined by the representation of the Chevalley generators $(e_i^\rho,f_i^\rho,h_i^\rho)$, canonically induces a representation of its subalgebra $K(E_{10})$, generated by $J_{\alpha_i}^\rho:=e_i^\rho-f_i^\rho$ (\ref{JDefi}).} Indeed, a reshuffling (\`a la Baker-Campbell-Hausdorff), needed to pass from one form to the other, only involves commutators of the $sl_2$ subalgebra generated by $e_i$, $f_i$ and $h_i=[e_i,f_i]$, so that it should suffice to check it in the natural representation \cite{K95} $e_i^{\textit{sl}_2}=\binom{0\,\,1}{0\,\,0}$, $f_i^{\textit{sl}_2}=\binom{0\,\,0}{1\,\,0}$ and $J_{\alpha_i}^{\textit{sl}_2}=e_i^{\textit{sl}_2}-f_i^{\textit{sl}_2}= \binom{\,0\,\,\,\,1}{-\!1\,\,0}$. For this case, one finds $r_{\alpha_i}^{\textit{sl}_2}=\binom{0\,\,-\!1}{1\,\,\,\,0}$ that indeed agrees with $\exp(-\frac{\pi}{2}\binom{\,0\,\,\,\,1}{-\!1\,\,0})=-\binom{\,0\,\,\,\,1}{-\!1\,\,0}$.\footnote{Note that one would have obtained $\exp(+\frac{\pi}{2}\binom{\,0\,\,\,\,1}{-\!1\,\,0})$, if one had changed $f_i\mapsto -f_i$, $e_i\mapsto -e_i$ in Kac's definition (\ref{rKac}).} Note in this context that an extension of the Weyl group of a Kac--Moody algebra, together with generalized Coxeter relations, also appears in the construction of the underlying Kac--Moody \textit{group} structure in the work of Kac and Peterson \cite{KP85}. However, Ref. \cite{KP85} introduces an extension of the Weyl group by $(\Z_2)^{\textit{r}}$ (where $\textit{r}$ is the rank), which means that the generalized Coxeter generators are fourth roots of unity (contrary to our `spinorial' extension that involves eighth roots of unity).\\
	
It is clear that there are similarities between the statements made in the propositions 1-3. In particular, the group $\mathcal{W}^\rho$ of proposition 3 also is an extension of the Weyl group. We wish however to emphasize that our results in the propositions 1 and 2 can in no way be deduced from proposition 3 for several deep reasons. First, our propositions concern $K(E_{10})$ representations and not $E_{10}$ representations. Although any $E_{10}$ representation also is a $K(E_{10})$ representation by restricting the action of $E_{10}$ to its subalgebra $K(E_{10})$, the reverse is clearly false in the case we consider in our propositions, because both the vector-spinor representation $\textit{vs}$ (\ref{Jexpl}) and the spinor representation $\textit{s}$ (\ref{SDefi}) of $K(E_{10})$ being unfaithful (and finite-dimensional) can certainly \textit{not} be lifted to representations of $E_{10}$ [indeed, $E_{10}$, being simple, does not admit unfaithful representations (apart from the trivial one)]. Second, the structure of the normal subgroups that are factored out is different: while in our propositions the normal subgroups $\SWeylN,\DWeylN$  are \textit{non-abelian}, $\mathcal{D}^\rho$ is abelian in the case of proposition 3.\\

Note, however, that Kac has used the setting of his proposition 3 to prove a result which is closely related to the one we found above for the \textit{bosonic} billiard. Indeed, by applying the proposition to the adjoint representation $\rho=\textit{ad}$, he showed that the adjoint action of $r_{\alpha_i}^{\textit{ad}}$ (\ref{rKac}) to the Cartan subalgebra $\mathfrak{h}\subset E_{10}$ reproduces the action of the fundamental Weyl reflection $r_{\alpha_i}$ \cite{K95}. Given, for $E_{10}$ representations $\rho$, the link $r_\alpha^{\rho}\sim e^{\pm\frac{\pi}{2}J_{\alpha}^\rho}$ mentioned above, this agrees with our observation in section \ref{one} in which we evaluated the $K(E_{10})$-rotation $\cR_{\alpha_i,\e_{\alpha_i}}^{\textit{Ad}}$ (\ref{RDefi2}) on the element of the Cartan subalgebra that corresponded to an incoming Kasner geodesic (\ref{Pexpl},\,\ref{Reflection2}).\\

In the next section, we will investigate the truncation of SUGRA$_{11}$ to $D=4$ $\cN=1$ supergravity that can be linked to the Kac--Moody algebra $AE_3$ in the BKL-limit as mentioned in section \ref{subsystem}. The analysis of this simpler case will provide a useful `toy setting' in which some aspects of our $E_{10}$ results become easier to analyze in detail. In particular, we shall be able to solve explicitly the supersymmetry constraint $\mathcal{S}$ (\ref{susyC}) and to explicitly exhibit the remaining degrees of freedom of the gravitino in terms of two Dirac spinors.
\end{subsection}

\end{section}

\begin{section}{Fermionic billiard of $D=4$ $\cN=1$ supergravity and its $K(AE_{3})$ structure}\label{KAE3}
\begin{subsection}{$AE_3$ and $D=4$ $\cN=1$ supergravity}\label{AE31}
In order to facilitate the link to the $E_{10}$-description of maximal supergravity, we want to describe $D=4$ $\cN=1$ supergravity as a truncation of the former theory reduced on a seven-torus $T^7$. The analysis of section \ref{subsystem} reviewed the known fact that the hyperbolic Kac--Moody algebra $AE_3$ naturally appears in a description of the BKL-limit of $d=4$ gravity \cite{BKL,DHN02}. We start this section with a review of the bosonic dynamics. Since the gravitational walls are linked to the structure functions $C^a{}_{bc}$ introduced in section \ref{subsystem}, it is sufficient to focus on their dominant contribution to the supergravity Lagrangian in the BKL-limit, which is
\be\label{LAE3}
\cL_{C}&=&-\cV_g\nn\\
&=& +\det(g_{mn}) R^{\text{spatial}}\nn\\
&=& -\frac14\sum\limits_{a,b,c=1}^3 e^{-2\alpha^g_{abc}(\beta)}(C^a{}_{bc})^2+\text{subleading terms}
\ee
where the gravitational wall $\alpha^g_{abc}$ was defined in Eq. (\ref{gravwalls}) and where we used as usual the gauge choice $N=\sqrt{\det(g_{mn})}$ (\ref{lapse}) \cite{DHN02} with $m,n=1,2,3$. It was shown in \cite{DHN02b,DHN02,dBD07} that the (Iwasawa frame) structure functions $C^a{}_{bc}$ have limits on the singularity, and that they can be identified with the conserved conjugate momenta $\Pi_\alpha$ (\ref{constmom}) of coset variables $\nu_\alpha$ parametrizing a level-$1$ coset variable $\Phi_{ab}=\Phi_{ba}$. Here, we are interested in extracting from all the gravitational walls the \textit{dominant} one near the singularity. In view of the ordering among the $\beta$'s imposed by the symmetry walls ($\alpha^{\textit{s}}_{(12)}(\beta)>0$ and $\alpha^{\textit{s}}_{(23)}(\beta)>0$ implying $\beta^1<\beta^2<\beta^3$), it is easily seen that the dominant gravitational wall will be $\alpha^g_{123}$, which is connected to $C^1{}_{23}$. Then, either by using the coset-gravity dictionary given in \cite{DHN02b,DHN02}, or by identifying (in the gauge $n=1$) the one wall potential $e^{-2\alpha_*(\beta)}\Pi_{\alpha_*}^2$ of Eq. (\ref{energy2}) to the dominant gravitational potential $\frac14 e^{-2\alpha^g_{123}(\beta)}(C^1{}_{23})^2$ (\ref{LAE3}), we find that the dominant gravitational wall $\alpha_*\equiv\alpha^g_{123}$ contributes the following term to the coset connection $\cQ|_{\alpha}$ (\ref{PQtrunc}):
\be\label{QAE3}
\left.\cQ\right|_{\alpha_*}&=&\frac12 e^{-\alpha_*(\beta)}C^1{}_{23}J_{\alpha_*}.
\ee
This equation will be crucial to obtain the correct normalization of the $K(AE_3)$-generator $J_{\alpha_*}$ in the next section.
\end{subsection}

\begin{subsection}{$K(AE_3)$ and the fermions of $D=4$ $\cN=1$ supergravity}\label{fermi2}
The dynamics of the gravitino $\psi$ of $D=4$ $\cN=1$ supergravity can e.g. be taken from \cite{FN76} (neglecting third order terms in fermions and with $\p_t=N\p_0$)
\be\label{sugraF2}
0&=&\sum\limits_{\alpha=0}^3 \g^{\alpha}\nabla_{[\alpha}\psi_{\beta]}
\ee
with the standard Levi--Civita connection $\nabla$, the Clifford algebra $\{\g^\alpha,\g^\beta\}=\eta^{\alpha\beta}$ and $\eta=\text{diag}(-+++)$. We use the same redefinitions (\ref{fermTrafo}) as for the $E_{10}$-case (i.e. the same rescaling, but with indices restricted to $a=1,2,3$) and adopt the supersymmetry gauge $\psi^0=0$ (\ref{susyfix}). 
This leads to the following explicit form for the time evolution of the (coset) fermionic variable $\psi_a$ (using Einstein's summation convention)
\be\label{gravExpl}
-\p_t\psi_a&=&N\left(\omega_{0ab}\psi^b +\frac14\omega_{0ef}\g^{ef}\psi_a
\right.\\
&&\left.
+(\omega_{abc}-\omega_{bac})\g^0\g^b\psi^c +\frac12\omega_{abc}\g^0\g^{bcd}\psi_d -\frac14\omega_{bcd}\g^0\g^{bcd}\psi_a
 \right)\nn
\ee
Our task now is to single out on the RHS of this equation the dominant terms in the BKL-limit: i.e. the ones corresponding to the two symmetry walls $\alpha^{\textit{s}}_{(12)},\,\alpha^{\textit{s}}_{(23)}$ and the gravitational wall $\alpha_*=\alpha^g_{123}$. First, we notice that the temporal derivatives of the metric coefficients completely follow the scheme used in the $E_{10}$-case (\ref{ke10ferm},\,\ref{Jexpl}): They give rise to the symmetry walls (\ref{symmW}). One then easily finds that they lead to the same results (restricted to indices $a=1,2,3$) for the corresponding `compact' generators $J_{\alpha^{\textit{s}}_{(12)}},\,J_{\alpha^{\textit{s}}_{(23)}}$. Denoting for simplicity $\alpha_1:=\alpha_{(12)}^{\textit{s}}$ and $\alpha_2:=\alpha_{(23)}^{\textit{s}}$ as in (\ref{symmW}), one finds the following explicit expressions for the two putative $K(AE_3)$ generators $J_{\alpha_1}$ and $J_{\alpha_2}$ corresponding to these symmetry walls:
\be\label{JexplAEs}
J_{\alpha_{i}}^{\textit{vs}}(\psi)^a&=& \frac12 \g^{i}\g^{i+1}\psi^a +2\delta^{a[i}\psi^{i+1]} \quad\text{for }i\,=\,1,2.
\ee
If we now extract from the RHS of Eq. (\ref{gravExpl}) the contribution proportional to the dominant gravitational wall $\alpha_*:=\alpha^g_{123}$, we must be careful to collect several different connection terms. Indeed, $C^1{}_{23}$ is proportional to $\Omega_{231}$ in the notation of \cite{DKN06}. Then the formula giving the `spin connection' $\omega_{abc}$ in terms of $\Omega_{abc}$ (see Eq. (\ref{Omega})) shows that $C^1{}_{23}\propto \Omega_{231}$ will enter in $\omega_{123}$, $\omega_{231}$ and $\omega_{312}$. Finally, we get that the dominant gravitational wall contribution to the equation of motion reads after substituting (\ref{lapse},\,\ref{vielbein}):
\beg
-\p_t\psi_a&=&\frac12e^{-2\beta^1}C^1{}_{23}\g^0\left(
 4\delta_{a}^{[2} \g^{3]}\psi^1  +\g^{123}\left[\frac12\psi^a -2\delta_a^1\psi^1 
 \right]\right).
 \eeg
By rewriting this equation in the universal coset-model form (\ref{ke10ferm})
\be\label{kAE3ferm}
\p_t \psi&=& \cQ^{\textit{vs}}(\psi),
\ee
we can fix (by comparison with our previous result (\ref{QAE3}), and remembering that the dominant gravitational wall $\alpha_{*}(\beta)= 2\beta^1$ (\ref{gravwalls}) does indeed give a factor $e^{-\alpha_*(\beta)}=e^{-2\beta^1}$) the form of the putative $K(AE_3)$ vector-spinor representation $\textit{vs}$ of the generator $J_{\alpha_*}$ in terms of $\g$-matrices:
\be\label{JexplAE}
J_{\alpha_*}^{\textit{vs}}(\psi)^a 
&=&
-\g^0\left(
 4\delta_{a}^{[2} \g^{3]}\psi^1  +\g^{123}\left[\frac12\psi^a -2\delta_a^1\psi^1 
 \right]\right).
\ee
At this stage, we extracted from the SUGRA$_{4}$ fermionic equations of motion explicit $\g$-matrix expressions for the generators $J_{\alpha_1}^{\textit{vs}}$, $J_{\alpha_2}^{\textit{vs}}$ and $J_{\alpha_*}^{\textit{vs}}$ that should give rise to a `vector-spinor' representation of $K(AE_3)$. The proof that these generators indeed form a representation of the compact subalgebra $K(AE_3)$ of $AE_3$ crucially depends on the correct normalization of the `gravitational wall' generator $J_{\alpha_*}^{\textit{vs}}$ (which is fixed by the precise coefficient in Eq. (\ref{QAE3})). We start by recalling that $K(AE_3)$ is defined by the free Lie algebra generated by the compact generators $J_{\alpha_i}$ (\ref{JDefi}) for $i=*,1,2$ \textit{modulo} Serre-like relations \cite{DKN06} that are inherited from the Serre relations of $AE_3$. In particular, the condition $\text{ad}^3_{E_{\alpha_*}}(E_{\alpha_{1}})=0$ that is encoded in the Dynkin diagram (figure 2) gives rise to the following relation for $J_{\alpha_i}$ involving the exceptional node:
\be\label{Jrel}
\text{ad}^3_{J_{\alpha_*}}(J_{\alpha_{1}}) + 4\text{ad}_{J_{\alpha_*}}(J_{\alpha_{1}})&=&0.
\ee
As in the $K(E_{10})$-case treated in \cite{DKN06}, the manifest $SO(d)$ invariance of the gauge-fixed gravitino equations of motion (which follow from their original, local Lorentz covariance) ensures that the condition (\ref{Jrel}) is the only Serre-like relation that needs to be checked. By doing the explicit $\g$-algebra calculation involved in Eq. (\ref{Jrel}) when using the expressions (\ref{JexplAEs},\,\ref{JexplAE}) for $J_{\alpha}^{\textit{vs}}$ in terms of $\g$-matrices, we have verified that Eq. (\ref{Jrel}) is fulfilled and hence, that these generators $J_{\alpha}^{\textit{vs}}\in \R^{12\times 12}$ indeed form a matrix representation of $K(AE_3)$. 
\\

This result was expected because the fact that the equations of motion of SUGRA$_{4}$ admits a $K(AE_3)$-covariant coset reformulation (in the BKL-like approximation where one keeps the leading effect of all the gravitational walls $\alpha^{\textit{g}}_{abc}$ with $a,b,c$ all different), actually follows from three known results:
\begin{enumerate}[(1)]
	\item The fermionic dynamics of SUGRA$_{11}$ admits a $K(E_{10})$-covariant reformulation up to a level of approximation (``$l=3^{-}$'') which includes all the gravitational walls associated to real roots \cite{DHN02b}.
	\item $AE_3$ is a subalgebra of $E_{10}$ obtained by keeping only a fraction of the generators of $E_{10}$ \cite{K95,KN05b}. This truncation, being compatible with the Chevalley involution, also implies that $K(AE_3)$ is a subalgebra of $K(E_{10})$.
	\item The dynamics of SUGRA$_4$ can be obtained by reducing the dynamics of SUGRA$_{11}$ (truncating away, in particular, both the spatial dependence upon a seven-torus $T^7$ and the corresponding gravitino components).
\end{enumerate}
It is moreover interesting to note that the `gravity truncation' mentioned in fact (3) does canonically determine an embedding of $AE_3$ within $E_{10}$. Indeed, this truncation shows that, besides the two dominant symmetry walls $\alpha_{(12)}^{\textit{s}}$, $\alpha_{(23)}^{\textit{s}}$, surviving the reduction $10+1\rightarrow 3+1$, the third simple root of $AE_3$ should come from the $D=11$ (or $E_{10}$) gravitational wall associated with $C^1{}_{23}$, whose dual form (entering the $E_{10}$ gravity-coset dictionary \cite{DHN02,DHN02b,DN04}\footnote{In the notation of \cite{DHN02b}, this is $DA^{1|14\dots 10}$, and in the one of \cite{DN04}, it is $P^{(3)}_{1|14\dots 10}$. It also corresponds to the conjugate momentum $\Pi^{11}$ of $\phi_{11}$ in \cite{DHN02}.}) is $\Pi^{1|1\,4\,5\,6\,7\,8\,9\,10}$, and which is therefore associated to the level-3 $E_{10}$ generator $E^{1|14\dots 10}$. In other words, the gravity truncation $11\rightarrow 4$ predicts that the subset of $E_{10}$ generators\footnote{These conventions for the labelling of the generators is directly linked with the fact that the Iwasawa variable $\cN$ was chosen to be upper-triangular, which implied that e.g. $\alpha_{(12)}^{\textit{s}}=\beta^2-\beta^1$. Note that these conventions differ from the ones used in \cite{DKN06}, e.g. $e_1^{\text{[DKN]}}=K^1{}_2$. With our conventions, the exceptional generator in the $E_{10}$ context is $e_0=E^{123}$ instead of $e_0^{\text{[DKN]}}=E^{8\,9\,10}$.} $e_1:=K^2{}_1$ (associated to $\alpha_{(12)}^{\textit{s}}$), $e_2:=K^3{}_2$ (associated to $\alpha_{(23)}^{\textit{s}}$) and $e_*:=E^{1|14\dots 10}$ (associated to $\alpha_{123}^{\textit{g}}$) together with their Chevalley duals $f_i$ and the corresponding $h_i=[e_i,f_i]$ should be identifiable with the three basic Serre--Chevalley generators of $AE_3$. The fact that the Cartan matrix associated to $\alpha_1$, $\alpha_2$ and $\alpha_*$ coincides with the Cartan matrix of $AE_3$ is known and has already been reviewed in section \ref{subsystem} above. It is less evident that $e_1$, $e_2$ and $e_*$ satisfy the required Serre relations of $AE_3$, and notably the delicate one
\be\label{serreae3}
\text{ad}^3_{e_{*}}(e_{{1}})=0.
\ee
One can directly prove that (\ref{serreae3}) is satisfied within the $E_{10}$ Lie algebra. Either one can use the general theorem proven in \cite{FN03} or notice that the putative $E_{10}$-root $3\alpha_*+\alpha_1$ associated to the LHS of (\ref{serreae3}) has a squared norm $(3\alpha_*+\alpha_1)^2=+8$, which is too large for the LHS of (\ref{serreae3}) to be an $E_{10}$-generator.\footnote{We thank P.~Cartier for suggesting this test.}\\

This reasoning shows that the third $K(AE_3)$ generator $J_{\alpha_*}=E_{\alpha_*}+\omega(E_{\alpha_*})$ (with $E_{\alpha_*}=e_*$) can be embedded within $K(E_{10})$ as a generator of the form\footnote{Note that for the truncation of SUGRA$_{11}$ to supergravities in $d+1$ dimensions (associated to the Kac--Moody algebra $AE_d$), the $K(E_{10})$ generator linked to the dominant gravitational wall is $J_{\alpha_*}=c_{\alpha_*}J^{1|1\dots d-2\, d+1\dots 10}$.}
\be\label{E10vgl}
J_{\alpha_*}&=&c_{\alpha_*}J^{1|1\,4\dots 10}\,\,=\,\,c_{\alpha_*}\left(E^{1|1\,4\dots 10} -F_{1|1\,4\dots 10}\right).
\ee 
By starting from the unfaithful $K(E_{10})$ representation of $J^{a_0|a_a\dots a_8}$ given in Eq. (2.30) of \cite{DKN06}, and decomposing the Clifford(10) algebra according to $\G^{a'}=\g^{a'}\otimes \id_8$ (for $a'=1,2,3$ and setting $\psi^{\bar{a}}=0$ for $\bar{a}=4,\dots,10$), we have checked the consistency of (\ref{E10vgl}) with the Clifford(3)-algebra representation (\ref{JexplAE}) and determined that the numerical coefficient $c_{\alpha_*}$ in Eq. (\ref{E10vgl}) must be equal to $c_{\alpha_*}=\pm \frac13$. [This factor $\frac13$ also follows from the normalization adopted in \cite{DKN06} for the level-3 generators which implies $(E^{1|1\,4\dots 10}|F_{1|1\,4\dots 10})=9$. The sign of $c_{\alpha_*}$ is unimportant in the present context and can be fixed at will.]
\end{subsection}

\begin{subsection}{Factorized structure of the $K(AE_3)$ fermionic billiard}\label{Factor}
Let us now describe the SUGRA$_4$ fermionic dynamics in the billiard limit. The treatment is parallel to the SUGRA$_{11}$ case discussed in section \ref{Ferm2} and consists in combining the effects on $\psi$ coming from successive chaotic collisions on the three dominant walls $\alpha_1$, $\alpha_2$ and $\alpha_*$ of $AE_3$. As for maximal supergravity in Eq. (\ref{ke10ferm3}), we focus on a one-wall system and integrate the equation of motion (\ref{kAE3ferm}) as before. The computations from section \ref{one} still hold for the $AE_3$-case, with (\ref{QAE3}) replacing (\ref{PQtrunc}) for the dominant gravitational wall $\alpha_*$. Hence, the net effect of a collision with a wall $\alpha$ is a $K(AE_3)$-rotation about the axis $J_\alpha$ with an angle $\e_\alpha\frac{\pi}{2}$. This rotation turns the incoming gravitino $\psi$ into an outgoing one $\psi'{}$:
\be\label{kAEferm3}
\psi'{} &=&\cR_{\alpha,\e_\alpha}^{\textit{vs}}(\psi)
\ee
with $\cR_{\alpha,\e_\alpha}^{\textit{vs}}= e^{\e_{\alpha}\frac{\pi}{2}J_\alpha^{\textit{vs}}}$ being a $12\times 12$ matrix acting in the vector-spinor representation space of $\psi^{a A}$ (where $a=1,2,3$, while $A=1,2,3,4$ is a real Majorana spinor index). Similarly to the SUGRA$_{11}$ treatment above, we found that the transformation law of the gravitino simplifies very much, if we replace $\psi^a$ by the following redefined fermionic variable:
	\be\label{phiDefiAE}
	\phi^a&:=&\g_0\g^a\psi^a\quad \text{(no sum)}.
	\ee
	With this redefinition, we find again, in complete analogy to (\ref{ke10ferm4}), that the transformation (\ref{kAEferm3}) in the vector-spinor representation \textit{factorizes} as the tensor product of a Weyl reflection $r_\alpha$ (now acting in a 3-dimensional vector space $a=1,2,3$) with a Dirac-spinor rotation $e^{\e_\alpha\frac{\pi}{2}J^{\textit{vs}}_{\alpha}}$ (now acting in the 4-dimensional Majorana spinor space). In formul\ae, we have:
\begin{subequations}\label{kAEferm4}
\be
\phi'{}^{a\,A} &=&\sum\limits_{b=1}^{3}\sum\limits_{B=1}^{4}\left[r_{\alpha}\right]^a{}_b\left[\cR_{\alpha,\e_\alpha}^{\textit{s}}\right]^A{}_B \phi^{b\,B}\\
\text{where}\quad \left[r_{\alpha}(\phi)\right]^a&=&\phi^a -\frac{2\alpha(\phi)}{(\alpha|\alpha)}\alpha^{\# a}
\\
\text{and}\quad \cR_{\alpha,\e_\alpha}^{\textit{s}}&:=& e^{\e_{\alpha}\frac{\pi}{2}J_\alpha^{\textit{s}}}
\ee
\end{subequations}
with $J_\alpha^{\textit{s}}$ being Dirac-spinor generators acting in the 4-dimensional Majorana spinor space. For symmetry wall rotations associated to $\alpha_1$ and $\alpha_2$, this result follows from a straightforward truncation of the $K(E_{10})$-results, with the Dirac-spinor representation
\be\label{Jsym2}
J^{\textit{s}}_{\alpha_{i}}&=&\frac{1}{2}\g^{i}\g^{i+1} \quad\text{for }i=1,2.
\ee
By contrast, the gravitational wall case needs a special treatment that we now sketch. We start with the observation that the redefinition (\ref{phiDefiAE}) simplifies the Lie algebra transformation (\ref{JexplAE}) (with $J^{\textit{vs}}_{\alpha_{*}}(\phi)^a=\g_0\g^a J^{\textit{vs}}_{\alpha_{*}}(\psi)^a$) to
\be
J^{\textit{vs}}_{\alpha_{*}}(\phi)^a &=& \frac12\g_0\g^{123}\big[\phi^a-4(\delta_1^a-\delta_2^a-\delta_3^a)\phi^1\big].
\ee
The orthogonal and parallel components $\phi_\bot^a$ and $\phi_{||}^a$ (\ref{decomp2}) of the gravitino $\phi^a$ turn out again to be `eigendirections' and hence, the $J^{\textit{vs}}_{\alpha_{*}}$-action simplifies in the same way as in (\ref{Jgut}):
\begin{subequations}\label{Jgut2}
\be
J^{\textit{vs}}_{\alpha_{*}}\phi_{||}^a &=& J^{\textit{s}}_{\alpha_{*}}\phi_{||}^a\\
J^{\textit{vs}}_{\alpha_{*}}\phi_\bot^a &=& -3J^{\textit{s}}_{\alpha_{*}}\phi_\bot^a
\ee
\end{subequations}
where $J^{\textit{s}}_{\alpha_{*}}$ is the Dirac-spinor action now defined as
\be\label{JAE}
J^{\textit{s}}_{\alpha_{*}}&=&\frac12 \g_0\g^{123}.
\ee
In order to evaluate the group rotation $\cR_{\alpha,\e_\alpha}^{\textit{vs}}$ (\ref{kAEferm3}) on the eigenspaces spanned by $\phi_\bot$ and $\phi_{||}$ (\ref{Jgut2}), we have to evaluate the exponential series defined in (\ref{kAEferm4}) for the gravitational wall $\alpha_*$:
\be\label{RDefi4AE}
\cR_{{\alpha_{*}},\e_{\alpha_{*}}}^{\textit{s}}&=&e^{\e_\alpha \frac{\pi}{2}J_{\alpha_{*}}^{\textit{s}}}.
\ee
Since the difference of the action on the eigenspaces (\ref{Jgut2}) again is only a sign due to $e^{-\frac{3i\pi}{4}}=-e^{\frac{i\pi}{4}}$, we finally obtain after recombining $\phi=\phi_\bot+\phi_{||}$ the factorized transformation of the gravitino that we quoted above (and that is completely similar to (\ref{ke10ferm4})). Let us also note (for future use) that the explicit form of the Weyl reflection $r_{\alpha_*}$ at the gravitational wall $\alpha_*=\alpha^g_{123}=2\beta^1$ reads:
\be\label{gravWeyl}
r_{\alpha_*}(\phi)^a 
&=&\phi^a-2(\delta_1^a-\delta_2^a-\delta_3^a)\phi^1.
\ee
We will continue the investigation of this gravitino billiard in the next section, taking into account the supersymmetry constraint $\mathcal{S}$.
\end{subsection}

\begin{subsection}{`Longitudinal' and `transversal polarizations' of the `$\beta$-photon'}\label{Long}
Concerning the supersymmetry constraint $\mathcal{S}$, the $AE_3$ model, describing the BKL-limit of $\cN=1$ $d=4$ supergravity, exhibits a qualitative difference with respect to the $E_{10}$-case of maximal supergravity: the constraint $\mathcal{S}\approx 0$ can be solved explicitly, which entails that the gravitino or `polarization vector' $\phi^a$ can be split into a `longitudinal' and a `transversal' part, both being Dirac-spinor valued.\footnote{Note however that these components do not transform as Dirac spinors under $K(AE_3)$ as we shall see in Eq. (\ref{gravTrafoAE}) below.}\\

As a first step, recall that the supersymmetry constraint in the billiard limit (far away from any wall) only depended on the velocities $\dot{\beta}^a$, e.g. (\ref{susy1}) for the $E_{10}$-case. Since $\cN=1$ $d=4$ supergravity can be obtained by a truncation of maximal supergravity as explained in section \ref{subsystem}, it is not a surprise that the equation of motion of the gravitino component $\psi^0$ (\ref{sugraF2}) in the present discussion (after the standard rescaling (\ref{fermTrafo}) and gauge-fixing $\psi^0=0$ (\ref{susyfix})) reduces to the obvious truncation of the formula (\ref{susy1}):
\be\label{susy1AE}
0\,\approx\,\mathcal{S}_{\text{BKL}} &=& \sum\limits_{a=1}^{3}\big(\dot{\beta}^a-\sigma\big)\g_0\g^a\psi^a\\
\text{with}\quad \sigma&:=&\sum\limits_{c=1}^{3}\dot{\beta}^c.
\label{sigmaDefi}
\ee
As before, the formulation in terms of the redefined gravitino $\phi^a$ (\ref{phiDefiAE}) allows one to write $\mathcal{S}$ as an orthogonality constraint with respect to the Lorentzian $\beta$-space metric $G_r$ (\ref{supermetric}):
\be\label{susyCAE}
0\,\approx\,\mathcal{S}_{\text{BKL}}&=& \sum\limits_{a,b=1}^{3} G^r_{ab}\dot{\beta}^a\phi^b.
\ee
From the point of view of the 3-dimensional Lorentzian geometry of $\beta$-space (forgetting about the hidden spinor index), the supersymmetry constraint $\mathcal{S}\approx 0$ (\ref{susyCAE}) is a scalar constraint on the 3-dimensional vector $\phi^a$. It therefore restricts the gravitino $\phi^a$ to live in a 2-dimensional hyperplane (actually, the `null' hyperplane tangent along $\dot{\beta}$ to the $\beta$-light-cone). One can then explicitly parametrize $\phi^a$ in terms of two (spinor-valued) scalars, say $\eta$ and $\Phi$, by using the existence of the cross product of vectors in three dimensions:
\be\label{paramAE}
\phi^a &=&  \frac{1}{\sigma}\dot{\beta}^a\eta + \frac{1}{\sigma}\sum\limits_{b,c=1}^3\e^{abc}u^b\dot{\beta}^c\Phi.
\ee
Here, $\sigma$ is the quantity (\ref{sigmaDefi}), $\e^{123}=+1$, and $u^a$ denotes the timelike vector in $\beta$-space which has unit (contravariant) components: $u=(1,1,1)$. The vector $u^a$ plays a somewhat special role in the $AE_3$ billiard in that it naturally relates the Lorentzian metric $G_r$ (\ref{supermetric}) to the Euclidean metric, say $v\cdot w:=\sum_{a=1}^3 v^a w^a$. Indeed, we have the identity:
\be\label{Gr}
\sum\limits_{a,b=1}^3 G^r_{ab}v^a w^b&=& v\cdot w -(u\cdot v)(u\cdot w).
\ee
Keeping in mind that $\dot{\beta}^a$ is lightlike (\ref{null}), the parametrization of $\phi^a$ in terms of $\Phi$ and $\eta$ is easily seen to span the general solution of the supersymmetry constraint $\mathcal{S}_{\text{BKL}}\approx 0$ (\ref{susyCAE}). It is also easy to invert the parametrization (\ref{paramAE}) (constrained by (\ref{susyCAE})) and to express the two `scalars' $\eta$ and $\Phi$ in terms of $\phi^a$ (using $\sigma=u\cdot\dot{\beta}$ (\ref{sigmaDefi}) and $\sigma^2=\dot{\beta}\cdot\dot{\beta}$ (\ref{null})):
\begin{subequations}\label{etaPhi}
\be
\eta&=& \frac{1}{\sigma}\sum\limits_{a=1}^3\dot{\beta}^a\phi^a \quad\stackrel{(\ref{susyCAE})}{=}\quad \sum\limits_{a=1}^3 \phi^a\\
\Phi&=& \frac{1}{2\sigma}\sum\limits_{a,b,c=1}^3\e^{abc}u^a \dot{\beta}^b\phi^c
\ee
\end{subequations}
Pursuing the analogy of considering the gravitino $\phi^a$ as a polarization vector, $\eta$ and $\Phi$ can be thought of as its longitudinal and transversal part, respectively. This point of view is reinforced by the fact that the bilinear form $(\cdot|\cdot)$ (\ref{bilin}) for any two `polarizations' $\phi_1,\phi_2$, having the same null velocity $\dot{\beta}^a$, only depends on their transversal parts $\Phi_1,\Phi_2$ (\ref{paramAE}): 
\be\label{bilinAE}
(\phi_1|\phi_2)&=&2\Phi_1^T\Phi_2
\ee
where $\Phi_1^T\Phi_2=\sum_{A=1}^4\Phi_1^A\Phi_2^A$ involves a trace over the hidden spinor indices. Note also that the `gauge transformation' of the gravitino $\phi^a$ along the lightlike direction $\dot{\beta}^a$ preserves this bilinear form (\ref{gauge}) and correspondingly only affects the longitudinal part $\eta$. At this point, we want to emphasize again that the decomposition of $\phi^a$ into two scalar-like components (\ref{paramAE}) is a peculiarity of the three dimensions of the $\beta$-space of SUGRA$_{4}$ (or $AE_3$). In higher dimensions (i.e. higher \textit{rank}), one will still have $\dot{\beta}$ and $u$ as special directions in $\beta$-space, but they will not provide enough structure for selecting special directions in the null hyperplane orthogonal to $\dot{\beta}^a$ (besides the `longitudinal' direction $\dot{\beta}^a$ itself). Note also that instead of parametrizing $\phi^a$ (in the $AE_3$ case) by two scalars, one could have also parametrized it by a lightlike vector $\lambda^a$ $(\sum_{a,b=1}^3 G^r_{ab}\lambda^a\lambda^b=0)$, with $\lambda^a\neq\dot{\beta}^a$ by writing $\phi=\dot{\beta}\times \lambda$, i.e. $\phi^a=\sum_{b,c=1}^3\e^a{}_{bc}\dot{\beta}^b\lambda^c$ (with $\e^a{}_{bc}=\sum_{b',c'=1}^3 G^r_{bb'} G^r_{cc'} \e^{ab'c'}$). We shall come back to this possibility below.\\

As a next step, we will investigate the transformation of the parameters $\eta$ and $\Phi$ under a collision with a dominant wall. At first, recall that the dynamics of the velocities $\dot{\beta}$ is described by the bosonic billiard, i.e. for any dominant wall $\alpha$ with $(\alpha|\alpha)=2$, we obtain the Weyl reflection (\ref{Reflection}) 
\be\label{Weyl2}
\dot{\beta}'\,\,=\,\,r_{\alpha}(\dot{\beta}) &=& \dot{\beta} -\alpha(\dot{\beta})\alpha^{\#}.
\ee
For the dominant symmetry walls $\alpha_1$ and $\alpha_2$ (\ref{dominant}), the reflection $r_{\alpha_i}$ (\ref{Weyl2}) reduces to a mere permutation:
\be\label{SymBeta}
r_{\alpha_1}\hspace{-3pt}\left(\hspace{-5pt}\begin{tabular}{c} $\dot{\beta}^1$\\$\dot{\beta}^2$\\$\dot{\beta}^3$\end{tabular}\hspace{-5pt}\right)
\,=\,
\left(\hspace{-5pt}\begin{tabular}{c} $\dot{\beta}^2$\\$\dot{\beta}^1$\\$\dot{\beta}^3$\end{tabular}\hspace{-6pt}\right)
&\text{and}&
r_{\alpha_2}\hspace{-3pt}\left(\hspace{-5pt}\begin{tabular}{c} $\dot{\beta}^1$\\$\dot{\beta}^2$\\$\dot{\beta}^3$\end{tabular}\hspace{-5pt}\right)
\,=\,
\left(\hspace{-5pt}\begin{tabular}{c} $\dot{\beta}^1$\\$\dot{\beta}^3$\\$\dot{\beta}^2$\end{tabular}\hspace{-6pt}\right).
\ee
Since these permutations leave the unit vector $u=(1,1,1)$ as well as $\sigma$ (\ref{sigmaDefi}) invariant and since their determinant is $-1$, the relations (\ref{kAEferm4},\,\ref{etaPhi}) immediately fix the transformation of $\eta$ and $\Phi$ for the symmetry walls $\alpha_i$ with $i=1,2$ in terms of the Dirac spinor action $\cR_{\alpha_i,\e_{\alpha_i}}^{\textit{s}}$ (\ref{kAEferm4}):
\begin{subequations}
\be
\eta'&=&\cR_{{\alpha_i},\e_{\alpha_i}}^{\textit{s}}\eta
\\
\Phi'&=&-\cR_{{\alpha_i},\e_{\alpha_i}}^{\textit{s}}\Phi.
\label{symTrafoAE2}
\ee
\end{subequations}
On the other hand, a collision with the gravitational wall $\alpha_*$ induces the Weyl reflection $r_{\alpha_*}$ (\ref{gravWeyl})
\be\label{GravBeta}
\dot{\beta}'
\,=\,r_{\alpha_*}\hspace{-3pt}\left(\hspace{-5pt}\begin{tabular}{c} $\dot{\beta}^1$\\$\dot{\beta}^2$\\$\dot{\beta}^3$\end{tabular}\hspace{-5pt}\right)
&=&
\left(\hspace{-5pt}\begin{tabular}{c} $-\dot{\beta}^1$\\$\dot{\beta}^2+2\dot{\beta}^1$\\$\dot{\beta}^3+2\dot{\beta}^1$\end{tabular}\hspace{-6pt}\right).
\ee
It implies $\sigma'=\sigma+2\dot{\beta}^1$ (\ref{sigmaDefi}) and it allows one to deduce the induced transformations for $\Phi$ and $\eta$ using the Hamiltonian constraint $\dot{\beta}\cdot\dot{\beta}=\sigma^2$ (\ref{null}):
\begin{subequations}\label{gravTrafoAE}
\be\label{gravTrafoAE1}
\eta'&=&\cR_{\alpha_*,\e_{\alpha_*}}^{\textit{s}}\left(\left(1+\frac{2}{\sigma}\dot{\beta}^1\right)\eta +\frac{2}{\sigma}(\dot{\beta}^3-\dot{\beta}^2)\Phi\right)
\\
\Phi'&=&-\cR_{\alpha_*,\e_{\alpha_*}}^{\textit{s}}\Phi\label{gravTrafoAE2}
\ee
\end{subequations}
The gauge direction or longitudinal polarization $\eta$ is changed in an inhomogeneous way, in contradistinction to the transversal polarization $\Phi$. This implies that neither $\eta$ nor $\Phi$ transform in the Dirac spinor representation of $K(AE_3)$, the former due to the inhomogeneity and the latter due to the sign in (\ref{symTrafoAE2},\,\ref{gravTrafoAE2}).\footnote{$\Phi$ could, however,  be described as a pseudo-Dirac spinor, in analogy to a pseudo-scalar in particle physics.} The homogeneous transformation behaviour of the transversal polarization $\Phi$ was expected due to the preservation of the Euclidean norm $(\phi|\phi)=2\Phi^T\Phi$ (\ref{bilinAE}). By contrast, as this norm does not contain $\eta$ it does not provide any means to control the evolution of the longitudinal polarization $\eta$ and is thereby compatible with the inhomogeneous character of (\ref{gravTrafoAE1}). 
\end{subsection}

\begin{subsection}{Fermionic billiard dynamics and BKL-eras}
Let us now briefly discuss the qualitative evolution of $\eta$ and $\Phi$ near a singularity. We recall that, qualitatively speaking, the 3-dimensional BKL chaotic behaviour can be thought of as made of \textit{eras} that collect together (possibly long) sequences of Kasner `epochs' (where an epoch is an intervall between two successive wall collisions) \cite{BKL}. To explain the meaning of these eras, we start by recalling the `u-parametrization' of the `Kasner circle' \cite{BKL}. First, it is convenient to work with the \textit{Kasner exponents} $p^a$ which are projective coordinates of the (homogeneous) Lorentzian $\beta$-space velocities $\dot{\beta}^a$, namely
\be\label{pDefi3}
p^a&:=&\frac{\dot{\beta}^a}{\sigma}\quad \text{with } \sigma\,=\,\sum_{a=1}^{3}\dot{\beta}^a \text{ and }a\,=\,1,2,3.
\ee
By construction, these lie in the plane
\begin{subequations}\label{C12}
\be\label{C1}
p^1+p^2+p^3&=&1.
\ee
Furthermore, the Hamiltonian constraint (\ref{null}) also restricts them to a sphere
\be\label{C2}
(p^1)^2+(p^2)^2+(p^3)^2&=&1.
\ee
\end{subequations}
The intersection is a circle that is parametrized by $u\in \R\cup \{\infty\}$ following \cite{BKL}:
\be\label{uDefi}
p^1(u)\,=\,\frac{-u}{\sigma_u},\quad p^2(u)&=&\frac{1+u}{\sigma_u},\quad p^3(u)\,=\,1-\frac{1}{\sigma_u}\\
\text{with}\quad \sigma_u&=& 1+u+u^2.
\nn
\ee

The transformation of the projective velocities $p^a$ (\ref{pDefi3}) under a collision with a dominant wall directly follows from the action on the velocities $\dot{\beta}^a$ (\ref{SymBeta},\,\ref{GravBeta}). These induce the following transformations on the parameter $u$:\footnote{As an example, the symmetry wall $\alpha_1$ interchanges $p^1$ and $p^2$ and leaves $p^3$ inert (\ref{SymBeta}). This implies $r_{\alpha_1}(p^1)(u)=p^2(u)=p^1(r_{\alpha_1}(u))$, $r_{\alpha_1}(p^2)(u)=p^1(u)=p^2(r_{\alpha_1}(u))$ and $r_{\alpha_1}(p^3)(u)=p^3(u)=p^3(r_{\alpha_1}(u))$.}
\begin{subequations}\label{ruDefi}
\be
r_{\alpha_*}(u)&=&-u
\\
r_{\alpha_1}(u)&=&-u-1
\\
r_{\alpha_2}(u)&=&\frac{1}{u}
\ee
\end{subequations}
Next, we recall that the Weyl reflections $r_{\alpha_i}$ for $i=*,1,2$ fulfill the Coxeter relations that follow from the Dynkin diagram of $AE_3$ (figure 2). Apart from $r_{\alpha_i}^2=1$, these read
\begin{subequations}\label{CoxAE}
\be\label{CoxAE1}
\left(r_{\alpha_*}r_{\alpha_2}\right)^2&=&1
\\
\left(r_{\alpha_1}r_{\alpha_2}\right)^3&=&1
\\
\left(r_{\alpha_1}r_{\alpha_*}\right)^\infty&=&1
\ee
\end{subequations}
The exponents $k=2,3,\infty$ correspond to the angles $\frac{\pi}{k}$ between the dominant walls that confine the billiard motion of the scale factors $\beta^a$ to a polyhedron, which intersects the unit hyperboloid $\sum G_{ab}\beta^a\beta^b=-1$ into a hyperbolic triangle of finite volume. The special geodesics with velocity $u=+\infty$ correspond to the ones that end at the fixed point $(p^1,p^2,p^3)=(1,0,0)$ (which corresponds to the flat Milne space-time), and hence do not exhibit a chaotic behaviour. Chaos is restored for generic values of $u\in \R$, however. In particular note that this fixed point is unstable, because any finite, but large value $u$ decreases in a sequence of consecutive collisions with the walls $\alpha_*$ and $\alpha_1$ (\ref{ruDefi}) due to:
\be\label{uMinus}
r_{\alpha_1}r_{\alpha_*} (u)&=& u-1
\ee
This sequence is called a BKL-era of length $n=[ u]\in \N$ \cite{BKL}.\footnote{Here, $n_0:=[u]$ denotes the integer part of $u$. The generic chaotic dynamics can in fact be understood as a sequence of BKL-eras, whose lengths $n_i$ are fixed by a continuous fraction decomposition of $u$, i.e. $u=n_0+\frac{1}{n_1+\frac{1}{n_2+\cdots}}$ \cite{BKL}.
} 
For long eras $n>>1$, we can neglect contributions of order $u^{-2}$, which simplifies the parametrization of the projective velocities $p^a$ in terms of $u$ (\ref{uDefi}) to
\beg
p^1(u)\,\approx\,-\frac{1}{u},\quad
p^2(u)\,\approx\,\frac{1}{u},\quad
p^3(u)\,\approx\,1.
\eeg
Substituting this in the transformation for $\eta$ (\ref{gravTrafoAE1}) yields for $\eta_{u-1}=r_{\alpha_1}r_{\alpha_*}(\eta)_{u}$:
\beg
\eta_{u-1}
&=&
\cR_{\alpha_1,\e_{\alpha_1}}^{\textit{s}}\cR_{\alpha_*,\e_{\alpha_*}}^{\textit{s}}\left(\left(1-\frac{2}{u}\right)\eta_{u} +2(1-\frac{1}{u})\Phi_{u}\right)
+\mathcal{O}\big(\frac{1}{u^2}\big).
\eeg
We iterate this transformation $n=[ u]$ times and obtain for large $n$ an approximate asymptotic relation linking the values at the beginning $\eta_u$ to the ones at the end $\eta_{u-n}$ of the BKL-era:
\be\label{asymp}
\eta_{u-n}&\approx& \left(\cR_{\alpha_1,\e_{\alpha_1}}^{\textit{s}}\cR_{\alpha_*,\e_{\alpha_*}}^{\textit{s}}\right)^n
\left(\frac{1}{n^2}\eta_{u} +2n\Phi_{u}\right).
\ee
As for $\Phi$, its evolution during the considered era is simply given by
\beg
\Phi_{u-n}&=& \left(\cR_{\alpha_1,\e_{\alpha_1}}^{\textit{s}}\cR_{\alpha_*,\e_{\alpha_*}}^{\textit{s}}\right)^n
\Phi_{u}.
\eeg
Qualitatively, we see that while the `spinor indices' of both $\eta$ and $\Phi$ continuously `rotate' under the spinor actions $\cR_{\alpha,\e_{\alpha}}^{\textit{s}}$'s, their `scalar magnitudes' evolve in very different ways: the $\Phi$ one remains constant, while the $\eta$ one is made of two parts: One part, $\frac{1}{n^2}\eta_u$, \textit{decreases} during an era, while the other part, $2n\Phi$, linearly \textit{increases} during an era. When considering successive eras, this leads us to expect that the (gauge-like) `longitudinal' part $\eta$ will perform very large `excursions', whose `directions' (in spinor space) are restricted to the orbit of the Dirac spinor action $\cR_{\alpha,\e_{\alpha}}^{\textit{s}}$ on $\Phi$. In a `baby problem', where we would replace the `spinor direction' by a direction in a complex plane, we conceive of $\eta$ as performing a kind of Brownian motion in the complex plane under a randomly oriented force ($\propto \Phi$) that grows linearly in time. In the next section, we shall study in detail the evolution of the $\Phi$ degrees of freedom, i.e. the rotations in spinor space generated by the product of many $K(AE_3)$ spinor rotations $\cR_{\alpha,\e_{\alpha}}^{\textit{s}}$.\\

But before doing so let us indicate that if one had used the parametrization of $\phi^a$ by means of two lightlike vectors $(\dot{\beta}^a,\lambda^b)$, the action of the fermionic billiard on the pair $(\dot{\beta}^a,\lambda^b)$ (where only $\lambda^b$ has an extra spinor index) would be a ``double BKL billiard'', namely
\begin{subequations}\label{bl}
\be
\dot{\beta}' &=& r_\alpha(\dot{\beta})\\
\lambda' &=&-\cR_{\alpha,\e_{\alpha}}^{\textit{s}}\otimes r_\alpha (\lambda)
\ee
\end{subequations}
with the \textit{same} vector reflection $r_\alpha$ (\ref{Reflection}) acting on the vector index of both light-like vectors. We already know that the simple billiard $\dot{\beta}' = r_\alpha(\dot{\beta})$ has interesting chaotic properties; we shall see in the next section that the $\cR_{\alpha,\e_{\alpha}}^{\textit{s}}$-billiard is \textit{not} chaotic, but rather periodic (finite multiplicative group). This means that the `chaotic' part of the `doubled billiard' (\ref{bl}) will essentially come from a double bosonic BKL billiard: $\dot{\beta}' = r_\alpha(\dot{\beta}),\lambda' = r_\alpha(\lambda)$. We leave to further work the study of these dynamics.
\end{subsection}

\begin{subsection}{Spinorial extension of the Weyl group $\mathcal{W}_{AE_3}$}
An important link between the cosmological billiard dynamics and hyperbolic Kac--Moody algebras $\mathfrak{g}$ consists of the fact that any sequence of billiard collisions corresponds to a \textit{word} in the associated Weyl group $\mathcal{W}_{\mathfrak{g}}$. We have transferred this idea to the fermionic dynamics in section \ref{Weyl} by defining the spin extension $\mathcal{W}^{\textit{spin}}_{\mathfrak{g}}$ of the Weyl group $\mathcal{W}_{\mathfrak{g}}$. For $\mathfrak{g}=E_{10}$, we have found two concrete realizations $\mathcal{W}^{\textit{vs}}_{\mathfrak{g}}$ and $\mathcal{W}^{\textit{s}}_{\mathfrak{g}}$ in terms of matrix groups (generated by the matrices $\cR_{\alpha_i}^{\textit{vs}}$ (\ref{Wvs}) and $\cR_{\alpha_i,\e_{\alpha_i}}^{\textit{s}}$ (\ref{Ws}) respectively) that were characterized by the propositions 1 and 2. For $\mathfrak{g}=AE_3$, we can perform a similar analysis. Since $AE_3$ is a subalgebra of $E_{10}$, the structure of the matrix realizations $\mathcal{W}^{\textit{vs}}_{AE_3}$ and $\mathcal{W}^{\textit{s}}_{AE_3}$ of the spin extension of the Weyl group shows great similarity to the $E_{10}$-case, as exemplified by the following statements:
\begin{subsubsection}*{Proposition 4}
\textit{
\begin{itemize}
	\item The vector-spinor realization $\SWeyl_{AE_3}$ of $\mathcal{W}^{\textit{spin}}_{AE_3}$, i.e. the group multiplicatively generated by the three $12\times 12$ matrices $\cR_{\alpha_i}^{\textit{vs}}=e^{\frac{\pi}{2}J_{\alpha_i}^{\textit{vs}}}$ (\ref{kAEferm3}) for $i=*,1,2$, is infinite.
	\item The generators $\cR_{\alpha_i}^{\textit{vs}}$ of $\SWeyl_{AE_3}$ satisfy the following generalized Coxeter relations:
	\begin{subequations}\label{GenCoxAE3}
	\be
\big(\cR_{\alpha_i}^{\textit{vs}}\big)^4&=&-\id,\\
\big(\cR_{\alpha_1}^{\textit{vs}}\cR_{\alpha_2}^{\textit{s}}\big)^3&=&-\id
\ee
\end{subequations}
and $\cR_{\alpha_*}^{\textit{vs}}$ commutes with $\cR_{\alpha_2}^{\textit{vs}}$.
		\item The squares of the matrices $(\cR_{\alpha_i,+}^{\textit{vs}})^2$ generate a normal subgroup $\SWeylN_{AE_3}$ of $\SWeyl_{AE_3}$, which is \textbf{non-abelian} and whose cardinality is $2\times 8$.
	\item The Weyl group $\mathcal{W}_{AE_3}$ is \textbf{isomorphic} to the quotient group $\SWeyl_{AE_3}/\SWeylN_{AE_3}$:
	\be\label{homo2AE}
	\mathcal{W}_{AE_3}&\simeq&\SWeyl_{AE_3}/\SWeylN_{AE_3}.
	\ee
	\end{itemize}
}
\end{subsubsection}

The discussion of the Dirac spinor action for the $AE_{3}$-case can be made more precise than the corresponding one for $E_{10}$ in proposition 2:
\begin{subsubsection}*{Proposition 5}
\textit{
\begin{itemize}
 \item The spinor realization $\DWeyl_{AE_3}$ of $\mathcal{W}^{\textit{spin}}_{AE_3}$, i.e. the group multiplicatively generated by the three $4\times 4$ matrices $\cR_{\alpha_i}^{\textit{s}}=e^{\frac{\pi}{2}J_{\alpha_i}^{\textit{s}}}$ (\ref{Jsym2},\,\ref{JAE}) with $i=*,1,2$, forms a finite group of cardinality $4\times 48$.
	\item The generators $\cR_{\alpha_i}^{\textit{s}}$ of $\DWeyl_{AE_3}$ fulfill an additional generalized Coxeter relation compared to the ones of proposition 4: The matrix $\cR_{\alpha_*}^{\textit{s}}$ \textbf{commutes with both} $\cR_{\alpha_1}^{\textit{s}}$ and $\cR_{\alpha_2}^{\textit{s}}$, so that the group $\DWeyl_{AE_3}$ is the direct product of two separate groups. 
	\item The squares of the matrices $(\cR_{\alpha_i}^{\textit{vs}})^2$ form a normal subgroup $\DWeylN_{AE_3}$ of $\DWeyl_{AE_3}$. It is isomorphic to the group $\SWeylN_{AE_3}$ defined for the vector-spinor representation.
	\item The quotient group $\DWeyl_{AE_3}/\DWeylN_{AE_3}$ is isomorphic to $\Z_2\times S_3$. (Here $S_3\simeq \textit{SL}(2,\Z_2)$ denotes the permutation group $S_3$ of order $3!$, which is the Weyl group of $sl_3$.)
	\item There is a homomorphism between the Weyl group $\mathcal{W}_{AE_3}$ and the quotient group $\DWeyl_{AE_3}/\DWeylN_{AE_3}$
	\be\label{homo4AE}
	\mathcal{W}_{AE_3}&\rightarrow&\DWeyl_{AE_3}/\DWeylN_{AE_3}.
	\ee
	Its kernel forms a \textbf{normal subgroup} of the Weyl group $\mathcal{W}_{AE_3}$. This normal subgroup is isomorphic to the principal congruence subgroup, defined by the kernel of the canonical homomorphism
	\be\label{canon}
	\textit{PSL}(2,\Z)&\rightarrow& \textit{PSL}(2,\Z_2).
	\ee
\end{itemize}
}
\end{subsubsection}
The proof of theorems 3 and 4 closely follows the one for the $E_{10}$-case provided in Appendix \ref{appendix2}: the factorizability of the gravitino action $\cR_{\alpha_i,\e_{\alpha_i}}^{\textit{vs}}$ into a Weyl reflection $r_{\alpha_i}$ and a Dirac spinor action $\cR_{\alpha_i,\e_{\alpha_i}}^{\textit{s}}$ allows one to consider proposition 4 as a corollary of proposition 5 as before. The latter discusses the property of the group $\DWeyl_{AE_3}$, whose generators can be written in the form (\ref{Jsym2},\,\ref{JAE}):
\beg
\cR_{\alpha_*,\e_{\alpha_*}}^{\textit{s}}&=& \frac{1}{\sqrt{2}}\big(\id + \g_0\g^{123}\big),
\\
\cR_{\alpha_1,\e_{\alpha_1}}^{\textit{s}}&=& \frac{1}{\sqrt{2}}\big(\id + \g^{12}\big),
\\
\cR_{\alpha_2,\e_{\alpha_2}}^{\textit{s}}&=& \frac{1}{\sqrt{2}}\big(\id + \g^{23}\big).
\eeg
In this form, it is clear that $\cR_{\alpha_*,\e_{\alpha_*}}^{\textit{s}}$ commutes with both $\cR_{\alpha_1,\e_{\alpha_1}}^{\textit{s}}$ and $\cR_{\alpha_2,\e_{\alpha_2}}^{\textit{s}}$ and hence can be discussed separately from the other two. It is interesting to phrase this fact in terms of Dynkin diagrams: for $AE_3$, it is straightfoward to check that all elements in the quotient group $\DWeyl_{AE_3}/\DWeylN_{AE_3}$ fulfill the Coxeter relations corresponding to the \textit{disconnected} diagram
\begin{center}
\begin{tabular}{cl}
&
\scalebox{.5}{
\begin{picture}(90,15)(0,0)

\put(94,5){ \circle*{15}}

\put(52,5){ \circle*{15}}
\put(60,5){\line(1,0){35}}

\put(10,5){ \circle*{15}}

\end{picture}
}
\\
&\scalebox{.7}{
\begin{picture}(90,10)(0,0)

\put(64,8){$\alpha_{2}$}
\put(34,8){$\alpha_{1}$}
\put(4,8){$\alpha_{*}$}
\end{picture}
}
\end{tabular}\\
\vspace{-4.0pt}
\small{figure 3}
\end{center}
This is, however, not the Dynkin diagram of $AE_3$ (figure 2), but the one of the finite dimensional product group $\textit{SL}(2)\times \textit{SL}(3)$. Its associated Weyl group is $\Z_2\times S_3$ being isomorphic to the group of equivalence classes $\DWeyl_{AE_{3}}/\DWeylN_{AE_{3}}$. Note that this statement is in sharp contrast to the $E_{10}$-case of section \ref{Weyl}, for which there is no additional Coxeter relation for the quotient group. The associated Coxeter diagrams for $E_{10}$ and for the quotient group $\DWeyl_{E_{10}}/\DWeylN_{E_{10}}$ are the same but nevertheless, the latter was of finite cardinality. We have checked that the `disconnection' (figure 3) that occurs for $AE_3$ is non-generic in that the Dirac-spinor representations of the higher rank cases $AE_d$ lead to generalized Coxeter relations rooted in the full corresponding \textit{connected} Dynkin diagram.\\

The statement that the kernel of the homomorphism (\ref{homo4AE}) is a congruence subgroup (\ref{canon}) is related to the observation that the Weyl group of $AE_3$ is isomorphic to $\textit{PGL}(2,\Z)$ \cite{FF83}. We can make this precise within the context of the cosmological billiard dynamics. Using the parameter $u$ (\ref{uDefi}) for their description, the transformations $r_{\alpha_i}$ of $u$ (\ref{ruDefi}) provide a natural link between the generators of the Weyl group $r_{\alpha_i}$ (for $i=*,1,2$) and $2\times 2$ matrices $A_i:=\binom{a_i\,\,b_i}{c_i\\d_i}$ with integer coefficients $a_i,b_i,c_i,d_i$ and determinant $\pm 1$ tantamount to
\be\label{frac}
r_{\alpha_i}(u)&=&\frac{a_iu +b_i}{c_iu+d_i}.
\ee
The three matrices $A_*,A_1,A_2$ generate the group $\textit{PGL}(2,\Z)$ \cite{FF83,HPS07}.\footnote{The $\Z_2$-factor in $\textit{PGL}(2,\Z)=\textit{GL}(2,\Z)/\Z_2$ corresponds to the subgroup $(\pm \id_2)$ that clearly corresponds to the same fractional transformation (\ref{frac}) keeping in mind that the matrix product agrees with the concatenation of two transformations (\ref{frac}).
} As a next step, consider the two group homomorphisms\footnote{The canonical homomorphism $\text{C}$ amounts to mapping any element $\binom{a\,\,b}{c\,\,d}\in \textit{PGL}(2,\Z)$ to the matrix of $\Z/2\Z$-equivalence classes $\binom{[a]\,\,[b]}{[c]\,\,[d]}\in \textit{PGL}(2,\Z_2)$.}
\beg
\det:\textit{PGL}(2,\Z) &\rightarrow& \Z_2 \quad\text{and}\\
\text{C}:\textit{PGL}(2,\Z) &\rightarrow& \textit{PGL}(2,\Z_2)\,=\,\textit{SL}(2,\Z_2).
\eeg
The kernel of $\det$ is the modular group $\textit{PSL}(2,\Z)$ that hence is a normal subgroup of $\textit{PGL}(2,\Z)$. The same is true for the kernel of $\text{C}$. Furthermore, the intersection of two normal subgroups again is a normal subgroup and it is non-trivial in this case: On the one hand, we have $r_{\alpha_*}\in \text{Ker}(\text{C})$ and $r_{\alpha_*}\notin \text{Ker}(\det)$, whereas on the other hand $r_{\alpha_*}r_{\alpha_1}\notin \text{Ker}(\text{C})$ and $r_{\alpha_*}r_{\alpha_1}\in \text{Ker}(\det)$ (\ref{ruDefi}). Therefore, there is a homomorphism with kernel $\text{Ker}(\text{C'})=\text{Ker}(\text{C})\cap\text{Ker}(\det)$ of the form
\beg
\text{C'}:\textit{PGL}(2,\Z) &\rightarrow& \Z_2\times \textit{SL}(2,\Z_2).
\eeg
Due to the isomorphisms $\textit{PGL}(2,\Z)\simeq \mathcal{W}_{AE_3}$ and $\textit{SL}(2,\Z_2)\simeq S_3$, the kernel of the homomorphism in (\ref{homo4AE}) is isomorphic to $\text{Ker}(\text{C'})=\text{Ker}(\text{C})\cap \textit{PSL}(2,\Z)$. Due to the identity $\textit{SL}(2,\Z_2)=\textit{PSL}(2,\Z_2)$, it is furthermore isomorphic to the principal congruence subgroup, defined by the kernel of the canonical homomorphism (\ref{canon})
	\beg
	\textit{PSL}(2,\Z)&\rightarrow& \textit{PSL}(2,\Z_2).
	\eeg
This completes the proof of theorem 4.\qed\\

We can also make this more precise in terms of $2\times 2$-matrices. A comparison of (\ref{ruDefi}) with (\ref{frac}) yields:
\beg
A_*\,=\,\left(\begin{tabular}{cc} $-1$&$0$\\$0$&$1$\end{tabular}\right),\quad
A_1\,=\,\left(\begin{tabular}{cc} $-1$&$-1$\\$0$&$1$\end{tabular}\right),\quad
A_2\,=\,\left(\begin{tabular}{cc} $0$&$1$\\$1$&$0$\end{tabular}\right).
\eeg
It is an easy exercise to verify the Coxeter relations (\ref{CoxAE}) in terms of these matrices.\footnote{Note that the relation $(r_{\alpha_*}r_{\alpha_2})^2=1$ (\ref{CoxAE1}) (i.e. that the two matrices $A_*$ and $A_2$ commute in $PGL(2,\Z)$) follows from the equivalence relation $\id_2\sim -\id_2$.} The additional Coxeter relation from proposition 5, saying that the matrix $\cR^{\textit{s}}_{\alpha_*}$ commutes with $\cR^{\textit{s}}_{\alpha_1}$, corresponds to the fact that the images of the canonical homomorphism $C(A_*)$ and $C(A_1)$ commute in $\textit{PSL}(2,\Z_2)$:
\beg
C(A_*A_1)\,=\,\left(\begin{tabular}{cc} $[1]$&$[1]$\\$[0]$&$[1]$\end{tabular}\right)&\sim&
C(A_1A_*)\,=\,\left(\begin{tabular}{cc} $[1]$&$[-1]$\\$[0]$&$[1]$\end{tabular}\right).
\eeg
It may also be interesting to investigate whether this pattern linking principal congruence subgroups to the description of the billiard dynamics of a Dirac spinor in terms of the finite group $\DWeyl_{\mathfrak{g}}$ allows for a generalization to other Kac--Moody algebras $\mathfrak{g}$. A promising result in this respect was recently provided in \cite{FKN08}, where the Weyl group of $E_{10}$ was shown to be related to the `modular group' $\textit{PSL}(2,O)$ of octonionic integers $O$ (octavians).\footnote{Concerning the subtleties in defining a `group' over the non-associative division algebra of octonions $\Oc$, we refer the reader to \cite{FKN08}.} This might hint at a possibility of identifying the kernel of the homomorphism $\mathcal{W}_{E_{10}}\rightarrow \DWeyl_{E_{10}}/\DWeylN_{E_{10}}$ (\ref{homo4}) in proposition 2 with a suitably generalized principal congruence subgroup involving the octavians.
\end{subsection}
\end{section}

\begin{section}{Conclusions}
We have studied the ``fermionic billiards'', i.e. the chaotic dynamics of the gravitino, that arise in the near-spacelike-singularity limit (or Belinski--Khalatnikov--Lifshitz, BKL limit) of supergravity. We have focussed on eleven-dimensional supergravity (whose bosonic cosmological billiard takes place in the Weyl chamber of $E_{10}$), and have also considered in detail $\cN=1$ $D=4$ supergravity (whose bosonic billiard takes place in the Weyl chamber of $AE_3=A_1^{++}$). We have shown that a useful tool for studying the fermionic side of the near-singularity cosmological billiards is the Kac--Moody coset reformulation of the supergravity dynamics. In this reformulation, the bosonic coset velocity $\cP$, and the gravitino variable $\psi$, are both parallel-transported by the same abstract connection $\cQ$ belonging to the maximal compact subalgebra of the relevant Kac--Moody algebra, namely $K(E_{10})$ for maximal supergravity and $K(AE_3)$ for its $D=4$ $\cN=1$ truncation. The time evolution, near a spacelike singularity, of the, say, $K(E_{10})$-valued `angular velocity' $\cQ(t)$ was shown to consist of a succession of well-separated inverse-$\cosh$ `spikes' associated to the collision on a `Toda wall' (itself associated to a certain root $\alpha$ of $E_{10}$). Each such $K(E_{10})$-spike takes place along the `rotational axis' $J_\alpha=E_{\alpha}-E_{-\alpha}=E_{\alpha}+\omega(E_{-\alpha})$ associated (within $K(E_{10})$) to the corresponding wall root $\alpha$.\\

We found that the integral over the collision of each `angular velocity' $\cQ(t)$ generates a finite $K(E_{10})$ rotation given by the universal formula
\be\label{c1}
\cR_{\alpha,\pm}&=&e^{\pm\frac{\pi}{2}J_\alpha}\,\,=\,\,e^{\pm\frac{\pi}{2}(E_\alpha-E_{-\alpha})}
\ee
where the normalization of $E_\alpha$ and $E_{-\alpha}=-\omega(E_\alpha)$ ($\omega$ denoting the Chevalley involution) is such that $E_\alpha$, $E_{-\alpha}$ and $h_\alpha:=[E_\alpha,E_{-\alpha}]$ form a standardly normalized $sl_2$-algebra (see Appendix \ref{appendix1}). When evaluated within the `adjoint' representation which acts on the coset velocity $\cP$ (and in the billiard limit where $\cP\in \mathfrak{h}$, the Cartan subalgebra of $E_{10}$), the $K(E_{10})$-rotation (\ref{c1}) reproduces the standard geometrical Weyl reflection $r_\alpha$, see Eq. (\ref{Reflection}). The effect of each collision on the fermionic variable $\psi$ is obtained by evaluating the general abstract $K(E_{10})$ rotation (\ref{c1}) within the `vector-spinor' representation of $K(E_{10})$ to which $\psi$ belongs. As a consequence, the billiard dynamics of the gravitino can be described as a `word', i.e. a product, of discrete vector-spinor $K(E_{10})$ rotations, say (as in (\ref{wf}))
\be\label{c2}
w^{\textit{vs}}&=&\cR_{\alpha_{i_N},\e_{\alpha_{i_N}}}^{\textit{vs}}\cdots
\,\,\cR_{\alpha_{i_2},\e_{\alpha_{i_2}}}^{\textit{vs}}
\cR_{\alpha_{i_1},\e_{\alpha_{i_1}}}^{\textit{vs}}
\ee
where $\alpha_{i_1}\rightarrow\alpha_{i_2}\rightarrow\dots\rightarrow \alpha_{i_N}$ denotes an unbounded sequence of collisions encountered by the billiard particle moving in the Weyl chamber of $E_{10}$. Here, $(\alpha_{i_1},\dots,\alpha_{i_N})$ is a sequence of simple roots (the signs $\e_{\alpha_i}=\pm$ associated to each simple root are fixed in the BKL-limit and could here be all conventionally replaced by $+$). The vector-spinor word $w^{\textit{vs}}$ (\ref{c2}) is the fermionic side of the corresponding bosonic billiard result, described by a word in the Weyl group, say
\be\label{c3}
w&=&r_{\alpha_{i_N}}\cdots r_{\alpha_{i_2}}r_{\alpha_{i_1}}.
\ee
We found that the `vector-spinor reflections' (\ref{c2}) happen to factorize as the tensor product of the usual vector Weyl reflections (\ref{c3}) and of a corresponding $K(E_{10})$ `Dirac-spinor' reflection as in  (\ref{ws}), induced by the same sequence of collisions $\alpha_{i_1}\rightarrow\alpha_{i_2}\rightarrow\dots\rightarrow \alpha_{i_N}$:
\be\label{c4}
w^{\textit{s}}&=&\cR_{\alpha_{i_N},\e_{\alpha_{i_N}}}^{\textit{s}}\cdots
\,\,\cR_{\alpha_{i_2},\e_{\alpha_{i_2}}}^{\textit{s}}
\cR_{\alpha_{i_1},\e_{\alpha_{i_1}}}^{\textit{s}}.
\ee
We considered (\ref{c2}) and (\ref{c4}) as particular representations of an abstract `spin' extension $\mathcal{W}^{\textit{spin}}$ of the Weyl group of $E_{10}$ (or more generally of any Kac--Moody algebra), formally defined as the discrete subgroup of a suitable covering of the `group' $K(E_{10})$, which is multiplicatively generated by the formal exponentials (\ref{c1}). We found that the generators of the two representations (\ref{c2}) and (\ref{c4}) of $\mathcal{W}^{\textit{spin}}$ satisfy generalized Coxeter relations (see propositions 1 and 2 in section \ref{Weyl}). We also found that its realization $\mathcal{W}^{\textit{vs}}$ in the vector-spinor representation is infinite and contains a quotient group $\SWeyl/\SWeylN$ (where $\SWeylN$ is a \textit{normal} subgroup of finite cardinality) that is isomorphic to the usual Weyl group $\mathcal{W}$. Therefore, we can think of $\mathcal{W}^{\textit{vs}}$ as a finite-index extension of $\mathcal{W}$. By contrast, we found that the Dirac-spinor realization $\mathcal{W}^{\textit{s}}$ of $\mathcal{W}^{\textit{spin}}$ is of finite cardinality. Physically, this means that a Dirac-billiard, described by the action of the words (\ref{c4}) on some initial $\mathbf{32}$-valued Dirac-spinor variable $\ep_0$, has only a finite orbit in $\R^{32}$ (though, due to the chaotic nature of the underlying Weyl-chamber billiard, the sequence of Dirac variables $\ep_1,\ep_2,\dots,\ep_n,\dots$ will run chaotically over the finite orbit $\DWeyl\ep_0$).\\

Although our results have relied on the use of the only currently known spinorial representations of $K(E_{10})$ (which are both finite-dimensional and unfaithful), we think that these results could be more general, and will, in particular, apply to the representations of $\mathcal{W}_G^{\textit{spin}}$ (where $G$ is a `Kac--Moody group') that are defined as the tensor product of a representation of the Weyl group $\mathcal{W}_G$ and a `pure spin $\frac12$' representation of $K(G)$ (where each $K(G)$ rotation generator has eigenvalue $\pm\frac12$).\\

As a particular example of another Kac--Moody algebra of interest for physics we considered $AE_3$, whose Weyl group describes the billiard of pure supergravity in $D=4$ dimensions. We explicitly studied the fermionic billiard that arises in $\cN=1$ $D=4$ supergravity. It is a simpler version of the $E_{10}$ case, though it retains the main features mentioned above: e.g. factorizability, existence of generalized Coxeter relations, finite cardinality of $\DWeyl_{AE_3}$. A special feature of the $AE_3$ case is, however, that $\mathcal{W}^{\textit{s}}_{AE_3}$ can be decomposed into two \textit{commuting} quotient groups, because the rotation generator $J_{\alpha_*}^{\textit{s}}\in K(AE_3)$ (associated to the dominant gravitational wall) happens to commute with the two other simple generators $J_{\alpha_1}^{\textit{s}}$ and $J_{\alpha_2}^{\textit{s}}$. This situation is non-generic in that it does not occur for the $AE_d$ case (which is associated to pure gravity in $D=d+1$ dimensions).\\

Finally, we found that the `super-billiard' obtained by combining the bosonic and fermionic billiards exhibits (for $E_{10}$ as well as for $AE_3$) a striking analogy with the dynamics of a `polarized photon' bouncing on the Lorentzian mirrors corresponding to the Weyl chamber of $E_{10}$ (or $AE_3$). It has a `momentum' $v^a$ (given by the Cartan-space velocity $v^a=\dot{\beta}^a$), a `polarization vector' $\phi^a$ (linked to the vector-index of the gravitino, after a suitable redefinition (\ref{phiDefi}) or (\ref{phiDefiAE})), two on-shell constraints (`masslessness', $\sum_{a,b}G_{ab}v^a v^b=0$, and `transversality, $\sum_{a,b}G_{ab}v^a \phi^b=0$, where $G_{ab}$ is the flat Lorentzioan metric in Cartan space), and a Maxwell-like gauge invariance ($\phi'{}^a=\phi^a+\xi v^a$). This analogy exhibits some intriguing metamorphoses of gauge symmetries (local supersymmetry $\leftrightarrow$ Maxwell-like gauge symmetry), as well as of group symmetries ($SO(d)\leftrightarrow SO(d-1,1)$). It may hint at new ways of using the conjectural gravity-coset correspondence for illuminating the hidden symmetries of maximal supergravity.
\end{section}

\begin{subsection}*{Acknowledgements}
We thank Fedor Bogomolov, Christophe Breuil, Pierre Cartier, Ofer Gabber, Victor Kac, Axel Kleinschmidt, Laurent Lafforgue and Pierre Vanhove for informative discussions.
\end{subsection}

\begin{section}*{Appendices}
\renewcommand{\thesubsection}{A}
  \renewcommand{\theequation}{A.\arabic{equation}}
  \setcounter{equation}{0}
\begin{subsection}{$D=11$ supergravity}\label{appendix0}
Following the conventions in \cite{DKN06}, we use a real representation of the Clifford algebra 
\be\label{Clifford}
\{\G^A,\G^B\}&=&\eta^{AB}
\ee
with $\eta=\text{diag}(-+\cdots+)$ and $A,B=0,\dots,10$. The Lagrangian of $D=11$ supergravity \cite{CJS78} in the vielbein frame (\ref{vielbein}) reads modulo higher fermionic terms using Einstein's summation convention\footnote{Note that in most of the text of the paper, we suppress the use of Einstein's summation convention to avoid ambiguities.}
\be\label{Lagr11}
	e^{-1}\mathcal{L} &=& \frac{1}{4}R_{11} -\frac{i}{2}\bar{\psi}^{(11)}_{A}\G^{ABC}\nabla_{B}\psi^{(11)}_{C} -\frac{1}{48}F_{ABCD}F^{ABCD}\nn
\\
	&& -\frac{i}{96}\left( \bar{\psi}^{(11)}_{E}\G^{ABCDEF}\psi^{(11)}_{F} +12\bar{\psi}_{(11)}^{A}\G^{BC}\psi_{(11)}^{D}\right)F_{ABCD} \nn\\
	&& +\frac{2e^{-1}}{12^4}\e^{ABCDEFGHIJK}F_{ABCD}F_{EFGH}A_{IJK}.
\ee
Here $e$ denotes the determinant of the vielbein $e_M^A$ in $11$ dimensions and $A,B$ are `flat' frame indices. We reserve $M,N,\dots$ for $D=11$ coordinate indices. The field strength is $F_{ABCD}=4\nabla_{[A}A_{BCD]}$ with the Levi--Civita connection $\nabla$. The supersymmetry variations are with $\G_M:=e_M^A\G_A$ 
\begin{subequations}\label{Trafo1neu}
\be
	 \delta_{\ep} {e_{M}}^{A} &=& i\bar{\ep}\G^{A}\psi^{(11)}_{M}
	\\
	\delta_{\ep} \psi^{(11)}_{M} &=& \nabla_{M}\ep +\frac{1}{144}\left(\G_{M}{}^{NPQR}-8\delta_{M}^{N}\G^{PQR}\right)\ep F_{NPQR} \\
	\delta_{\ep} A_{MNP} &=& -\frac{3}{2}i\bar{\ep}\G_{[MN}\psi_{P]}^{(11)}.
\ee
\end{subequations}
The equations of motion following from the Lagrangian (\ref{Lagr11}) are as in \cite{DKN06} to lowest order in fermions:
\begin{subequations}\label{eom}
\be
	 R_{AB} &=& \frac13 F_{ACDE}F_{B}{}^{CDE} -\frac1{36}\eta_{AB}F_{CDEF}F^{CDEF}
	\\
		\nabla_A F^{ABCD} &=& \frac{1}{576}\e^{BCDE_1\dots E_4F_1\dots F_4} F_{E_1\dots E_4}F_{F_1\dots F_4} \\
		\G^B\nabla^{(11)}_{[A}\psi_{B]}&=&-\frac{1}{144}\G^B\left(\G_{[A}{}^{CDEF}-8\delta_{[A}^C\G^{DEF}\right)\psi^{(11)}_{B]}F_{CDEF}
	\ee
\end{subequations}
We also use the decomposition of the spin connection $\omega_{ABC}=\omega_{A[BC]}$ in terms of the anholonomy coefficients $\Omega_{ABC}=\Omega_{[AB]C}$ that reads
\be\label{Omega}
\omega_{ABC}&=&\frac12\left(\Omega_{ABC}+\Omega_{CAB}-\Omega_{BCA}\right)\\
\text{with}\quad \Omega_{ABC}&:=&2E_A^ME_B^N\p_{[M}E_{N]C}.
\nn
\ee
\end{subsection}

\renewcommand{\thesubsection}{B}
  \renewcommand{\theequation}{B.\arabic{equation}}
  \setcounter{equation}{0}
\begin{subsection}{General (non-simply laced) case for the one wall limit}\label{appendix1}
Let us indicate what are the modifications to bring to the reasonings of section \ref{one} in the case of a non-simply laced algebra where the various real roots that one might have to consider do not all have the same norm. First, note that, in the context of the gravity-coset correspondence, the Cartan subalgebra is naturally endowed with the symmetric bilinear form $G_{ab}$ associated to Eq. (\ref{supermetric}). The metric $G_{ab}$ (together with its inverse $G^{ab}$) then defines the norm of any root: $(\alpha|\alpha):=\sum_{ab}G^{ab}\alpha_a\alpha_b$. $G^{ab}$ also provides one with the general definition of the co-root $\check{\alpha}\equiv h_\alpha\in \mathfrak{h}$ that is associated to a general (non-null) root $\alpha\in \mathfrak{h}^*$, namely (generalizing the simple root case (\ref{hNorm})):
\be\label{aLcheck}
\check{\alpha}^a&:=&\frac{2}{(\alpha|\alpha)}\alpha^{\# a}
\ee
where $\alpha^{\# a}:=\sum_b G^{ab}\alpha_b$. The definition, and in particular the normalization (with the factor $2/(\alpha|\alpha)$) of $h_\alpha\equiv \check{\alpha}^aH_a\in \mathfrak{h}$ (where $H_a$ is the $\beta$-coordinate basis in $\mathfrak{h}$; $G_{ab}=(H_a|H_b)$) is such that
\be\label{aLh}
\alpha(h_\alpha)&=&\alpha(\check{\alpha})\,\,=\,\,2.
\ee
Kac \cite{K95} has shown that the symmetric bilinear form $(\cdot|\cdot)$ defined on $\mathfrak{h}$ (and $\mathfrak{h}^*$) admits a unique extension to the full Kac--Moody algebra $\mathfrak{g}$ when requiring its invariance ($([x,y]|z)=(x|[y,z])$ for any $x,y,z\in \mathfrak{g}$). In particular, he has shown that two generators $E_\alpha$ (with associated root $\alpha>0$) and $E_{-\alpha}$ (with associated opposite root $-\alpha<0$) satisfy $[E_\alpha,E_{-\alpha}]=(E_{\alpha}|E_{-\alpha})\alpha^{\#}$. In view of this result, it is convenient to normalize the generator $E_\alpha$ (and its Chevalley associated $E_{-\alpha}=-\omega(E_\alpha)$) such that
\be\label{norm2}
(E_\alpha |E_{-\alpha})&:=&\frac{2}{(\alpha|\alpha)}\,\,=:\,\,\sigma_\alpha.
\ee
Indeed, with this normalization the three generators $E_\alpha$, $E_{-\alpha}$ and $h_\alpha\equiv \check{\alpha}\in \mathfrak{h}$ satisfy
\be\label{HDefi2}
h_\alpha &=&[E_\alpha,E_{-\alpha}]
\ee
together with
\be\label{sLb}
\left[h_\alpha,E_{\pm \alpha}\right]&=& \pm \alpha(h_\alpha) E_{\pm\alpha}
\,\,\stackrel{(\ref{aLh})}{=}\,\,\pm 2E_{\pm\alpha}.
\ee
In other words, the normalization (\ref{norm2}) ensures that the three generators $E_\alpha,E_{-\alpha},h_\alpha$ satisfy the commutation relations of a standardly  normalized $sl_2$ algebra. [When $(\alpha|\alpha)=2$, Eq. (\ref{norm2}) reduces to $(E_\alpha |E_{-\alpha})=1$ which is the normalization we had used in section \ref{one}.]\\

Let us now see how the use of the normalization (\ref{norm2}) changes, when $(\alpha|\alpha)\neq 2$, i.e. $\sigma_\alpha\neq 1$, the calculations done in section \ref{one}. The first formul\ae{} of section \ref{one} that are affected by the generalization  allowing for $\sigma_\alpha\neq 1$, are the Lagrangian $\mathcal{L}|_{\alpha}$ (\ref{Lagra2}) and the conjugate momentum $\Pi_\alpha$ (\ref{constmom}), which now read:
\be\label{Lagra3}
\left.\mathcal{L}\right|_{\alpha}&=&  \frac12 \sum\limits_{a,b=1}^{10} G_{ab}\dot{\beta}^a\dot{\beta}^b + \frac{\sigma_\alpha}{4} e^{2\alpha(\beta)}\dot{\nu}_{\alpha}^2
\\
\label{constmom2}
\Pi_\alpha &=& \frac{\sigma_\alpha}{2}e^{2\alpha(\beta)}\dot{\nu}_{\alpha}
\ee
Then, when decomposing the motion of $\beta$ in parallel and orthogonal pieces, we also need to remember that the kinetic energy of the orthogonal part is no longer $\frac14\alpha(\dot{\beta})^2$, but $\frac{\alpha(\dot{\beta})^2}{2(\alpha|\alpha)}=\frac{\sigma_\alpha}{4}\alpha(\dot{\beta})^2$. The conservation of orthogonal energy (\ref{energy2}) gets modified into
\beg
E_{||}&=&\frac{\sigma_\alpha}{4}\alpha(\dot{\beta})^2 + \frac{1}{\sigma_\alpha}e^{-2\alpha(\beta)}\Pi_\alpha^2.
\eeg
Then, the solution (\ref{solution}) takes the form
\beg
e^{\alpha(\beta)}&=&\frac{|\Pi_{\alpha}|}{\sqrt{\sigma_\alpha E_{||}}}\cosh(c_0(t-t_c))\\
\text{with}\quad c_0^2 &=& \frac{4 E_{||}}{\sigma_\alpha},
\eeg
and the angular velocity $\dot{\theta}$ (\ref{theta}) reads
\beg
\dot{\theta}&=&\frac{\Pi_{\alpha}}{\sigma_\alpha}e^{-\alpha(\beta)}
\eeg
while its integral becomes
\beg
\theta(t) &=& \e_{\alpha}\arctan\left(e^{c_0(t-t_c)}\right)+\theta_{-\infty}
\eeg
with the sign $\e_\alpha:=\frac{\Pi_{\alpha}}{|\Pi_{\alpha}|}\in \{\pm 1\}$ as before. As we see, the final result is not affected by this generalization: the kink in $\theta(t)$ is $\pm \frac{\pi}{2}$ independently of the initial data (apart from the sign of $\Pi_\alpha$), and of the norm of $\alpha$. The crucial point is that the `rotation generator' $J_\alpha=E_\alpha-E_{-\alpha}$ entering the final result (\ref{RDefi2}) is normalized in the standard $sl_2$ way, see Eqs. (\ref{HDefi2},\,\ref{sLb}). This is, e.g., what we had used in the explicit $\textit{SL}(2,\R)$ calculation (\ref{RTH}).
\end{subsection}

\renewcommand{\thesubsection}{C}
  \renewcommand{\theequation}{C.\arabic{equation}}
  \setcounter{equation}{0}
\begin{subsection}{Proof of the propositions of section \ref{Weyl}}\label{appendix2}
The factorizability of the vector-spinor action $\cR_{\alpha,\e_{\alpha}}^{\textit{vs}}$ into the tensor product of a Weyl reflection $r_\alpha$ (\ref{Reflection}) and a Dirac spinor action (\ref{ke10ferm4}) will allow us in fact to consider proposition 1 as a corollary of proposition 2. The proof of the latter will be divided into several parts. We start with proving the generalized Coxeter relations:

\begin{subsubsection}*{Lemma 1}
\textit{
The matrices $\cR_{\alpha_i}^{\textit{s}}\in \R^{32\times 32}$ (\ref{SDefi},\,\ref{Ws}), associated to the dominant walls $\alpha_i$ with $i=0,\dots,9$ (\ref{dominant}), fulfill the following generalized Coxeter relations:
\be\label{root}
\big(\cR_{\alpha_i}^{\textit{s}}\big)^4&=&-\id
\ee
If two nodes $\alpha_{i},\alpha_{j}$ in figure 1 are not linked, the associated matrices commute, otherwise these fulfill
\be\label{triple}
\big(\cR_{\alpha_i}^{\textit{s}}\cR_{\alpha_j}^{\textit{s}}\big)^3&=&-\id.
\ee
The ten elements $\cR_{\alpha_i}^{\textit{s}}$ generate a group $\DWeyl$. Note that this implies in particular that the generators $\cR_{\alpha_i,\e_{\alpha_i}}^{\textit{s}}$ (\ref{SDefi},\,\ref{RDefi4}) would have generated the same group for all choices of signs $\e_{\alpha_i}$.
}
\end{subsubsection}

\begin{subsubsection}*{Proof}
We start with the observation that the matrices $\G^{\alpha_{i}}=2J_{\alpha_{i}}^{\textit{s}}$ (\ref{GaW}) with $i=0,\dots,9$ square to the negative identity matrix. This entails the relation (\ref{root}) after an evaluation of a matrix exponential series (equivalent to $e^{i\pi}=-1$) with the unit matrix $\id$:
\beg
\big(\cR_{\alpha_i}^{\textit{s}}\big)^4&\stackrel{(\ref{RDefi4})}{=}& e^{\pi\G^{\alpha_{i}}}\\
&=& -\id
\eeg
It also allows us to rewrite the group elements in a different manner:
\be\label{Rrep}
\cR_{\alpha_i}^{\textit{s}}&=& \frac{1}{\sqrt{2}}\big(\id + \G^{\alpha_{i}}\big).
\ee
Given any two nodes $\alpha_{i},\alpha_{j}$ in figure 1 that are not linked, the identification of $\G^{\alpha_{i}}$ with $\G$-matrices (\ref{GaW}) then immediately implies that the corresponding group elements commute, taking into account the Clifford algebra relation $\{\G^a,\G^b\}=2\delta^{ab}$. For the case of linked nodes $\alpha_{i},\alpha_{j}$, the product $\mathbf{k}$ of the anticommuting matrices $\mathbf{i}:=\G^{\alpha_{i}}$ and $\mathbf{j}:=\G^{\alpha_{j}}$ is a square root of $-\id$ again. $\mathbf{k}:=\mathbf{i}\mathbf{j}$ also anticommutes with $\mathbf{i}$ and $\mathbf{j}$. The elements $\mathbf{i},\mathbf{j},\mathbf{k}$ generate a multiplicative group with 8 elements (namely $\{\pm 1,\pm \mathbf{i},\pm\mathbf{j},\pm\mathbf{k}\}$). This allows for an easy evaluation of the product of the matrices $\cR_{\alpha_i}^{\textit{s}}$ (\ref{Rrep}):
\be\label{root3} 
\cR_{\alpha_i}^{\textit{s}}\cR_{\alpha_j}^{\textit{s}}
&=&\frac{1}{2}\big(\id + \mathbf{i}+\mathbf{j}+\mathbf{k}\big)\nn\\
&=& \frac{1}{2}\big(\id + \sqrt{3}\mathbf{I}\big)\\
\text{with}\quad m\mathbf{i}+n\mathbf{j}+p\mathbf{k}&=:& \sqrt{m^2+n^2+p^2}\mathbf{I}\quad\text{for }m,n,p\in \R,\nn
\ee
which is a standard formula for anticommuting quaternions $\mathbf{i},\mathbf{j},\mathbf{k}$ with $\mathbf{I}^2=-\id$. Since (\ref{root3}) is a third root of $-\id$, the relation (\ref{triple}) follows.\\

Another consequence of the relation (\ref{root}) is that any matrix $\cR_{\alpha_i}^{\textit{s}}$ is an eighth root of unity. This in particular implies that its inverse also is contained in the set of words $\DWeyl$ generated by arbitrary products of the matrices $\cR_{\alpha_i}^{\textit{s}}$. Thus, $\DWeyl$ is endowed with a group structure. Since the inverse of $\cR_{\alpha_i,\e_{\alpha_i}}^{\textit{s}}$ (\ref{RDefi4}) is $\cR_{\alpha_i,-\e_{\alpha_i}}^{\textit{s}}$, the group $\DWeyl$ is not sensitive to the choices of $\e_{\alpha_i}\in \{\pm1\}$.\qed
\end{subsubsection}

\begin{subsubsection}*{Lemma 2}
\textit{
The group $\DWeyl$ generated by the ten matrices $\cR_{\alpha_i}^{\textit{s}}$ (\ref{SDefi},\,\ref{Ws}) is a finite subgroup of $SO(32)$.
}
\end{subsubsection}

\begin{subsubsection}*{Proof} 
First, we recall from \cite{DKN06} that the algebra generators $J_{\alpha_i}^{\textit{s}}$ (\ref{SDefi}) are particular cases of $K(E_{10})$-generators acting unfaithfully on a $32$ dimensional space endowed with an invariant norm $Q^{\textit{s}}$ which reads for any $\ep\in \R^{32}$:
\be\label{quadrDS}
Q^{\textit{s}}(\ep_1,\ep_2)&=& \ep_1^T\ep_2\,\,=\,\,\sum_{A=1}^{32}\ep^A\ep^A.
\ee
This implies $J_{\alpha_i}^{\textit{s}}\in so_{32}$ and for the group generators $\cR_{\alpha_i}^{\textit{s}}=e^{\frac{\pi}{2}J_{\alpha_i}^{\textit{s}}}$ (\ref{Ws}) $\cR_{\alpha_i}^{\textit{s}}\in SO(32)$. Hence, the group $\DWeyl$ is a subgroup of $SO(32)$. We will show next that $\DWeyl$ is finite. Squaring the equation (\ref{Rrep}) results for any $i\in \{0,\dots,9\}$ in the identity
\be\label{square}
\big(\cR_{\alpha_i}^{\textit{s}}\big)^2 &=& \G^{\alpha_i}
\nn\\
&=&\cR_{\alpha_i,\e_{\alpha_i}}^{\textit{s}}\sqrt{2} - \id.
\ee
We have already shown in lemma 1 that two rotations $\cR_{\alpha_i}^{\textit{s}}$ commute, if the associated nodes in the Dynkin diagram of $E_{10}$ in figure 1 are not connected with a line. For all the remaining cases, it is a straightforward calculation with $\G$-matrices to verify the following anticommutation relation:
\be\label{anticom}
\left\{\cR_{\alpha_i}^{\textit{s}},\cR_{\alpha_j}^{\textit{s}}\right\}
&=& \cR_{\alpha_i}^{\textit{s}}\sqrt{2} + \cR_{\alpha_j}^{\textit{s}}\sqrt{2} -\id.
\ee
Hence, any matrix product of generators $\cR_{\alpha_i}^{\textit{s}}$ can be written as a sum of ordered matrix products, for which there only are a finite number of possibilities due to the relation (\ref{square}). Furthermore, the coefficients of the ordered matrix products are contained in the Galois extension $\Z[\sqrt{2}]$ of integers. In other words, the group action in $\DWeyl$ stabilizes on a $\Z[\sqrt{2}]$-lattice, spanned by the ordered products of generators $\cR_{\alpha_i}^{\textit{s}}$. In formul\ae, this means that any word $g\in \DWeyl$ constructed from these generators can be associated to constants $c_*\in\Z[\sqrt{2}]$ such that 
\be\label{lattice}
g&=&c_0\id +\sum\limits_{i=0}^{9}c_i \cR_{\alpha_i}^{\textit{s}} + \sum\limits_{i_1<i_2} c_{i_1i_2} \cR_{\alpha_{i_1}}^{\textit{s}}\cR_{\alpha_{i_2}}^{\textit{s}}
+\dots\nn\\
&&
+\sum\limits_{i_1<\cdots<i_{10}}\!\!\!\!\! c_{i_1\dots i_{10}} \cR_{\alpha_{i_1}}^{\textit{s}}\cdots\cR_{\alpha_{i_{10}}}^{\textit{s}}.
\ee
In order to prove the finiteness of $\DWeyl$, it is hence sufficient to show that the constants $c_*\in\Z[\sqrt{2}]$ may only take a finite number of values. As a next step, we substitute the $\G$-matrices for $\cR_{\alpha_i}^{\textit{s}}$ (\ref{GaW},\,\ref{Rrep}) in this expansion. This implies that a general element $g\in \DWeyl$ has the form
\beg
g&=&\frac{1}{32}\left(d_0 \id+ \sum\limits_{a=1}^{10}d_a \G_a + \sum\limits_{a_1<a_2} \!\!d_{a_1a_2} \G_{a_1a_2}+\dots
+\sum\limits_{a_1<\cdots<a_{10}} \!\!\!\!\! d_{a_1\dots a_{10}} \G_{a_1\dots a_{10}}\right).
\eeg
The important observation is that there is a maximal denominator which is $32=\sqrt{2}^{10}$ and that the coefficients $d_*\in \Z[\sqrt{2}]$ are still Galois extended integers. Since the group $\DWeyl$ is a subgroup of the special orthogonal group $SO(32)$ (\ref{quadrDS}), any $g\in \DWeyl$ has to fulfill the orthogonality constraint $g^Tg=\id$. Taking the trace of this constraint (i.e. considering the part $\propto \id$ in the expansion of $g^Tg$ along the ordered $\G$-basis) leads us to the following diophantine equation:
\be\label{dio1}
32 &=&d_0^2 + \sum\limits_{a=1}^{10}d_a^2  + \sum\limits_{a_1<a_2} d_{a_1a_2}^2 +\dots
+\sum\limits_{a_1<\cdots<a_{10}}d_{a_1\dots a_{10}}^2
\ee
Now, the substitution $d_*=:d'_*+d''_*\sqrt{2}$ of the Galois extended integers $d_*\in \Z[\sqrt{2}]$ for ordinary integers $d'_*,d''_*\in \Z$ produces two equations from (\ref{dio1}):
\be\label{dio2}
32 =\!&&\!\!\!\!\!\!\!\!\!d'_0{}^2 +2d''_0{}^2+ \!\sum\limits_{a=1}^{10}\big(d'_a{}^2 +2d''_a{}^2\big)  +\dots
+\!\!\!\!\!\!\!\!\sum\limits_{a_1<\cdots<a_{10}}\!\!\!\!\!\!\!\!\big(d'{}^2\!\!\!\!_{a_1\dots a_{10}} +2d''{}^2\!\!\!\!\!\!_{a_1\dots a_{10}}\!\big)\\
\nn
0=\!&&\!\!\!\!\!\!\!\!\!d'_0d''_0 + \sum\limits_{a=1}^{10}d'_a d''_a  +\dots
+\sum\limits_{a_1<\cdots<a_{10}}d'_{a_1\dots a_{10}}d''_{a_1\dots a_{10}}
\ee
Since all terms on the RHS of (\ref{dio2}) are positive and the coefficients $d'_*,d_*''$ are integer, the number of solutions of the equation (\ref{dio2}) is finite. As the collective of integers $d'_*,d_*''$ parametrize a general word $g\in\DWeyl$, this proves that $\DWeyl$ is of finite cardinality.\qed
\end{subsubsection}

\begin{subsubsection}*{Lemma 3}
\textit{
The squares of the generating elements $\cR_{\alpha_i}^{\textit{s}}$ (\ref{SDefi},\,\ref{Ws}) of the group $\DWeyl$ for $i=0,\dots,9$ form a normal subgroup $\DWeylN$ of $\DWeyl$:
\be\label{normal}
\DWeylN&:=&\big\langle\big\langle\big(\cR_{\alpha_i}^{\textit{s}}\big)^2\big|i=0,\dots,9\big\rangle\big\rangle\\
\forall g\in \DWeyl, \,\forall h\in \DWeylN && g\cdot h\cdot g^{-1}\in \DWeylN
\label{normal2}
\ee
The subgroup $\DWeylN$ is not abelian and its cardinality is $2048$.
}
\end{subsubsection}

\begin{subsubsection}*{Proof}
Since the ten generating elements $(\cR_{\alpha_i}^{\textit{s}})^2$ (\ref{SDefi},\,\ref{Ws}) for $i=0,\dots,9$ are fourth roots of unity (\ref{root}), their free product indeed forms a subgroup $\DWeylN$ of $\DWeyl$. Furthermore, the relations (\ref{SDefi}) and (\ref{square}) in fact imply that the group $\DWeylN$ (\ref{normal}) is generated by products of the following Clifford matrices:
\beg
H_{DS}&=&\big\langle\big\langle \G^{1\,2},\G^{2\,3},\dots,\G^{9\,10},\G^{1\,2\,3}\big\rangle\big\rangle.
\eeg
A short computation proves that the generating Clifford matrices $\G^a$ for $a=1,\dots,10$ are contained in $\DWeylN$ and hence, we have
\be\label{total}
\DWeylN&=&\left\{\pm\id,\pm\G^a,\pm \G^{a_1a_2},\dots,\pm\G^{a_1\dots a_{10}}|a_i\in \{1,\dots,10\}\right\}.
\ee
Therefore, the cardinality of $\DWeylN$ is $2^{10+1}=2048$ and it obviously is not abelian. To prove that $\DWeylN$ is a normal subgroup, it is sufficient to show that the relation (\ref{normal2}) is fulfilled for the generators $\cR_{\alpha_i}^{\textit{s}}\in  \DWeyl$ (\ref{Rrep}) (with $i=0,\dots,9$) and the generators $h\in\{\pm \id,\pm \G^a\}$ (for $a=1,\dots,10$)
\beg
 \cR_{\alpha_i}^{\textit{s}}\cdot h\cdot \left(\cR_{\alpha_i}^{\textit{s}}\right)^{-1}&\in& \DWeylN,
\eeg
which can be easily checked by a direct computation.\qed
\end{subsubsection}

\begin{subsubsection}*{Lemma 4}
\textit{
There is a homomorphism $\Phi$ between the Weyl group $\mathcal{W}$ of $E_{10}$ and the quotient group $\DWeyl/\DWeylN$:
\be\label{homo}
\Phi :\mathcal{W}&\rightarrow& \DWeyl/\DWeylN
\ee
}
\end{subsubsection}
\vspace{-0.8cm}
\begin{subsubsection}*{Proof}
As shown in lemma 3, $\DWeylN$ is a normal subgroup of $\DWeyl$. Hence, the coset $\DWeyl/\DWeylN$ in (\ref{homo}) is well-defined. Next, we define the mapping $\Phi$ (\ref{homo}) by linking the fundamental reflections $r_{\alpha_i}\in \mathcal{W}$ to the $\DWeylN$-equivalence classes $[\cR_{\alpha_i}^{\textit{s}}]$ of generating elements of $\DWeyl$:
\be\label{homo3}
\Phi :\mathcal{W}&\rightarrow& \DWeyl/\DWeylN\\
r_{\alpha_i}&\mapsto &[\cR_{\alpha_i}^{\textit{s}}]\,\,=:\,\,\bar{r}_{\alpha_i}
\nn
\ee
The mapping $\Phi$ can then be extended to the entire Weyl group $\mathcal{W}$ by the usual relation for any two fundamental reflections $r_{\alpha_i}, r_{\alpha_j}\in \mathcal{W}$ with $i,j\in \{0,\dots,9\}$
\beg
\Phi(r_{\alpha_i}r_{\alpha_j}) &:=&\Phi(r_{\alpha_i})\Phi(r_{\alpha_j}).
 \eeg
To complete the proof that $\Phi$ (\ref{homo3}) defines a homomorphism, one has to check that the identity $1\in \mathcal{W}$ can be mapped to the $\DWeylN$-equivalence class $[\id]$ of the identity in $\DWeyl$ in a consistent way. This is equivalent to checking that the Coxeter relations that the Weyl group $\mathcal{W}$ is subject to and that are encoded in the Dynkin diagram of $E_{10}$ (figure 1), are preserved by the mapping $\Phi$:
\begin{subequations}\label{properties}
\be
\bar{r}_{\alpha_i}^2&\stackrel{!}{=}&[\id],\label{properties1}\\
\big(\bar{r}_{\alpha_i}\bar{r}_{\alpha_j})^3&\stackrel{!}{=}&[\id] \quad \text{for adjacent nodes }\alpha_i,\alpha_j\text{ and}\\
\big(\bar{r}_{\alpha_i}\bar{r}_{\alpha_j})^2&\stackrel{!}{=}&[\id] \quad \text{for non-adjacent nodes }\alpha_i,\alpha_j.
\ee
\end{subequations}
The first equation is an immediate consequence of the definition of the normal subgroup $\DWeylN$, the second one follows from the relations (\ref{triple},\,\ref{total}) and the third one can be deduced from the first one (\ref{properties1}) taking into account that the matrices $\bar{r}_{\alpha_i}$ and $\bar{r}_{\alpha_j}$ commute for the case of non-adjacent nodes as shown in lemma 1.\qed\\
\end{subsubsection}

The proof of theorem 2 is then completed by combining the four lemmas: Since the group $\DWeyl$ is finite (lemma 2), the cardinality of the quotient group $\DWeyl/\DWeylN$ also is finite. Combining this observation with the infinite cardinality of the Weyl group $\mathcal{W}$ of $E_{10}$, any homomorphism $\Phi$ (\ref{homo}, lemma 4) must have a non-trivial kernel, corresponding to a normal subgroup of the Weyl group $\mathcal{W}$ of $E_{10}$.\qed\\

As mentioned above, the proof of proposition 1 follows from a combination of proposition 2 and the factorization of the vector-spinor action $\cR_{\alpha}^{\textit{vs}}$ into the tensor product of a Weyl reflection $r_\alpha$ (\ref{Reflection}) (acting on the vector part of the gravitino $\phi^a$) and a Dirac spinor action $\cR_{\alpha_i}^{\textit{s}}$ (acting on its spinor part) as exhibited in Eq. (\ref{ke10ferm4}). In formul\ae, for all dominant walls $\alpha_i$ (\ref{dominant}) with $r_{\alpha_i}\in \mathcal{W}$ and $\cR_{\alpha_i}^{\textit{s}}\in \DWeyl$ we have
\be\label{prod}
\cR_{\alpha_i}^{\textit{vs}}&=& r_{\alpha_i}\otimes\cR_{\alpha_i}^{\textit{s}}.
\ee
The group structure of $\SWeyl$ is then inherited from the group structure of the Weyl group $\mathcal{W}$ and of the Dirac-spinor group $\DWeyl$. Since the fundamental reflections $r_{\alpha_i}$ of the Weyl group $ \mathcal{W}$ are subject to the Coxeter relation $r_{\alpha_i}^2=1$, the squares of the generators of $\SWeyl$ have the form
\beg
(\cR_{\alpha_i,\e_{\alpha_i}}^{\textit{vs}})^2&=& 1\otimes (\cR_{\alpha_i,\e_{\alpha_i}}^{\textit{s}})^2.
\eeg
This immediately implies that these matrices generate a normal subgroup $\SWeylN$ of $\SWeyl$ that is isomorphic to $\DWeylN$ and hence shares all its properties. The generalized Coxeter relations stated in proposition 1 then also directly follow from the tensor product structure (keeping in mind lemma 1 and the standard Coxeter relations for the Weyl group $\mathcal{W}$ of $E_{10}$ (\ref{properties})). Furthermore, a homomorphism $\Phi$ linking the Weyl group $\mathcal{W}$ of $E_{10}$ to the quotient group $\SWeyl/\SWeylN$ can be constructed in a similar way as in (\ref{homo3}). Hence, we are left with proving that the homomorphism $\Phi$ linking the Weyl group $\mathcal{W}$ of $E_{10}$ to the quotient group $\SWeyl/\SWeylN$ is an isomorphism (\ref{homo4}):
\beg
\Phi :\mathcal{W}&\rightarrow& \SWeyl/\SWeylN.
\eeg
The tensor product structure of $\cR_{\alpha_i}^{\textit{vs}}=r_{\alpha_i}\otimes\cR_{\alpha_i,\e_{\alpha_i}}^{\textit{s}}$ (\ref{prod}) together with the isomorphism of the quotient groups $\SWeylN\simeq \DWeylN$ and the Dirac-spinor mapping (\ref{homo3}) implies that a general word $w$ (with $w=r_{\alpha_1}r_{\alpha_2}\cdots r_{\alpha_n}$) in the Weyl group $\mathcal{W}$ is mapped by $\Phi$ to:
\be\label{homo5}
\Phi :\mathcal{W}&\rightarrow& \SWeyl/\SWeylN\nn\\
w& \mapsto & w\otimes\bar{r}
\ee
The issue is now to know whether one can \textit{uniquely} reconstruct the pre-image $w\in \mathcal{W}$, when one is given some equivalence class of matrices in the tensor product $\SWeyl$. Clearly given a matrix $\cR^a{}_b{}^A{}_B$ (\ref{ke10ferm4}), it determines the factor $w^a{}_b$ it comes from modulo a (real) factor, say $\lambda w^a{}_b$. But we also know that $w\in \mathcal{W}$ is an \textit{orthochronous Lorentz transformation} which preserves the Lorentzian norm $G_{ab}\dot{\beta}^a\dot{\beta}^b$. This implies that there is a unique $w$ within a ``line'' $\lambda w$. [Even $\lambda=-1$ is impossible, because $\mathcal{W}$ preserves the future light cone in $\beta$-space.] This completes the proof of the propositions of section \ref{Weyl}.\qed
\end{subsection}

\end{section}

\end{document}